\documentclass[aps,prb,twocolumn,nofootinbib,longbibliography,citeautoscript,superscriptaddress]{revtex4-2}
\usepackage{amsfonts}
\usepackage{amscd}
\usepackage{amsmath}
\usepackage{amssymb}
\usepackage{times}
\usepackage{here}
\usepackage[dvipdfmx]{graphicx}
\usepackage{dcolumn}
\usepackage{bm}
\usepackage{xcolor}
\usepackage[%
  colorlinks=true,
  urlcolor=blue,
  linkcolor=blue,
  citecolor=blue
]{hyperref}
\usepackage{tabularx}
\usepackage{etoolbox}
\usepackage{braket}
\usepackage[caption=false]{subfig}
\newcommand{\nn}{\nonumber}
\newcommand{\beq}{\begin{equation}}
\newcommand{\eeq}{\end{equation}}

\makeatletter
\let\cat@comma@active\@empty
\makeatother

\allowdisplaybreaks

\usepackage[normalem]{ulem}

\renewcommand\sout{\bgroup \color{blue} \ULdepth=-.5ex \ULset}

\usepackage{hyperref}
\begin{document}

\title{
Collective modes in Fulde-Ferrell-Larkin-Ovchinnikov superconductors: The role of long-range Coulomb interaction and signatures in density response}

\author{Ryoga Miwa}
\email{miwa@blade.mp.es.osaka-u.ac.jp}
\affiliation{Department of Materials Engineering Science, Osaka University, Toyonaka, Osaka 560-8531, Japan}
\author{Ryoi Ohashi}
\affiliation{Department of Materials Engineering Science, Osaka University, Toyonaka, Osaka 560-8531, Japan}
\author{Satoshi Fujimoto}
\affiliation{Department of Materials Engineering Science, Osaka University, Toyonaka, Osaka 560-8531, Japan}
\affiliation{Center for Quantum Information and Quantum Biology, Osaka University, Toyonaka 560-8531, Japan}
\author{Takeshi Mizushima}
\email{mizushima@mp.es.osaka-u.ac.jp}
\affiliation{Department of Materials Engineering Science, Osaka University, Toyonaka, Osaka 560-8531, Japan}
\date{\today}

\begin{abstract}
We theoretically investigate collective excitations in the Fulde-Ferrell-Larkin-Ovchinnikov (FFLO) states of Pauli-limited superconducting films. When the long-range Coulomb interaction is absent,  excitation spectra consist of two gapless and three gapped modes. The gapless modes are the Nambu-Goldstone modes associated with the spontaneous breaking of the ${\rm U}(1)$ symmetry and the translational symmetry. The gapped modes include the Higgs mode and the twofold degenerate modes that cause the oscillation of the domain width and grayness of FFLO nodal planes. We find that the long-range Coulomb interaction only gaps out the gapless phase mode through the Anderson-Higgs mechanism, while the other modes remain unaffected. Furthermore, the field evolution of the dispersion of the gapless elastic mode, the Nambu-Goldstone mode associated with the translational symmetry breaking, is associated with that of the bandwidth of the mid-gap Andreev bound states. We demonstrate that the signature of the elastic mode can be detected by measuring the density-density response function. 
\end{abstract} 

\maketitle

\section{Introduction}

In 1964, Fulde and Ferrell~\cite{ff} and Larkin and Ovchinnikov~\cite{lo} independently predicted that the competition between the condensation energy and paramagnetic depairing effect due to the Zeeman splitting of electron bands leads to the formation of a nonuniform superconducting order. Among various possibilities, the Larkin-Ovchinnikov state, with a periodic modulation of the order parameter amplitude in space, is most stable in Pauli-limited superconductors, which is referred to as the Fulde-Ferrell-Larkin-Ovchinnikov (FFLO) state for simplicity.
In such a state, the Cooper pairs formed by electrons on mismatched Fermi surfaces possess a nonzero center-of-mass momentum. The realization requires sufficiently weak orbital depairing effect and disorders. Organic and heavy fermion superconductors have been extensively studied as potential platforms to satisfy these conditions. The former provides low dimensional systems with sufficiently suppressed orbital effect, while the latter has very large orbital limiting fields~\cite{mat07,bey13,wos18}. Since the seminal works in 1964, the study of the FFLO phase has expanded beyond superconducting materials to encompass various physics fields, including ultracold atomic gases~\cite{miz05,kin18}, nuclear matter~\cite{sed19}, and quantum chromodynamics~\cite{cas04}. 

Experimental signatures of the FFLO state have recently been reported in many superconducting materials, including Sr$_2$RuO$_4$~\cite{kin22}, KFe$_2$As$_{2}$~\cite{cho17}, FeSe~\cite{kas20,kas21,shi20}, CeCoIn$_5$~\cite{lin20,kit23}, CeCu$_2$Si$_2$~\cite{kit18}, NbS$_2$~\cite{cho21} and organic superconductors~\cite{sug19,ima22,kot22,mol24}. Moreover, the more exotic type of finite-momentum Cooper pairs has been observed in Ba$_6$Nb$_{11}$S$_{28}$~\cite{dev21} and transition metal dichalcogenides~\cite{soh18,wan23,zha23}. The FFLO state is characterized by a spatially modulated superconducting order, which consists of a periodic array of a superconductor/normal/superconductor junction with $\pi$ phase shift. Thus, the FFLO state is accompanied by the mid-gap Andreev bound states (ABSs) at each nodal plane. The existence of the ABSs leads to the significant enhancement of the quasiparticle density of states in low energies, the accommodation of excess spins, and paramagnetic response~\cite{mac84,vor05,ich07,suz11,ros16,miz18,suz20}. Hence, a signature of the FFLO order can be captured by measuring the field evolution of specific heat, magnetization, and the Knight shift in nuclear magnetic resonance measurements.

Collective modes provide an alternative signature of the FFLO order. These modes are the centerpieces of dynamics and transport phenomena, as they involve the coherent motion of a macroscopic fraction of particles with a long lifetime. A successful example is the superfluid $^3$He. A rich spectrum of collective modes reflects the breaking of ${\rm SO}(3)\times {\rm SO}(3)\times {\rm U}(1)$ symmetry by spin-triplet $p$-wave Cooper pairs~\cite{leg66,mak74,nag75,tew75,mak76,wol76,tew79}. For instance, the B phase spontaneously breaks the spin-orbit symmetry, and the collective excitations involve massive bosonic modes with a total angular momentum of Cooper pairs, $J=2$, which have been observed through acoustic probes~\cite{gia80,mas80,ave80,mov88,lee99,dav06,dav08,zav16}. Quantitative studies on mass shifts of the $J=2$ modes have unveiled spontaneous symmetry breaking in the superfluid vacuum~\cite{koc81,sau00,sau17}. These observations motivate exploring collective modes in FFLO superconductors that can serve as a direct probe for the FFLO order.

In this paper, we elucidate the collective mode spectra in FFLO superconductors. Collective modes in the FFLO state have been studied in neutral fermionic superfluids with population imbalance~\cite{rad09,edg09,edg10,rad11,koi11,hei11,rad12,dut17,boy17,agt20}. Recently, a microscopic calculation unveiled the existence of the gapless amplitude mode in the FFLO superconductor~\cite{hua22}, which is the Nambu-Goldstone (NG) mode associated with the translational symmetry breaking~\cite{sam10,sam11}. Higgs modes in pair-density-wave superconductors have also been discussed in Refs.~\cite{sot17,jia20}. However, the effect of the long-range Coulomb interaction and the contribution to observable quantities are yet to be explored. 

To understand the impact of Coulomb interaction and the signature of the collective modes, we here consider a simple model of Pauli-limited superconductors that consists of electrons with a one-dimensional Fermi surface, the BCS and long-range Coulomb interactions, and the Zeeman term. Using a microscopic theory, we unveil that when the Coulomb interaction is absent, the collective excitation spectra in the FFLO state consist of two gapless modes and three gapped modes. The gapless modes involve the phase oscillation and the elastic oscillation of the FFLO nodal plane, which are the NG modes associated with the spontaneous breaking of the ${\rm U}(1)$ symmetry and the translational symmetry, respectively. The gapped modes include the twofold degenerate modes that cause oscillation of the FFLO domain width and the {\it grayness} of the nodal plane, in addition to the Higgs mode. In one-dimensional Pauli-limited superconductors, the mean-field theory is mapped onto the nonlinear Schr\"{o}dinger equation for the superconducting order~\cite{bas1,bas2,yos11,dat12,dat13}, and the FFLO state is a soliton solution, that is, a periodic array of black solitons that have a zero of the order parameter amplitude with a $\pi$-phase shift across its zero. The {\it grayness} or {\it darkness} indicates the deviation of the phase shift from $\pi$, which fills in the zeros of the soliton chain~\cite{fra10}. Furthermore, we find that the long-range Coulomb interaction gaps out the phase mode through the Anderson-Higgs mechanism, while the other modes remain unchanged. Furthermore, the dispersion of the gapless elastic mode is related to the bandwidth of the mid-gap ABSs and becomes dispersionless as the magnetic field approaches the BCS-FFLO critical field. We discuss that the signatures of the collective modes can be captured by measuring the density-density response function. 

The organization of this paper is as follows. In Sec.~\ref{sec:theory}, we start with the microscopic BCS action and derive the effective action for the fluctuations of the superconducting order and the charge density using the functional integral formalism. Here, we also summarize the equilibrium properties of the FFLO states. In Sec.~\ref{sec:CM}, we present the whole dispersion of the collective modes in the FFLO state. We examine the impact of the long-range Coulomb interaction and the field evolution of the collective mode spectra. The contribution of the collective modes to density response is discussed in Sec.~\ref{sec:dns}. Section~\ref{sec:summary} is devoted to the conclusion and discussion. In Appendices~\ref{sec:gauge} and \ref{sec:chi}, we describe the symmetry of the quasiparticle dispersion in the Brillouin zone folded by the FFLO superlattice and the detailed expression of the bare correlation functions, respectively.

\section{Functional integral formalism}
\label{sec:theory}

In this section, we present the effective action for the bosonic fluctuations in superconductors and the equilibrium properties of the FFLO state based on the saddle-point action. We take into account the fluctuations of the superconducting order and the charge density, with the latter arising from the long-range Coulomb interaction. In this paper, we set $\hbar=k_{\rm B}=1$.

\subsection{Effective action}

We start with the BCS action of interacting electrons in the presence of a magnetic Zeeman field
\begin{align}
\mathcal{S}[\bar{\psi},\psi]=& \sum_{\sigma}\int dx_i\int dx_j \bar{\psi}_{\sigma}(x_i)\left( \delta_{ij}\partial _{\tau}+\xi_{ij} \right) \psi _{\sigma}(x_j) \nn \\
&-V\int dx_i \bar{\psi}_{\uparrow} (x_i)\bar{\psi}_{\downarrow}(x_i) 
\psi _{\downarrow}(x_i) \psi _{\uparrow}(x_i) \nn \\
&+\frac{1}{2}\int dx_i\int dx_j \delta n(x_i) U(x_i-x_j)\delta n(x_j),
\label{eq:action}
\end{align}
where $\psi_{\sigma}$ is the electron operator with spin $\sigma=\uparrow,\downarrow$. 
The $2\times 2$ matrix in the spin space, $\xi$, is the single-particle Hamiltonian density and includes the magnetic Zeeman energy.
In the expression, we have introduced the short-hand notation $x\equiv ({\bm x},\tau)$, where $\tau$ is the imaginary time. In the last term, $\delta n(x)= \sum_{\sigma=\uparrow,\downarrow}\bar{\psi}_{\sigma}(x)\psi_{\sigma}(x) - n_0$ is the density fluctuation operator from the background charge density $n_0$.
We consider electrons interact through the $s$-wave attractive potential with strength $V>0$ and the Coulomb interaction potential, $U(x_i-x_j)=U({\bm x}_{ij})\delta(\tau_{ij})$ with 
$U({\bm x}_{ij})=e^2/|{\bm x}_i-{\bm x}_j|$.

We then perform the Hubbard-Stratonovich transformation, introducing the auxiliary bosonic fields $\Delta(x)$ and $\phi(x)$, where $\Delta(x)$ is the complex field representing the superconducting order, and $\phi(x)$ is the internal electric potential produced by electrons. Integrating out the fermion fields, the effective action for the bosonic fields reads
\beq
{\mathcal{S}} = -{\rm Tr} \log \left(-G^{-1}\right)
+ \sum _Q \frac{|\Delta_Q|^2}{V}+\sum_Q\frac{\phi_Q\phi_{-Q}}{U_Q}.
\label{eq:action2}
\eeq
Here we introduce the Fourier transform of the superconducting order ($\Delta_Q$), the charge density ($\phi_Q$), and the Coulomb potential ($U_Q$), and $Q\equiv ({\bm q},i\omega_m)$ is the four-momentum, where $\omega_m = 2m\pi T$ ($m\in\mathbb{Z}$) is the bosonic Matsubara frequency and $\sum_{Q}=T\sum_m\sum_{\bm q}$. The inverse matrix of the Green's function is defined in the particle-hole space as 
\begin{align}
G^{-1}_{ij}
&= -\left\{ \delta_{ij}\partial _{\tau_i} +
\begin{pmatrix}
{\xi}_{ij} & i\sigma_y\Delta(x_i)\delta_{ij} \\ 
-i\sigma_y\bar{\Delta}(x_i)\delta_{ij}  & - {\xi}^{\ast}_{ij}
\end{pmatrix}
\right\},
\end{align}
where $\sigma_{x,y,z}$ are Pauli matrices in the spin space.
In this paper, we focus on spin-singlet $s$-wave superconducting states.

We consider the small fluctuations of bosonic fields about the thermal equilibrium. Let $\Delta_{\rm eq}({\bm x}_i)$ be the saddle-point value of the superconducting order at equilibrium. Collective modes in superconductors comprise the amplitude and phase fluctuations of the superconducting order. Then, we define the space-time fluctuations of the $s$-wave superconducting order around $\Delta_{\rm eq}$ as $\delta \Delta(x) = \Delta(x) - \Delta_{\rm eq}({\bm x})$. Then, the self-energy fluctuation is given by
\begin{align}
\delta\Sigma(x) = \frac{1}{2}\tau_x\delta\Delta^+(x) + \frac{1}{2}i\tau_y\delta\Delta^-(x) + ie\phi\tau_z,
\end{align}
where $\tau_{x,y,z}$ are the Pauli matrices in the particle-hole space and $\delta \Delta^{+}\equiv \delta\Delta + \delta\Delta^{\ast}$ and $\delta \Delta^{-}\equiv \delta\Delta - \delta\Delta^{\ast}$ correspond to the fluctuations of the real part (amplitude) and imaginary part (phase) of the gap function, respectively.
The exact Green's functions can be obtained by means of the Dyson equation,
$G^{-1}=G^{-1}_{\rm eq}-\delta\Sigma$, with the equilibrium Green's function, $G_{\rm eq}$.
Performing an expansion in $\mathcal{S}$ to the quadratic order in the bosonic fields ($\delta\Delta^{\pm}$ and $\phi$) leads to
$\mathcal{S}_{\rm eff} = \mathcal{S}_{\rm eq} + \mathcal{S}_{\rm fluc}$. 
The saddle-point action reads 
\beq
{\mathcal{S}}_{\rm eq} =  \int d{\bm x}\frac{|\Delta(x)|^2}{V}-\frac{1}{2}{\rm Tr}\log \left(- { G}^{-1}_{\rm eq}\right).
\label{eq:Seq}
\eeq
The action for the bosonic fluctuations is given by 
\begin{align}
\mathcal{S}_{\rm fluc} =& \frac{1}{4} {\rm Tr}\left[ G_{\rm eq}\delta\Sigma G_{\rm eq}\delta\Sigma\right]\nn \\
&+ \sum_Q\sum _{s=\pm} \frac{s\delta\Delta^{s}_Q\delta\Delta^{s}_{-Q}}{4V}+\sum_Q\frac{\phi_{Q}\phi_{-Q}}{U_Q}.
\label{eq:Sf}
\end{align}
The Gaussian fluctuation approximation incorporates the coupling of fluctuations of superconducting order to the long-range Coulomb potential and describes all branches of the collective excitations in the FFLO state. In Sec.~\ref{sec:CM}, we obtain the collective mode spectra by computing the spectral functions of the fluctuation propagators. 

\subsection{Saddle-point action}

Before presenting numerical results on collective excitations, we briefly summarize the equilibrium properties of the FFLO state in Pauli-limited superconductors. 
To compute the saddle-point action in Eq.~\eqref{eq:Seq}, which determines the thermodynamically stable structure of the FFLO state, we start to introduce the Bogoliubov-de Gennes (BdG) Hamiltonian in the basis of $[\psi_{\uparrow}({\bm x}_i),\psi_{\downarrow}({\bm x}_i),\bar{\psi}_{\uparrow}({\bm x}_i),\bar{\psi}_{\downarrow}({\bm x}_i)]$ as 
\beq
\mathcal{H}_{ij}
\equiv 
\begin{pmatrix}
{\xi}_{ij} & i\sigma_y{\Delta}_{\rm eq}(x_i)\\ 
-i\sigma_y{\Delta}_{\rm eq}(x_i) & -{\xi}_{ij}
\end{pmatrix}.
\label{eq:Heq}
\eeq
The gap function in the equilibrium is assumed to be invariant under the time-reversal symmetry, and thus $\Delta_{\rm eq}\in\mathbb{R}$. The normal-state Hamiltonian density is given by 
\beq
{\xi}_{ij}= \xi^{(0)}_{ij} + \mu_0 B \sigma_z,
\label{eq:xi}
\eeq
where $\xi^{(0)}_{ij}$ is the single-particle Hamiltonian at $B=0$, and the magnetic field is assumed to be applied along the $z$-axis. The magnetic moment of electrons is denoted by $\mu_0$. We also impose the charge neutrality condition that electron charges are balanced by background positive charges in the equilibrium, resulting in $\phi_{\rm eq}=0$.

The BdG Hamiltonian maintains the spin rotation symmetry about the $z$-axis, $[\mathcal{H},{S}_z]=0$, where $S_z\equiv\frac{1}{2}\sigma_z\tau_z$ is the spin of quasiparticles. Hence, the BdG Hamiltonian can be block-diagonalized in terms of the eigenstates with $S_z = \pm \frac{1}{2}$ as $\mathcal{H} = \mathcal{H}^{+}\oplus\mathcal{H}^{-}$. The BdG Hamiltonian in each sector is expressed as $\mathcal{H}^{+} = \mathcal{H}_0+\mu_0 B$ and $\mathcal{H}^{-} = \tau_z\mathcal{H}_0\tau_z-\mu_0 B$, where $\mathcal{H}_0$ is the BdG Hamiltonian in the basis of $(\psi_{\uparrow},\bar{\psi}_{\downarrow})$ at $B=0$. Let $E_{\alpha}$ and ${\bm \varphi}_{\alpha,i}\equiv {\bm \varphi}_{\alpha}({\bm x}_i)$ be the eigenvalues and eigenfunctions of $\mathcal{H}_0$, respectively, which satisfies the BdG equation,
\beq
\sum_j\mathcal{H}_{0,ij}{\bm \varphi}_{\alpha,j} = E_{\alpha}{\bm \varphi}_{\alpha,i}.
\label{eq:bdg0}
\eeq
Then, the equilibrium Green's functions in each spin sector, $G^{\pm}_{ij}\equiv G^{\pm}({\bm x}_i,{\bm x}_j,i\varepsilon_n)$, are expressed in terms of the eigenstates and eigenfunctions as 
\begin{align}
&G^{+}_{ij}
= \sum_{\alpha}{}^{\prime} \bigg[
\frac{{\bm \varphi}_{\alpha,i}{\bm \varphi}^{\dag}_{\alpha,j}}{i\varepsilon_n-E_{\alpha,+}}
+\frac{\tilde{\mathcal{C}}{\bm \varphi}_{\alpha,i}{\bm \varphi}^{\dag}_{\alpha,j}\tilde{\mathcal{C}}^{-1}}{i\varepsilon_n+{E_{\alpha,-}}}
\bigg],  
\label{eq:G+} \\
&G^{-}_{ij}
=\sum_{\alpha}{}^{\prime} \tau_z\bigg[
\frac{{\bm \varphi}_{\alpha,i}{\bm \varphi}^{\dag}_{\alpha,j}}{i\varepsilon_n-E_{\alpha,-}}
+\frac{\tilde{\mathcal{C}}{\bm \varphi}_{\alpha,i}{\bm \varphi}^{\dag}_{\alpha,j}\tilde{\mathcal{C}}^{-1}}{i\varepsilon_n+{E_{\alpha,+}}}
\bigg]\tau_z.
\label{eq:G-}
\end{align}
where $\varepsilon_n=(2n+1)\pi T$ is the fermionic Matsubara frequency ($n\in\mathbb{Z}$) and $\sum_{\alpha}{}^{\prime}$ implies the sum over $\alpha$ that satisfies $E_{\alpha}>0$. 
The quasiparticle energies at $B$, $E_{\alpha,\pm} = E_{\alpha}\pm \mu_0 B$, are the eigenvalues of $\mathcal{H}^{\pm}$. We have also introduced the operator $\tilde{\mathcal{C}}=i\tau_yK$ that exchanges the particle and hole components of the quasiparticle wavefunction in each spin sector, where $K$ is the complex conjugation operator.

The equilibrium pair potential is obtained from the stationary condition of the saddle-point action, $\delta {\mathcal{S}}_{\rm eq}/\delta \Delta^{\ast}=0$. This reduces to the gap equation
\beq
\Delta({\bm x}_i) = \frac{V}{2}\sum_{\sigma=\pm}\sum_{\alpha}{}^{\prime}u_{\alpha,i}v^{\ast}_{\alpha,i}
\tanh\left(
\frac{E_{\alpha,\sigma}}{2T}
\right),
\label{eq:gap0}
\eeq
where $u_{\alpha,i}$ and $v_{\alpha,i}$ are the particle and hole components of the quasiparticle wavefunction ${\bm \varphi}_{\alpha,i}$, respectively. In numerical calculations, we determine the coupling strength $V$ by solving the gap equation at $T=0$ and $B=0$,
\beq
\frac{1}{V} = \sum_{\bm p}\frac{\tanh(E_{\bm p}/2T)}{2E_{\bm p}},
\eeq
with a given value of the pair potential of a spatially uniform (BCS) state, $\Delta_0$.

In the FFLO state, the equilibrium gap function modulates with the period $L$. We take the $x$-axis along the modulation direction and assume spatial uniformity along the other directions. Then, we impose the following periodic boundary condition on the gap function 
\beq
\Delta_{\rm eq}(x+L/2) = e^{i\chi_0}\Delta_{\rm eq}(x),
\label{eq:delta_periodic2}
\eeq
where $\chi_0=\pi$ in the FFLO state and $\chi_0=0$ in the BCS state. The quasiparticle wavefunction is factorized as 
\beq
\bm{\varphi}_{\alpha}({\bm x}) =\frac{1}{\sqrt{2N_{\rm cell}}} e^{i{\bm K}\cdot {\bm x}} \bm{w}_{\alpha,{\bm K}}({x}).
\label{eq:bloch4}
\eeq
We impose the Born-von Karman boundary condition with the period $N_{\rm cell}L$ on the quasiparticle wavefunctions as 
${\bm{\varphi}}_{\alpha}(x+LN_{\rm cell},y,z)
={\bm{\varphi}}_{\alpha}({\bm x})$,
where $N_{\rm cell}\in \mathbb{Z}$. Then, the Bloch vector ${\bm K}\equiv (K_x,{\bm k}_{\parallel})$ is expressed as $K_x=2\pi m/LN_{\rm cell}$ ($m = - N_{\rm cell}, \cdots, N_{\rm cell}-1$) and ${\bm k}_{\parallel}\equiv(k_y,k_z) = \frac{2\pi}{a}(n_y,n_z)$ ($n_{y,z}\in\mathbb{Z}$), where $a$ is the lattice constant. The Bloch wave function obeys the periodic boundary condition,
\beq
\bm{w}_{\alpha,{\bm K}}(x+L/2) = U^{\dag}_0\bm{w}_{\alpha,{\bm K}}({x}).
\label{eq:w_bc}
\eeq
and the normalization condition, 
\beq
\int^{L/2}_0  |\bm{w}_{\alpha,{\bm K}}({ x})|^2dx = 1.
\label{eq:normw}
\eeq
The unitariy matrix $U_0$ represents the gauge transformation to compensate the phase factor in Eq.~\eqref{eq:delta_periodic2}, 
\beq
U_0\equiv e^{i\vartheta}
\begin{pmatrix}
e^{-i\chi_0/2} & 0 \\ 
0 & e^{i\chi_0/2}
\end{pmatrix}.
\label{eq:u0}
\eeq
Although the choice of the ${\rm U}(1)$ phase ($\vartheta$) does not change the self-consistent fields, thermodynamic properties, and the collective modes in the FFLO state, it directly affects the symmetry of the quasiparticle dispersion in the Brillouin zone folded by the FFLO superlattice. Below, we choose the gauge of $\vartheta=0$ (see Appendix~\ref{sec:gauge}).

\subsection{FFLO states in equilibrium}
\label{sec:equilibrium}

\begin{figure*}[t!]
\includegraphics[width=180mm]{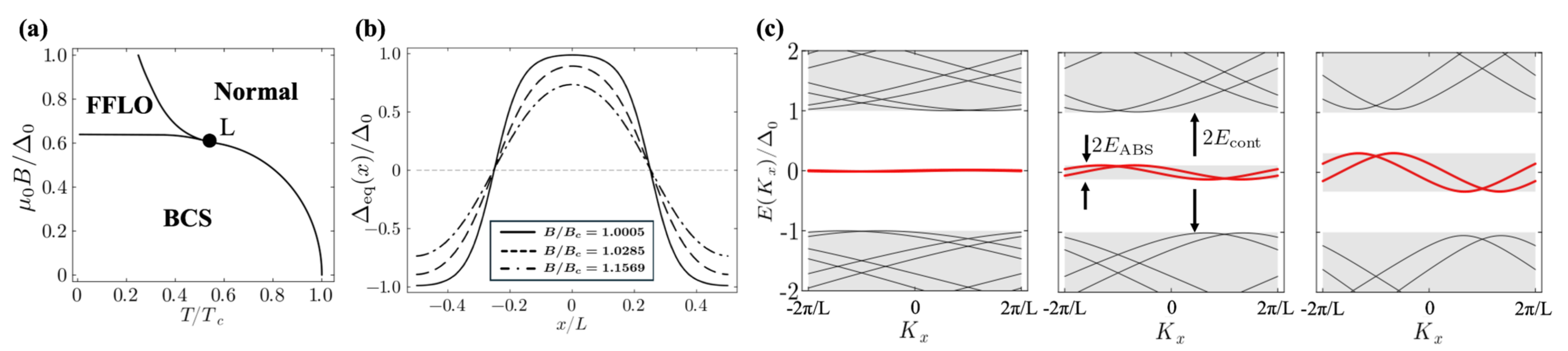}
\caption{(a) Phase diagram in a $B$-$T$ plane for Pauli-limited superconductors, where $\Delta_0$ is the superconducting gap in the BCS state at $T=B=0$. Three phase boundaries meet at the Lifshitz point ``L''. (b) Spatial profiles of $\Delta(x_i)$ in FFLO states for $B/B_{\rm c}= 1.001$ (solid line), $1.029$ (dashed line), and $1.157$ (dashed-dotted line) at $T/T_{\rm c}= 0.04$, where $\mu_0 B_{\rm c}/\Delta_0=2/\pi$ is the BCS-FFLO critical field. (c) Eigenvalues of the BdG equation \eqref{eq:bdg0} in the folded Brillouin zone, $E(K_x)$, which are related to the energies of quasiparticles with $S_z=\pm \frac{1}{2}$ as $E_{\pm}(K_x)=E(K_x)\pm \mu_0 B$. The applied magnetic fields are $B/B_{\rm c}= 1.001$ (left), $1.029$ (center), and  $1.157$ (right). The bandwidth of the Andreev bound states is denoted by $2E_{\rm ABS}$.}
\label{fig:bdg}
\end{figure*}

The equilibrium properties are determined by self-consistently solving the BdG equation \eqref{eq:bdg0} and the gap equation~\eqref{eq:gap0} in the supercell of the FFLO state, $x\in[-L/4,L/4]$, with the Bloch wavefunction in Eq.~\eqref{eq:bloch4}. The single-particle Hamiltonian density in the normal state is expressed in the momentum representation as $\xi^{(0)}_{\bm k}=-2t_{x}\cos(k_{x}a)-2t_{y}\cos(k_{y}a)-2t_{z}\cos(k_{z}a) - \mu$. To elucidate the collective modes in the FFLO state, we focus on systems with a one-dimensional Fermi surface, i.e., $t_y=t_z=0$. For numerical calculations, we fix the chemical potential to $\mu/t = -1$ and the temperature $T/T_{\rm c}=0.04$. The pair interaction $V$ is obtained by solving Eq.~\eqref{eq:gap0} with $\Delta_0/t=0.1$ in the BCS state. With these parameters, the coherence length is estimated as $\xi_0\equiv \hbar v_{\rm F}/\Delta_0\approx 17$ with the Fermi velocity $v_{\rm F}=\sqrt{3}ta/\hbar$. 

The $B$-$T$ phase diagram of Pauli-limited superconductors is displayed in Fig.~\ref{fig:bdg}(a). The FFLO phase becomes thermodynamically stable at the critical field, $\mu_0 B_{\rm c}/\Delta_0 = 2/\pi$. Three phase boundaries between the normal, BCS, and FFLO phases meet at the Lifshitz point ``L''. The temperature of the Lifshitz point is a fixed point at $T_{\rm L}/T_{\rm c}\approx 0.56$ in Pauli-limited superconductors. We compute the saddle-point action in Eq.~\eqref{eq:Seq} with the self-consistent solution of $\Delta(x_i)$. The optimal period, $L$, is determined by minimizing the equilibrium action, $\mathcal{S}_{\rm eq}$. For $B/B_{\rm c}=1.001$, $1.029$, and $1.157$, the optimal values are estimated as $L/a = 200$, $122$. and $86$, respectively. At the critical field, $B_{\rm c}$, the single kink $\Delta(x) = \Delta_0 \tanh(x/\xi_0)$ is accompanied by the mid-gap ABS at the nodal plane, $\bm{\varphi}_0^{\rm t}(x) \propto (\cos(k_{\rm F}x),\sin(k_{\rm F}x)){\rm sech}(x/\xi_0)$. The critical field results from the energy tradeoff between the nucleation energy of the single kink $E_{\rm kink} = 2 \Delta_0/\pi$ and the accumulation of a single excess spin in the Zeeman splitting of the ABS~\cite{tak80,mac84}. 

In Fig.~\ref{fig:bdg}(c), we plot $E_{\alpha}(K_x)$ in the Brillouin zone folded by the FFLO superlattice, which are the eigenvalues of $\mathcal{H}_0$. The energies of quasiparticles with spin $S_z=\pm \frac{1}{2}$, $E_{\alpha,\pm}(K_x)$, are obtained by shifting $E_{\alpha}(K_x)$ as $E_{\alpha,\pm}(K_x)=E_{\alpha}(K_x)\pm \mu_0 B$. The band structure consists of the continuum states within $|E(K_x)|> \Delta_0$ and the band of the mid-gap ABSs within $|E|<E_{\rm ABS}$. 
The low-energy excitations of the FFLO state are accompanied by the mid-gap ABSs residing in each FFLO nodal plane, forming the periodic lattice structure with the period $L/2$. In the vicinity of $B_{\rm c}$, the FFLO period diverges as $L/\xi_0 \gg 1$ and thus the lattice of the mid-gap ABSs results in the dispersionless flat band at the zero energy. The nearly zero-energy flat band leads to the pronounced zero-energy peak in the quasiparticle local density of states at each nodal plane. As $B$ increases from $B_{\rm c}$, the period of the FFLO superconducting gap, $L$, becomes shorter, and its amplitude is suppressed [Fig.~\ref{fig:bdg}(b)]. Then, the mid-gap ABS bound at each nodal plane starts to hybridize with the neighboring one and become dispersive on $K_x$. The bandwidth of the mid-gap ABS increases with increasing $B$, $2E_{\rm ABS}\propto e^{-L/\xi_0}$.
The Zeeman shift of the band of the mid-gap ABS gives rise to the accumulation of the excess spins at each nodal plane~\cite{mac84,miz05,ich07}. 

We note that Fig.~\ref{fig:bdg}(c) is asymmetric with respect to $K_x$, where $E(K_x)\neq E(-K_x)$, and this might seem to break the time-reversal symmetry. However, the symmetry of the quasiparticle dispersion in the folded Brillouin zone depends on the choice of the gauge in Eq.~\eqref{eq:u0}. Further details are discussed in Appendix~\ref{sec:gauge}.

\section{Collective modes in FFLO states}
\label{sec:CM}

Here, we examine the whole dispersion of the collective modes in FFLO states by calculating the spectral functions of the fluctuation propagators. In FFLO states, both the translational symmetry and the ${\rm U}(1)$ symmetry are spontaneously broken, resulting in the existence of two gapless modes: the elastic mode and the phase mode. We demonstrate that the long-range Coulomb interaction can gap out the phase mode through the Anderson-Higgs mechanism, while it does not alter the spectrum of the elastic mode. In addition, we observe the presence of twofold degenerate gapped modes that remain unaffected against the long-range Coulomb interaction.

\subsection{Propagators of bosonic fields}

The fluctuations of the bosonic fields, including the superconducting order $\delta\Delta^{\pm}$ and charge density $\phi$, are described by the effective action in Eq.~\eqref{eq:Sf}. The effective action for the bosonic fluctuation, ${\bm \Sigma}_j(i\omega_m)=[\delta\Delta^+(x_j,i\omega_m),\delta\Delta^-(x_j,i\omega_m),\phi(x_j,i\omega_m)]^{\rm t}$, reads
\begin{align}
\mathcal{S}_{\rm fluc} =& \frac{1}{2}T\sum_m\int dx_i\int dx_j 
\bm{\Sigma}^{\rm t}_i(i\omega_m) \mathcal{D}^{-1}_{ij}(i\omega_m) \bm{\Sigma}_j(-i\omega_m),
\label{eq:Sf2}
\end{align}
where $\mathcal{D}_{ij}\equiv\mathcal{D}(x_i,x_j)$ is the fluctuation propagator defined as
\begin{align}
\mathcal{D}_{ij} \equiv &
\begin{pmatrix}
\mathcal{D}^{++}_{ij} & \mathcal{D}^{+-}_{ij} & \mathcal{D}^{+\phi}_{ij} \\
\mathcal{D}^{-+}_{ij} & \mathcal{D}^{--}_{ij} & \mathcal{D}^{-\phi}_{ij} \\
\mathcal{D}^{\phi +}_{ij} & \mathcal{D}^{\phi -}_{ij} & \mathcal{D}^{\phi\phi}_{ij}
\end{pmatrix} \nn \\
=& 
\begin{pmatrix}
\frac{2}{V}\delta_{ij}+\chi^{++}_{ij} & \chi^{+-}_{ij} & \chi^{+\phi}_{ij} \\
\chi^{-+}_{ij} &  -\frac{2}{V}\delta_{ij}+\chi^{--}_{ij} & \chi^{-\phi}_{ij} \\
\chi^{\phi +}_{ij} & \chi^{\phi -}_{ij} & U^{-1}_{ij}+\chi^{\phi\phi}_{ij}
\end{pmatrix}^{-1}.
\label{eq:D}
\end{align}
The correlation functions $\chi^{ab}_{ij}$ ($a,b=\pm, \phi$) are given by 
\begin{align}
\chi^{ab}_{ij}(Q_{\parallel}) = \frac{1}{2}\sum_{{\bm k}_{\parallel}}{\rm tr}_2\left[
G_{ji}(k_{\parallel})\Lambda_a
G_{ij}(k_{\parallel}+Q_{\parallel})\Lambda_b
\right],
\label{eq:chi}
\end{align}
where we use the shorthand notation, $k_{\parallel}=({\bm k}_{\parallel},i\varepsilon_n)$ and $Q_{\parallel}=({\bm Q}_{\parallel}, i\omega_m)$. In the Coulomb sector, the inverse of the long-range Coulomb potential, $U^{-1}_{ij}$, is expressed as $U_{ij}^{-1}=\frac{-1}{4\pi a^2}(\delta_{i,j-1}+\delta_{i,j+1}-2\delta_{i,j})$, which reduces to $U_{Q}^{-1}=\frac{{\bm q}^2}{4\pi}$ in the momentum space for $qa\ll 1$. The bare vertex functions are defined as $(\Lambda_+,\Lambda_-,\Lambda_{\phi})=(\tau_x, i\tau_y,2ie\tau_z)/2$. The detailed expression of the bare correlation functions is described in Appendix~\ref{sec:chi}.

In Eq.~\eqref{eq:D}, the three diagonal components are the bare propagators of the bosonic fluctuations. In the spatially uniform BCS state, $\mathcal{D}^{++}$, $\mathcal{D}^{--}$, and $\mathcal{D}^{\phi\phi}$, represent the propagators of the Higgs boson, the NG boson (the phase mode), and the plasmon, respectively. In the long wavelength limit $v_{\rm F}q\ll \Delta_0$, the pole of the Higgs propagator, $\mathcal{D}^{++}({\bm q},i\omega_m\rightarrow \omega + i0_+)$, obtains the dispersion of the Higgs mode, 
\beq
\omega = \sqrt{(2\Delta_0)^2+(vq)^2}, 
\eeq
where $v_{\rm F}\equiv \partial \xi(k_{\rm F})/\partial k_{\rm F}$ is the Fermi velocity and the velocity is given by $v=v_{\rm F}/\sqrt{d}$ in $d$ spatial dimension. The propagators of the phase mode and the plasmon have a pole at $\omega = vq$ and $\omega = \omega_{\rm p}$, respectively, where $\omega_{\rm p}=\sqrt{4\pi e^2 n_0/m}$ is the plasma frequency and $n_0$ is the electron density. Note that introducing the coupling with the $\phi$ field ($\chi^{-\phi}$ and $\chi^{\phi-}$) gaps out the phase mode to the plasma frequency through the Anderson-Higgs mechanism. 

We expand the fluctuations of the bosonic fields in terms of the Bloch states of the FFLO superlattice as
\begin{align}
\delta\Delta^{\pm}(x) = \frac{1}{2N_{\rm cell}} \sum_{Q_x} e^{iQ_xx}\delta\Delta^{\pm}_{Q_x}(x).
\end{align}
The charge density $\phi$ is also expressed by the Bloch states in the same way. The expressions of the Green's functions in terms of the Bloch states are shown in Eqs.~\eqref{eq:Gbloch+} and \eqref{eq:Gbloch-}, which obey the periodic boundary condition in Eqs.~\eqref{eq:Gperiod1} and \eqref{eq:Gperiod2}. Using these expressions, one can reduce the fluctuation action in Eq.~\eqref{eq:Sf2} to the FFLO supercell within $x\in[-L/4,L/4]$ and obtain the folded Brillouin zone within $Q_x\in[-2\pi/L, 2\pi/L]$.

\subsection{Collective modes in FFLO states}

\subsubsection{Collective mode spectra and eigenmodes}
\label{sec:CM0}

\begin{figure}[t!]
\includegraphics[width=85mm]{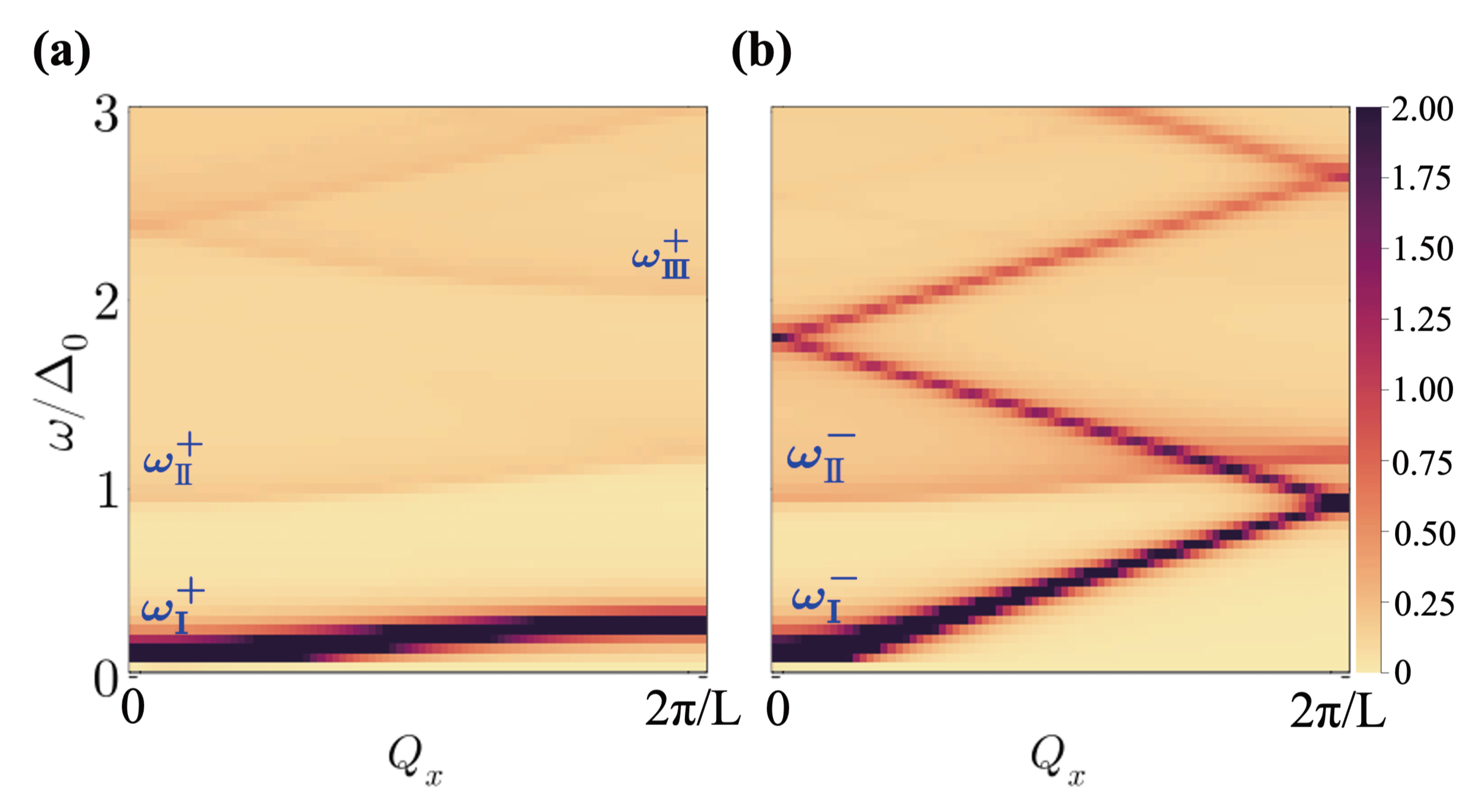}
\caption{Spectral functions of the bosonic excitations at $T=0.04$ and $B/B_{\rm c}=1.029$, where the long-range Coulomb interaction is neglected: (a) Amplitude fluctuations, $\rho^+(Q_x,\omega)$, and (b) phase fluctuations, $\rho^-(Q_x,\omega)$. In the amplitude sector, there are one gapless ($\omega^{+}_{\rm I}$) and two gapped modes ($\omega^+_{\rm II,III}$), while single gapless ($\omega^{-}_{\rm I}$) and gapped modes ($\omega^-_{\rm II}$) appear in the phase sector. The gapless modes in the amplitude and phase sectors are associated with the broken translational and ${\rm U}(1)$ symmetries, respectively. The gapped modes include the amplitude Higgs mode ($\omega^+_{\rm III}$) and twofold degenerate modes ($\omega^{\pm}_{\rm II}$).}
\label{fig:cm}
\end{figure}

We first present the excitation spectra in charge neutral systems, where the charge density fluctuation via the long-range Coulomb interaction is neglected. To clarify the collective mode dispersion, we calculate the spatially averaged spectral function of the bosonic excitations
\beq
\rho^{\pm}(Q_x,\omega) = -\frac{1}{\pi}\sum_i {\rm Im}\mathcal{D}^{(0)\pm}_{ii}(Q_x,i\omega_m\rightarrow \omega+i\eta),
\label{eq:rho}
\eeq
where $\rho^{+}$ ($\rho^-$) denotes the spectral function of the fluctuation propagators in the amplitude (phase) sector. The bare propagators are defined as $\mathcal{D}^{(0)+}_{ij}\equiv (\frac{2}{V}\mathbb{I}_N+\chi^{++})^{-1}_{ij}$ and $\mathcal{D}^{(0)-}_{ij}\equiv (-\frac{2}{V}\mathbb{I}_N+\chi^{--})^{-1}$ [see Eq.~\eqref{eq:D}], where $\mathbb{I}_N$ is the $N\times N$ unit matrix and $\chi^{\pm\pm}$ are the $N\times N$ matrix of $\chi^{\pm\pm}_{ij}$. In numerical calculation, we fix the infinitesimal constant to $\eta = 0.01\Delta_0$. 

Figure~\ref{fig:cm}(a) shows the spectral function in the amplitude sector, $\rho^+(Q_x,\omega) $. Owing to the anti-periodic structure of the equilibrium order in Fig.~\eqref{eq:delta_periodic2}, the spectra of the collective excitations exhibit a Brillouin zone
structure, where $Q_x\in[-2\pi/L,2\pi/L]$. There are three branches along which the spectral function peaks. These branches are denoted by $\omega^+_{\rm I}(Q_x)$, $\omega^+_{\rm II}(Q_x)$, and $\omega^+_{\rm III}(Q_x)$, respectively. One of them ($\omega^+_{\rm I}$) is the gapless mode, while the other two branches, $\omega^+_{\rm II}$ and $\omega^+_{\rm III}$, are dispersing from $(Q_x,\omega) \sim (0,\Delta_0)$ and $(Q_x,\omega) \sim (2\pi/L,2\Delta_0)$, respectively. The spectral function in the phase sector, $\rho^-(Q_x,\omega) $, is displayed in Fig.~\ref{fig:cm}(b), which has two branches denoted by $\omega^-_{\rm I}$ and  $\omega^-_{\rm II}$. The branch $\omega^-_{\rm I}$ is gapless and linearly dispersing from $(Q_x,\omega)=(0,0)$, which is well fitted by the dispersion of the NG mode associated with the ${\rm U}(1)$ symmetry breaking, $\omega = v_{\rm F}Q_x$, where the Fermi velocity is evaluated as $v_{\rm F}=\sqrt{3}ta/\hbar$ with the current parameters. Another branch, $\omega^-_{\rm II}$, is dispersing from $(Q_x,\omega) \sim (0, \Delta_0)$. The entire dispersion of the $\omega^-_{\rm II}$ mode is degenerate with that of the $\omega^{+}_{\rm II}$ mode.

To identify the peaks in the spectral functions as the poles of the fluctuation propagators, we introduce the inverse of the bosonic propagators in the matrix form, $\mathbb{D}^{-1}_{\pm}$, where $(\mathbb{D}^{-1}_{\pm})_{ij}\equiv (\mathcal{D}^{(0)\pm})^{-1}_{ij}$. We then diagonalize this matrix as 
\beq
\mathbb{D}^{-1}_{\pm}\ket{v_{\lambda}} = \lambda \ket{v_{\lambda}}. 
\label{eq:Dmat}
\eeq
It is confirmed that the lowest eigenvalue, $\min |{\rm Re}\lambda|$, tends toward zero along the peaks of the spectral function in the $\omega$-$Q_x$ plane. Thus, all the branches in $\rho^{\pm}(Q_x,\omega)$ coincide to the poles of the fluctuation propagators $\mathcal{D}^{\pm}(Q_x,\omega)$, implying that the spectral peaks represents the dispersion of the collective modes.

\begin{figure}[t!]
\includegraphics[width=85mm]{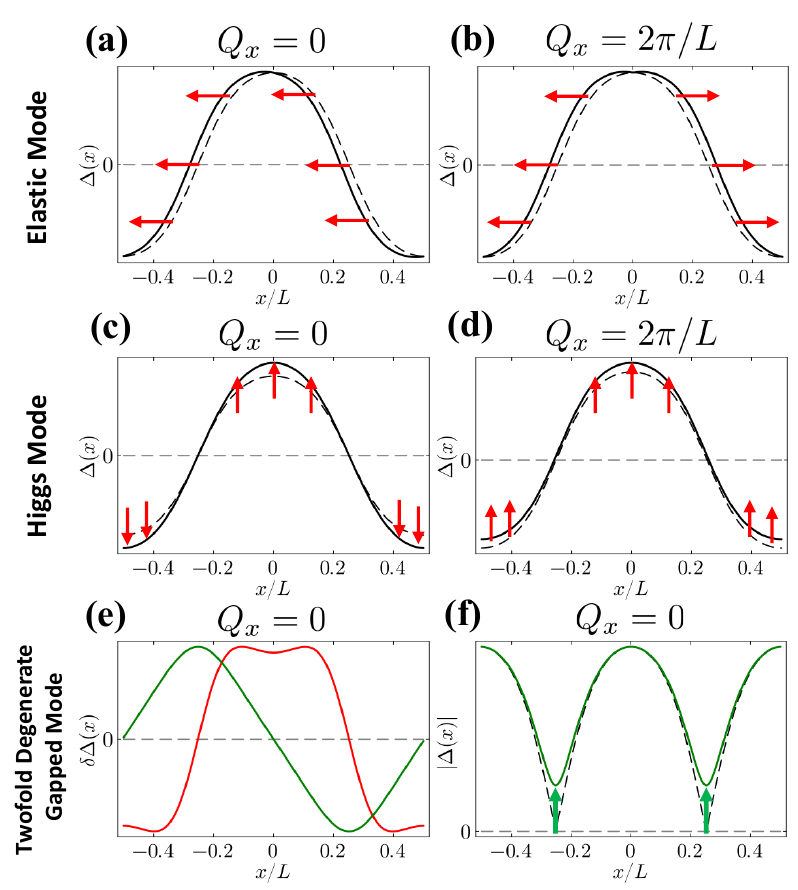}
\caption{(a,b) Spatial profiles of the elastic modes at $\omega = \omega^{+}_{\rm I}$: (a) $Q_x=0$ and (b) $Q_x=2\pi/L$. (c,d) Spatial profiles of the amplitude Higgs modes at $\omega = \omega^{+}_{\rm III}$: (c) $Q_x=0$ and (d) $Q_x=2\pi/L$. In (a-d), the dashed curves are the equilibrium gap profiles [$\Delta_{\rm eq}(x_i)$], and the solid curves represent the linear fluctuations from the equilibrium gap structure, $\Delta(x_i,t)=\Delta_{\rm eq}(x_i)+\epsilon\delta\Delta^+(x_i,t)$ at $t=0$. (e) The eigenstates of $\mathbb{D}^{-1}_+$ at $\omega = \omega ^{+}_{\rm II}$. (f) Amplitude of the linear fluctuation of the $\omega = \omega ^{\pm}_{\rm II}$ mode, $|\Delta_{\rm eq}(x_i)+\epsilon\delta\Delta^-(x_i,t)|$, at $t=0$. These are twofold degenerate gapped modes, which cause oscillations in both the width and grayness of the each FFLO nodal plane.}
\label{fig:modes}
\end{figure}

The eigenvector in Eq.~\eqref{eq:Dmat}, $\ket{v_{\lambda}(\omega)}$, at $\omega = \omega^{+}_{\rm I,II,III}$ provides information about the eigenmode of the fluctuation at each branch. Figure~\ref{fig:modes}(a) shows the profile of the gapless amplitude mode, $\Delta(x_i,t) = \Delta_{\rm eq}(x_i) + \epsilon e^{-i\omega t}\delta \Delta^{+}(x_i)$, at $t=0$, where $\epsilon$ is an arbitrary small constant. The fluctuation $\delta\Delta^+(x_i)$ is obtained by calculating the eigenvector in Eq.~\eqref{eq:D} at $\omega = \omega^{+}_{\rm I}$ and $Q_x=0$. As shown in Fig.~\ref{fig:modes}(a), this gapless mode causes the displacement of the nodes in $\Delta$, i.e., the elastic mode of the periodically arrayed nodal planes. This corresponds to the NG mode that recovers the translational symmetry spontaneously broken by the FFLO order. Figure~\ref{fig:modes}(b) shows the second lowest gapped mode ($\omega _{\rm III}^+$). This mode exhibits the oscillation of the gap amplitude without changing the nodal position, which is nothing but the amplitude Higgs mode in the FFLO state. While the dispersion of the elastic mode ends at the zone boundary, the dispersion of the Higgs mode is folded at the zone boundary and extends into the higher frequency region.  

The spatial profiles of gapless and gapped amplitude modes in Fig.~\ref{fig:modes} provide a key to understand their characteristic dispersions. As shown in Fig.~\ref{fig:cm}(a), the gapless mode ($\omega^+_{\rm I}$) disperses from $Q_x=0$, while the dispersion of the gapped $\omega^+_{\rm III}$ mode starts from $\omega=2\Delta_0$ and $Q_x=2\pi/L$. In the BCS state, the amplitude Higgs mode at $Q_x=0$ corresponds to a spatially uniform oscillation of the order parameter amplitude. In contrast, the $\omega^+_{\rm III}$ mode at $Q_x=0$ in the FFLO state, as shown in Fig.~\ref{fig:modes}(c), involves not a spatially uniform oscillation, but rather the oscillation of the superconducting amplitude within each superconducting domain, $-L/4<x<L/4$, $L/4<x<3L/4$, and so on. On the other hand, the amplitude mode at the zone boundary $Q_x=2\pi/L$ involves a spatially uniform oscillation [Fig.~\ref{fig:modes}(d)]. Therefore, the gapped amplitude mode ($\omega^+_{\rm III}$) in the FFLO state disperses from $Q_x=2\pi/L$ due to the anti-periodic boundary condition of $\Delta(x)$, which differs from that in the conventional BCS state. We note that the gapless elastic mode ($\omega^+_{\rm I}$) at $Q_x=0$ causes a uniform displacement of the FFLO nodal planes, and thus disperses from $Q_x=0$.

As mentioned above, the lowest gapped modes denoted by $\omega^{+}_{\rm II}$ and $\omega^-_{\rm II}$ are twofold degenerate in the amplitude and phase sectors. Figure~\ref{fig:modes}(c) and \ref{fig:modes}(d) depict the amplitude and phase fluctuations of the $\omega^{\pm}_{\rm II}$ modes. The fluctuations of the real and imaginary parts represent the oscillation of the FFLO domain width and the ``breathing'' oscillation of the nodal plane, respectively. The latter mode has been identified as the grayness oscillation in the context of soliton chains~\cite{dut17}. 
In one-dimensional Pauli-limited superconductors, the BdG equation~\eqref{eq:bdg0} is mapped onto the nonlinear Schr\"{o}dinger equation for the superconducting order~\cite{bas1,bas2,yos11,dat12,dat13}, and the FFLO state is one of the families of the soliton solutions. Indeed, the FFLO order parameter around $B=B_{\rm c}$ can be regarded as a periodic array of black solitons that have a zero of the order parameter amplitude with a $\pi$-phase shift across its zero. 
The single gray-soliton (or dark-soliton) state is represented by
\beq
\Delta(x) = \Delta_0 \left[ \cos\phi \tanh\zeta +i\sin\phi\right].
\label{eq:dark}
\eeq
The shape of the soliton is determined by $\zeta \equiv \cos\phi (x-x_0)/d$, where $x_0$ and $d$ are the soliton center and the domain width, respectively. In Eq.~\eqref{eq:dark}, the soliton angle is defined as $|\phi| < \pi/2$. The soliton with $\phi=0$ is called the black soliton, where the order parameter is zero at $x=x_0$. The equilibrium FFLO order corresponds to the black soliton. The {\it grayness} or {\it darkness} is defined as~\cite{fra10}
\beq
|\Delta(x)|^2 = \Delta_0^2 \left[ 
1-\cos^2\phi {\rm sech}^2\zeta
\right].
\eeq
The soliton becomes gray or dark when the soliton phase deviates from $\phi=0$, where the zeros are filled in as $|\Delta(x_0)|\neq 0$. Therefore, one of the twofold degenerate modes can be identified as the grayness oscillation of the FFLO nodes or the oscillation of the soliton angle around $\phi = 0$.

\begin{figure*}[t!]
\includegraphics[width=170mm]{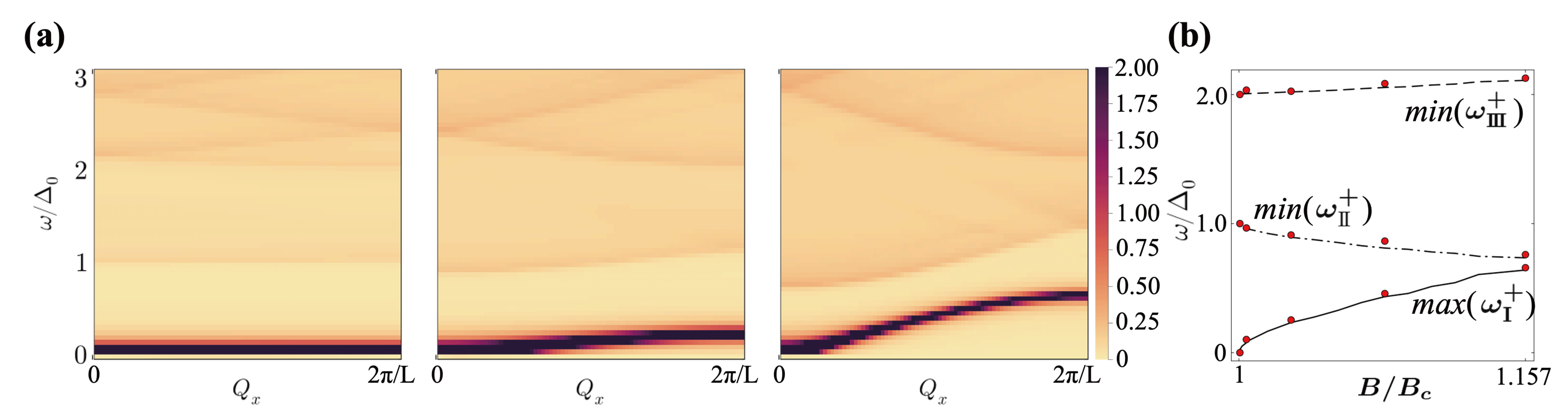}
\caption{Field evolution of the amplitude modes in the FFLO state. (a) Spectral function in the amplitude sector at $B/B_{\rm c}= 1.001$, $1.029$, and $1.157$, where the corresponding FFLO period is $L/a=200$, 122, and 86, respectively. The quasiparticle dispersions at each $B$ are shown in Fig.~\ref{fig:bdg}(c). (b) Field evolution of the bandwidth of the elastic mode dispersion, $\max(\omega^+_{\rm I})\equiv\omega^+_{\rm I}(Q_x=2\pi/L)$, and the energy gaps of the two gapped modes, $\min(\omega^{+}_{\rm II}) = \omega^{+}_{\rm II}(Q_x=0)$ and $\min(\omega^{+}_{\rm III})=\omega^+_{\rm III}(Q_x=2\pi/L)$. The solid, dashed-dotted, and dashed lines correspond to the characteristic energies in the quasiparticle spectrum, $2E_{\rm ABS}$, $E_{\rm cont}-E_{\rm ABS}$, and $2E_{\rm cont}$ [see Fig.~\ref{fig:bdg}(c)].}
\label{fig:cmb}
\end{figure*}

In Fig.~\ref{fig:cmb}, we summarize the field evolution of the amplitude mode spectrum, $\delta\Delta^+$, in the FFLO state. Figure~\ref{fig:cmb}(a) shows the spectral function of the amplitude modes at at $B/B_{\rm c}= 1.001$, $1.029$, and $1.157$, where the corresponding FFLO period is $L/a=200$, 122, and 86, respectively. At the BCS-FFLO critical field, $B=B_{\rm c}$, both the gapless elastic mode ($\omega^+_{\rm I}$) and twofold degenerate mode ($\omega^{+}_{\rm II}$) exhibit the dispersionless flat band at $\omega = 0$, while they become dispersive with increasing $B$. The field evolution of the collective mode spectrum is related to that of the quasiparticle dispersions. To clarify this, in Fig.~\ref{fig:cmb}(b), we plot the bandwidth of the elastic mode dispersion [$\omega^{+}_{\rm I}(Q_x=0)$] and the energy gaps of the two gapped modes [$\omega^{+}_{\rm II}(Q_x=0)$ and $\omega^+_{\rm III}(Q_x=2\pi/L)$] as a function of $B$. Figure~\ref{fig:cmb}(b) also shows the bandwidth of the mid-gap ABS ($2E_{\rm ABS}$), the energy gap between the mid-gap ABS and the continuum state ($E_{\rm cont}-E_{\rm ABS}$), and the gap between the continuum states ($2E_{\rm cont}$). It is seen that the bandwidth of the elastic mode coincides with the mid-gap ABSs, $\max(\omega^{\rm I})=2E_{\rm ABS}$. This indicates that the $\omega_{\rm I}^{+}$ mode is regarded as the elastic mode of the band that is constructed from the periodic array of the mid-gap ABSs. Figure~\ref{fig:cmb}(b) also shows that the energy of two gapped modes follows the field evolution of the quasiparticle energies, $\min(\omega^{+}_{\rm II})=E_{\rm cont}-E_{\rm ABS}$ and $\min(\omega^+_{\rm III})=2E_{\rm cont}$.

\subsubsection{Effect of long-range Coulomb interaction}

\begin{figure}[t!]
\includegraphics[width=85mm]{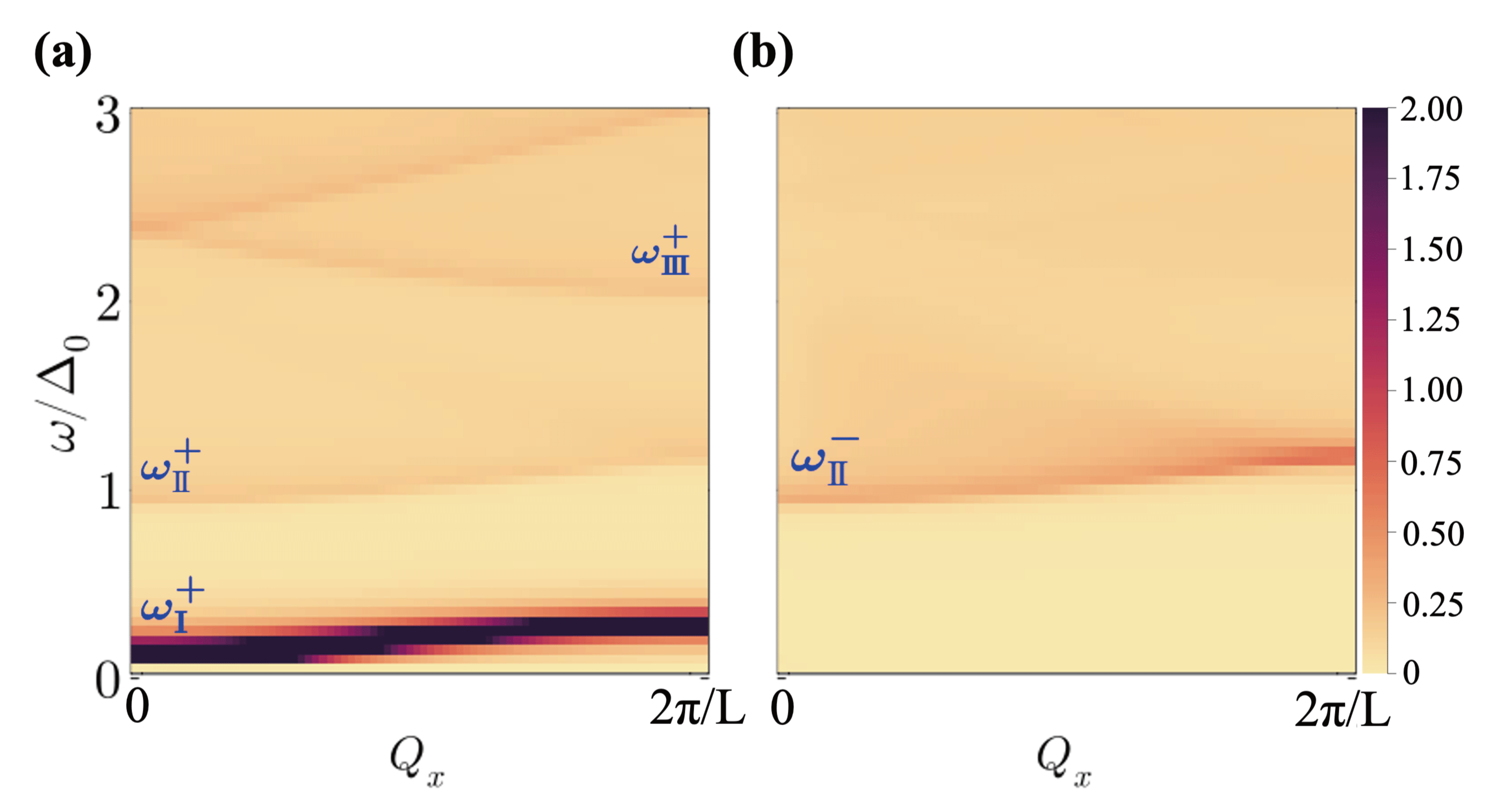}
\caption{Spectral functions of the excitations at $T/T_{\rm c}=0.04$ and $\mu_0 B/\Delta_0=1.029$, where the long-range Coulomb interaction is taken into account: (a) Amplitude fluctuations, $\tilde{\rho}^+(Q_x,\omega)$, and (b) phase fluctuations, $\tilde{\rho}^-(Q_x,\omega)$. The long-range Coulomb interaction only gaps out the gapless phase mode ($\omega^-_{\rm I}$) through the Anderson-Higgs mechanism, while the other modes remain unaffected.}
\label{fig:cm2}
\end{figure}

Let us now examine the impact of the long-range Coulomb interaction on the collective modes. In the conventional BCS state, the coupling of the long-range Coulomb potential to the phase fluctuation causes a significant change in the low-lying collective mode spectrum, known as the Anderson-Higgs mechanism~\cite{and63,higgs,varma}. For the calculation of the Coulomb potential, one needs to specify the material parameters, such as the hopping energy $t$ and the lattice constant $a$, as Eq.~\eqref{eq:D} contains the dimensionless parameter $\alpha\hbar c/ta$, where $\alpha = e^2/\hbar c$ is the fine structure constant and $c$ is the speed of light. In numerical calculations, we set $\alpha\hbar c/ta = 10$, corresponding to $t=1~{\rm eV}$ and $a=0.1~{\rm nm}$. This ensures that the plasma frequency is much higher than the energy scale of the superconducting collective modes.

In Fig.~\ref{fig:cm2}, we plot the spectral functions of the bosonic excitations,
\beq
\tilde{\rho}^{\pm}(Q_x,\omega) = -\frac{1}{\pi}\sum_i {\rm Im}{\mathcal{D}}^{\pm\pm}_{ii}(Q_x,i\omega_m\rightarrow \omega+i\eta),
\label{eq:rho2}
\eeq
where ${\mathcal{D}}^{++}_{ij}$ (${\mathcal{D}}^{--}_{ij}$) is the amplitude (phase) propagator renormalized by the long-range Coulomb interaction as seen in Eq.~\eqref{eq:D}. 
We plot the spectral functions, $\tilde{\rho}^{\pm}(Q_x,\omega)$, in Fig.~\ref{fig:cm2}, which are obtained by computing the inverse of the full $3N\times 3N$ matrix in Eq.~\eqref{eq:D}. Figure~\ref{fig:cm2}(a) demonstrates that all the amplitude modes, including the gapless elastic mode ($\omega^+_{\rm I}$), twofold degenerate mode ($\omega^+_{\rm II}$), and the Higgs mode ($\omega^+_{\rm III}$), remain unaffected even when the long-range Coulomb potential is taken into account. On the other hand, the gapless phase mode vanishes in the low frequency region, as shown in Fig.~\ref{fig:cm2}(b), which is similar to the Anderson-Higgs mechanism in the conventional BCS state. We also note that the lowest gapped mode ($\omega^{-}_{\rm II}$) in the phase sector, which is the twofold degenerate mode, remains unaffected by the long-range Coulomb potential.

The long-range Coulomb interaction does not alter the dispersion of the gapless elastic mode. To emphasize this, we evaluate the frequency shift and the damping rate of the elastic mode in the presence of the Coulomb interaction. Similarly with Sec.~\ref{sec:CM0}, we define the $N\times N$ matrix of the bare fluctuation propagators in the amplitude ($\mathbb{D}^{(0)}_+$) and Coulomb ($\mathbb{D}^{(0)}_{\phi}$) sectors. The bare propagator, $\mathbb{D}^{(0)}_+(Q_x,\omega)$, has poles at $\omega=\omega^+_{\rm I}$ as shown in Fig.~\ref{fig:cm}(a). The poles appear at the frequencies $\omega$ where the lowest eigenvalues of $[\mathbb{D}^{(0)}_+(Q_x,\omega)]^{-1}$ becomes zero [see Eq.~\eqref{eq:Dmat}]. In the vicinity of the pole, $\omega^+_{A}(Q_x)$, therefore, the bare fluctuation propagator is expressed in terms of the eigenvector ($\ket{v_0(Q_x,\omega)}$) associated with the lowest eigenvalue as 
\begin{align} 
\mathbb{D}^{(0)}_+(Q_x,\omega) \approx \frac{\ket{v_0(Q_x,\omega)}\bra{v_0(Q_x,\omega)}}{\omega - \omega^+_{\rm A}(Q_x)}.
\label{eq:D0approx}
\end{align}
It is obtained from Eq.~\eqref{eq:D} that the renormalized propagator of the amplitude fluctuation obeys the Dyson equation
\begin{align}
\left[\mathcal{D}^{++}(Q_x,\omega)\right]^{-1} = 
\left[\mathbb{D}^{(0)}_+(Q_x,\omega)\right]^{-1} - \Sigma^{+}(Q_x,\omega).
\label{eq:dyson}
\end{align}
When the couplings between the amplitude and phase sectors are negligible, i.e., $\chi^{+ -}\approx \chi^{-+}\approx0$, the effect of the Coulomb potential fluctuation is incorporated into the amplitude mode through the self-energy  $\Sigma^{+}(Q) = \chi^{+ \phi}(Q)\mathbb{D}^{(0)}_{\phi}(Q)\chi^{\phi +}(Q)$, where the $N\times N$ matrices, $\chi^{+ \phi}$ and $\chi^{\phi+}$, represent the coupling of the amplitude mode to the Coulomb propagator. By substituting Eq.~\eqref{eq:D0approx} into Eq.~\eqref{eq:dyson}, the pole of the renormalized propagator at $Q_x$ is expressed as $\omega^{+}_{\rm I}(Q_x)+\delta\omega^+_{\rm I}(Q_x)-i\gamma^+_{\rm I}(Q_x)$, where the frequency shift and the damping rate of the elastic mode are defined as $\delta\omega^+_A(Q_x)={\rm Re}[\bra{v_0(Q_x,\omega^+_{A})}\Sigma^+(Q_x,\omega^+_{A})\ket{v_0(Q_x,\omega^+_{A})}]$ and $\gamma^+_{\rm I}(Q_x)={\rm Im}[\bra{v_0(Q_x,\omega^+_{A})}\Sigma^+(Q_x,\omega^+_{A})\ket{v_0(Q_x,\omega^+_{A})}]$. Using the same parameters as shown in Fig.~\ref{fig:cm2}, we evaluate the frequency shift and the damping rate of the elastic and twofold degenerate modes as $(\delta\omega^{+}_{\rm I}/\Delta_0,\gamma^+_{\rm I}/\Delta_0)= (8.0\times 10^{-4}, 3.9\times 10^{-6})$ and 
$(\delta\omega^{+}_{\rm II}/\Delta_0,\gamma^+_{\rm II}/\Delta_0)= (7.3\times 10^{-4}, 6.0\times 10^{-4})$
at $Q_x=\pi/L$, respectively, which are significantly smaller than the resonant frequency $\omega^+_{\rm I}=0.15\Delta_0$ and $\omega^+_{\rm II}=1.0\Delta_0$. Therefore, incorporating the charge fluctuation into the self-energy does not alter the frequency and damping rates of these modes. These results reflect the absence of the pole in $\mathbb{D}^{(0)}_{\phi}$ in the low energy region $\omega \lesssim \Delta_0$ and the weak coupling $\chi^{+\phi}$. We note that for the phase mode the frequency shift estimated from Eq.~\eqref{eq:dyson} is comparable to the resonant frequency, $\delta\omega^{-}_{\rm I}/\omega^-_{\rm I}\sim  O(0.1)$, indicating the breakdown of the perturbative analysis in Eq.~\eqref{eq:dyson}.

\section{Signals of collective modes in density response}
\label{sec:dns}

Having shown the existence of gapless and gapped collective excitations that reflect the character of the FFLO order, we lastly discuss how the signatures of the collective modes can be captured by observables. As an example, in this paper, we focus on the density response, which can be measured by the electronic Raman spectroscopy~\cite{klein84,dev07}. In addition, the density-density response function is associated with the nonlinear current response via the diamagnetic coupling channel of electrons with light. The third harmonic generation by intense pulsed Gaussian beams essentially measures the density response~\cite{cea16,shi20}.

Within the functional integral formalism, the density response can be formulated by introducing in Eq.~\eqref{eq:action} a source term $V_{\rho,i}\equiv V_{\rho}(x_i)$ coupled to the density operator, $\rho(x)=\sum \bar{\psi}(x)\Lambda^{\rho}\psi(x)$. Here, $\Lambda^{\rho}=\tau_z/2$ is the density vertex function in the particle-hole space. Similarly with the procedure in Sec.~\ref{sec:theory}, one obtains the effective action for the fluctuations $(\delta\Delta^{\pm},\phi)$ and the source field $V_{\rho}$. By further integrating out the fluctuations, the effective action for $V_{\rho}$ reads
\begin{align}
\mathcal{S} [V_{\rho}] = &
\frac{1}{2}\sum_{ij}V_{\rho,i}(-\omega)\bigg[
\chi^{\rho\rho}_{ij}(\omega) \nn \\
&-\sum_{k,l}\chi^{\rho a}_{ik}(-\omega)
\mathcal{D}^{ab}_{kl}(\omega) \chi^{b\rho}_{lj}(\omega)
\bigg]V_{\rho,j}(\omega).
\end{align}
The bare density-density response function, $\chi^{\rho\rho}_{ij}$, is defined by substituting $\Lambda^{\rho}=\tau_z/2$ into Eq.~\eqref{eq:chi}. The density-density response function follows by the functional derivative of the effective action with respect to the source field, 
\beq
\tilde{\chi}^{\rho\rho}_{ij}= \frac{\delta^2 \mathcal{S}[V_{\rho}]}{\delta V_{\rho,i}\delta V_{\rho,j}}
= \tilde{\chi}^{\rm QP}_{ij}
+\tilde{\chi}^{\rm CM}_{ij}.
\eeq
The first term, $\chi^{\rm QP}_{ij} = \chi^{\rho\rho}_{ij}$, represents the density response through the quasiparticle excitations, while the second term corresponds to the density response mediated by the excitation of the amplitude modes, 
\beq
\tilde{\chi}^{\rm CM}_{ij}(\omega) = -\sum_{k,l}\chi^{\rho a}_{ik}(-\omega)\mathcal{D}^{ab}_{kl}(\omega) \chi^{b\rho}_{lj}(\omega).
\label{eq:chi_CM}
\eeq
The fluctuation propagators, $\mathcal{D}^{ab}(\omega)$ ($a,b=\pm,\phi$), which are defined in Eq.~\eqref{eq:D}, have poles at the eigenfrequencies of the collective modes. 

Let us briefly mention the contribution of the collective modes to the density response in the conventional BCS state. In the clean limit, two collective modes exist: the phase and amplitude Higgs modes. The former is gapped out by the long-range Coulomb interaction and does not affect the density response. The contribution of the Higgs mode is also negligible as $\chi^{+\rho}$ in Eq.~\eqref{eq:chi_CM}, which is the coupling of the Higgs propagator to the density fluctuation, is of the order $(k_{\rm F}\xi)^{-1}$ in the clean limit~\cite{cea16,shi20}. The density response is dominated by the pair excitations of quasiparticles, which peak at the threshold energy $\omega = 2\Delta_0$. Therefore, collective excitations cannot have a primary role in the clean limit of the BCS state.

\begin{figure}[t!]
\includegraphics[width=85mm]{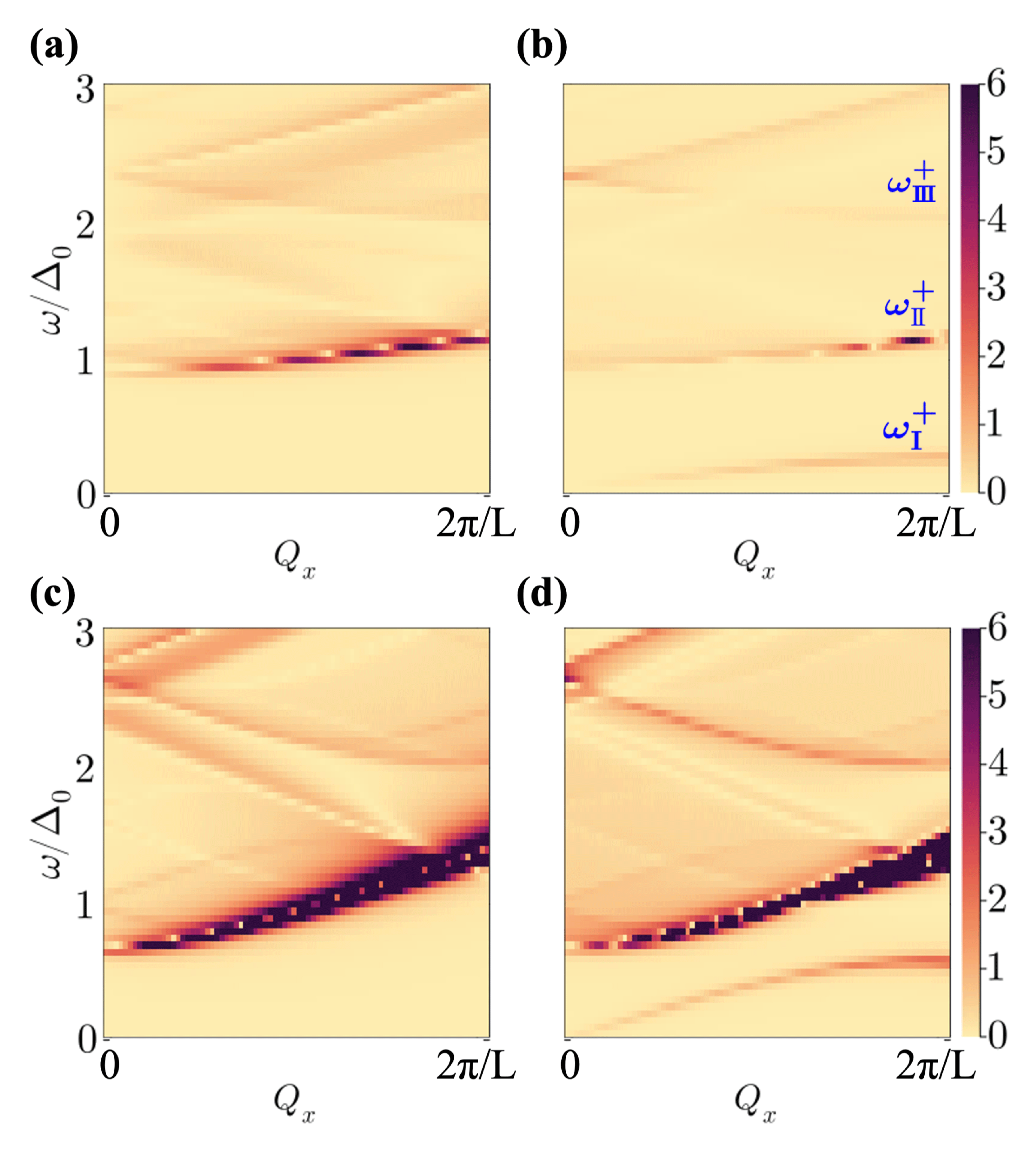}
\caption{Density-density response functions, $-{\rm Im}\tilde{\chi}^{\rho\rho}(Q_x,\omega)$, in the FFLO state: (a,c) the quasiparticle excitations ($\tilde{\chi}^{\rm QP}$) and (b,d) collective modes in the amplitude sector ($\tilde{\chi}^{\rm CM,+}$). We set $\mu_0 B/\Delta_0 = 1.029$ (a,b) and $1.157$ (c,d).}
\label{fig:dns}
\end{figure}

In Fig.~\ref{fig:dns}, we plot the imaginary parts of the density-density response functions, $-{\rm Im}\tilde{\chi}^{\rho\rho}(Q_x,\omega)$, in the FFLO state, where $\tilde{\chi}^{\rho\rho}(Q_x,\omega)\equiv \sum_{ij}\tilde{\chi}_{ij}^{\rho\rho}(\omega)e^{iQ_x(x_i-x_j)}$. The collective mode contributions consist of the amplitude mode ($\tilde{\chi}^{\rm CM,+}$), the phase mode ($\tilde{\chi}^{\rm CM,-}$), and the charge density ($\tilde{\chi}^{\rm CM,\phi}$). In Figs.~\ref{fig:dns}(b) and \ref{fig:dns}(d), we focus on the contribution of the amplitude modes to the density response, as the elastic mode is a reflection of the translational symmetry breaking by the FFLO order. It is clearly seen that density response via the collective excitations peaks along $(Q_x,\omega)$ resonant to the $\omega^{+}_{\rm I}$ and $\omega^+_{\rm II}$ modes. As for the $\omega^{+}_{\rm II}$ mode dispersing from $(Q_x,\omega)\approx(0,\Delta_0)$, however, the contribution of the quasiparticle excitations dominates the density response. We note that the contribution of the charge density is recast into 
\begin{align}
\tilde{\chi}^{\rm CM,\phi}= - \frac{\chi^{\rho\phi}\chi^{\phi\rho}}{\chi^{\phi\phi}}
+\frac{\chi^{\rho\phi}\chi^{\phi\rho}}{(\chi^{\phi\phi})^2}\tilde{\chi}^{\rm sc},
\end{align}
where the generalized density-density response function,
$\tilde{\chi}^{\rm sc} = \chi^{\phi\phi}/(1+V\chi^{\phi\phi})$ with the Coulomb potential $V$.
These terms, stemming from the long-range Coulomb interaction, represent the backflow to ensure particle number conservation of charge density fluctuations. The first term cancels the bare density response function, $\chi^{\rho\rho}$, and the quasiparticle contribution to the density response is represented by $\tilde{\chi}^{\rm sc}$.

It is noteworthy that the quasiparticle excitations within the band of the mid-gap ABSs do not contribute to $\chi^{\rm QP}$. As shown in Fig.~\ref{fig:bdg}(c), the mid-gap ABSs form the band structure within $|E_K|\le E_{\rm ABS}$. Thus, an external field coupled with the density is expected to induce the excitations of the mid-gap ABSs within the range of $\omega \lesssim 2E_{\rm ABS}$. As depicted in Fig.~\ref{fig:dns}, however, the mid-gap ABSs have no impact on the density response, contrary to expectations. On the other hand, the elastic mode ($\omega^+_{\rm I}$) exhibits the resonant contribution to the density response. 

Here we demonstrate that in the vicinity of the BCS-FFLO state, where $L\gg \xi_0$, the mid-gap ABSs in the FFLO state have no contribution to the density response. In the weak coupling limit $k_{\rm F}\xi_0\gg 1$, the mid-gap ABS for an isolated single FFLO nodal plane, $\Delta_{\rm eq}(x)=\Delta_0\tanh(x/\xi_{\rm d})$, is given by 
\beq
{\bm w}_+ (x) = \frac{1}{\sqrt{2\xi_0}}\begin{pmatrix}
\cos(k_{\rm F}x) \\ \sin(k_{\rm F}x)
\end{pmatrix}{\rm sech}\left( \frac{x}{\xi_{\rm d}}\right),
\eeq
and ${\bm w}_-=\tilde{\mathcal{C}}{\bm w}_+$ is the particle-hole symmetric state in the same spin sector. The domain size, $\xi_{\rm d}$, is an order of $\xi_0$. For $L\gg \xi_0$, the quasiparticle wavefunction in the periodic array of the FFLO nodal plane can be constructed as ${\bm \varphi}_{\pm, K_x}(x)=e^{iK_xx}{\bm w}_{\pm}(x)$. Then, the equilibrium Green's functions for the ABSs are obtained as 
\beq
G^{\rm ABS}_{ij}(i\varepsilon_n) = \frac{{\varphi}_{+}(x_i){\varphi}^{\dag}_{+}(x_j)}{i\varepsilon_n-\epsilon}
+ \frac{{\varphi}_{-}(x_i){\varphi}^{\dag}_{-}(x_j)}{i\varepsilon_n+\epsilon},
\eeq
where $\epsilon>0$ is the energy of the Andreev band. For $L\gg \xi_0$, the Andreev band is dispersionless, and $\epsilon$ is approximated by an infinitesimal constant independent of $K_x$. Using $G^{\rm ABS}$, we compute the ABS contribution to the density-density response function, $\chi^{\rm QP}(Q_x,\omega) \equiv \sum_{ij}\chi^{\rho\rho}_{ij}(\omega)e^{iQ_x(x_i-x_j)}$, which reads at $T=0$
\beq
\chi^{\rm QP}(Q_x,\omega) = \frac{1}{\xi_{\rm d}}\frac{2\epsilon\tanh(\epsilon/2T)}{\omega^2_+-(2\epsilon)^2}F_{Q_x}F_{-Q_x},
\eeq
where $F_Q$ is the Fourier transform of the coherence factor
\beq
F_Q\equiv \int dx \sin(2k_{\rm F}x){\rm sech}^2\left( \frac{x}{\xi_{\rm d}}\right)e^{iQx}.
\label{eq:FQ}
\eeq
Here we assume the long wavelength limit, $Q_x \lesssim L^{-1}\ll \xi_{0}^{-1}$, where the length scale of the external field $V_{\rm rho}$ varies slowly compared to the localization length of the mid-gap ABSs. Then, Eq.~\eqref{eq:FQ} vanishes, resulting in $\chi^{\rm QP}\rightarrow 0$ for $\omega \lesssim 2E_{\rm ABS}$. 

In summary, while the contribution of the twofold degenerate mode to the density response is smeared out by that of quasiparticle excitations, the elastic mode resonantly contributes to the density response. The measurement of the density response through the electronic Raman spectroscopy and the third harmonic generation by intense pulsed light may capture the signatures of the NG mode in the FFLO state that recovers the spontaneous breaking of the translational symmetry.

\section{Concluding remarks}
\label{sec:summary}

We have investigated collective excitations in the FFLO states of Pauli-limited superconductors. When the long-range Coulomb interaction is absent, the collective excitation spectra consist of two gapless and three gapped modes. Two gapless modes are the phase mode and the elastic mode, which reflect the spontaneous breaking of the ${\rm U}(1)$ symmetry and translational symmetry, respectively. Two of the gapped modes are degenerate and exhibit the oscillation of the FFLO domain width and the ``breathing'' oscillation of the FFLO nodal planes. We demonstrate that when the long-range Coulomb interaction is considered, only the phase mode vanishes in the low-frequency region as the Anderson-Higgs mechanism in the conventional BCS state, while other modes, including the gapped ($\omega_{\rm II}^-$) mode in the phase sector, remain unaffected. Furthermore, the dispersion of the gapless elastic mode is associated with the bandwidth of the mid-gap ABSs and sensitively depends on the applied magnetic field. As the magnetic field approaches the BCS-FFLO critical field, the gapless elastic mode becomes dispersionless, and the bosonic density of states sharply peaks at zero frequency. 

We have shown that the density-density response function can capture the clear signature of the elastic mode, which is the NG mode associated with the translational symmetry breaking by the FFLO order. We expect that the signal remains visible even in the presence of weak disorders, provided that the weak disorders are homogeneously distributed and do not explicitly break the translational symmetry.  It is also worth examining how disorders affect the response of the collective mode to external perturbations. In the clean limit of conventional BCS superconductors, it has been discussed that the coupling between the amplitude Higgs mode and the density fluctuation is negligibly small, implying that the Higgs mode is not detectable through density response~\cite{cea16}. In the dirty limit, however, disorders enhance the coupling of the Higgs mode to the density fluctuation, making the signal detectable in the density-density response function~\cite{juj18,sil19,tsu20}. Therefore, it is essential to understand how disorders impact the response of the collective modes of the FFLO state to external perturbations. 

In this work, we have also demonstrated that the quasiparticle excitations in the band of the mid-gap ABSs do not contribute to the density response near the BCS-FFLO critical field. It is necessary to thoroughly investigate the impact of the mid-gap ABSs across a wider region of the magnetic field. Furthermore, it is essential to clarify the influence of the mid-gap ABSs and collective modes on other observables, such as the spin-spin and current-current response functions. In this paper, we have also considered electrons with a one-dimensional Fermi surface as a simple model. The systematic study of the magnetic and optical responses of FFLO states and the impact of the dimensionality of the Fermi surface remain future challenges.

\begin{acknowledgments}
We thank Y. Masaki for fruitful discussions and for carefully reading the manuscript. This work was supported by JST CREST Grant No. JPMJCR19T5, Japan, a Grant-in-Aid for Scientific Research on Innovative Areas ``Quantum Liquid Crystals'' (Grant No.~JP19H05825 and No.~JP22H04480) from JSPS of Japan, and JSPS KAKENHI (Grant No.~JP20K03860, No.~JP21H01039, and No.~JP22H01221).
\end{acknowledgments} 

\appendix

\section{Gauge choice and symmetry in the folded Brillouin zone}
\label{sec:gauge}

The quasiparticle dispersion in Fig.~\ref{fig:bdg}(c) is asymmetric with respect to $K_x$ and might appear to break the time-reversal symmetry that the FFLO state maintains. However, we note that the choice of the ${\rm U}(1)$ gauge ($\vartheta$) in Eq.~\eqref{eq:u0} sensitively affects the structure and symmetry of the quasiparticle dispersion in the Brillouin zone folded by the FFLO superlattice, though the self-consistent fields, thermodynamics, and collective modes are independent of $\vartheta$.

To clarify this, let us consider the BdG equation for the Bloch wavefunction of the quasiparticles, ${\bm w}_K(x_i)$. As the quasiparticle wavefunctions are obtained from Eq.~\eqref{eq:bloch4}, the BdG equation for ${\bm w}_K(x_i)$ reads 
\begin{align}
\left(\mathcal{H}_K^{\vartheta}\right)_{ij} \equiv& e^{-iKx_i}\mathcal{H}_{0,ij}e^{iKx_j} \nn \\
=& \begin{pmatrix}
(\xi_{K}^{\vartheta_+})_{ij} & \Delta (x_i)\delta_{ij} \\ \Delta(x_i)\delta_{ij} & 
-(\xi_{-K}^{-\vartheta_-})^{\ast}_{ij}
\end{pmatrix}, 
\end{align}
where $\vartheta _{\pm} \equiv \vartheta \pm \chi_0/2$. The expression of the BdG Hamiltonian depends on the choice of the gauge $\vartheta$ through the boundary condition in Eq.~\eqref{eq:w_bc}; The elements of the BdG Hamiltonian include the gauge-dependent terms,  
$(\xi_{K}^{\vartheta_+})_{1,N} = -t e^{-i(Ka+\vartheta_+)}$ and 
$(\xi_{K}^{\vartheta_+})_{N,1} = -t e^{i(Ka+\vartheta_+)}$.
The particle-hole operator in each spin sector, $\tilde{\mathcal{C}}$, transforms the BdG Hamiltonian as 
\beq
\tilde{\mathcal{C}}\mathcal{H}_K^{\vartheta}\tilde{\mathcal{C}}^{-1} = - \mathcal{H}_{-K}^{-\vartheta}.
\eeq
This indicates that $\tilde{\mathcal{C}}$ flips the sign of the ${\rm U}(1)$ phase, $\vartheta$, as well as the Bloch momentum. The particle-hole symmetry guarantees the symmetry of the quasiparticle dispersion in the folded Brillouin zone
\beq
E^{\vartheta} (K_x)= -E^{-\vartheta}(-K_x),
\eeq
where $E^{\vartheta}(K_x)$ is the quasiparticle energy with $K_x$ for the choice of the gauge, $\vartheta$. 

\begin{figure*}[t]
\includegraphics[width=180mm]{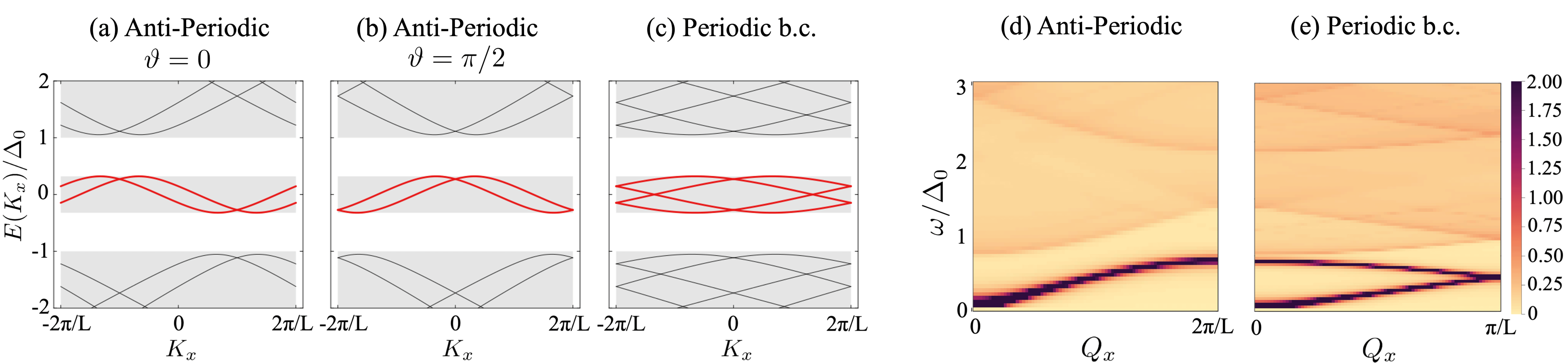}
\caption{(a-c) Eigenvalues of the BdG equation \eqref{eq:bdg0}, $E(K_x)$, in the FFLO state at $T = 0.04$ and $B/B_{\rm c} = 1.157$: (a) the gauge choice of $\vartheta=0$ and (b) $\vartheta=\pi/2$. In (c), we impose the periodic boundary condition, $\Delta(x+L)=\Delta(x)$, instead of the anti-periodic condition in Eq.~\eqref{eq:delta_periodic2}, which is free from the gauge choice. In (c), the Brillouin zone is folded to $K_x\in [-\pi/L,\pi/L]$. (d,e) Spectral functions in the amplitude sector, $\rho^+(Q_x,\omega)$, for the gauge choice of $\vartheta=0$ in the anti-periodic boundary condition (d) and for the periodic boundary condition (e). In (e), the Brillouin zone is folded to $Q_x\in [0,\pi/L]$. In both (d) and (e), the long-range Coulomb interaction is neglected.}
\label{fig:gauge}
\end{figure*}

Figures~\ref{fig:gauge}(a) and \ref{fig:gauge}(b) show the quasiparticle dispersion in the FFLO state for the gauge choice of $\vartheta=0$ and $\pi/2$, respectively. In the case of $\vartheta=0$, the dispersion obeys $E_{\alpha}(K_x)=-E_{\alpha}(-K_x)$, but $K_x=0$ is not the time-reversal invariant momentum as $E_{K_x} = -E_{-K_x}$. On the other hand, for $\vartheta=\pi/2$, $K_x=0$ is no longer the particle-hole symmetric point, while $K_x=0$ is the time-reversal invariant momentum, and the dispersion is symmetric on $K_x$ as $E_{K_x} = E_{-K_x}$. In the current work, we choose the gauge of $\vartheta=0$, since this choice enables the formulation of the collective excitations in a particle-hole symmetric manner and can reduce the computational cost. For comparison, in Fig.~\ref{fig:gauge}(c), we also plot the quasiparticle dispersion under the periodic boundary condition, $\Delta(x+L)=\Delta(x)$, instead of Eq.~\eqref{eq:delta_periodic2}. In this case, the unit cell for the periodic boundary condition is twice the size of that in Eq.~\eqref{eq:delta_periodic2}, leading to a folded Brillouin zone to $K_x\in [-\pi/L,\pi/L]$. In Figs.~\ref{fig:gauge}(d) and \ref{fig:gauge}(e), we show the spectral functions in the amplitude sector, $\rho^+(Q_x,\omega)$, for the gauge choice of $\vartheta=0$ in the anti-periodic boundary condition and under the periodic boundary condition, respectively. These two results are consistent, except for the fact that the momentum space is folded into $Q_x\in[0,\pi/L]$.

\begin{widetext}

\section{Derivation of the fluctuation action in Eq.~\eqref{eq:Sf2}}

In this section, we derive the effective action in Eq.~\eqref{eq:Sf2} from the quadratic form of the bosonic fluctuations in Eq.~\eqref{eq:Sf},
\begin{align}
\mathcal{S}_{\rm fluc} = \frac{1}{4} {\rm Tr}\left[ G\delta\Sigma G\delta\Sigma\right]
+ \sum_Q\sum _{s=\pm} \frac{s\delta\Delta^{s}_Q\delta\Delta^{s}_{-Q}}{4V}+\sum_Q\frac{\phi_{Q}\phi_{-Q}}{U_Q}.
\label{eq:Sf3}
\end{align}
Here, ${\rm Tr}[\cdots]$ denotes 
\begin{align}
{\rm Tr}\left( G\delta\Sigma \right)^2
=& T\sum_n T\sum_m \int d{\bm x}_i \int d{\bm x}_j
{\rm tr}_4\left[
\check{G}({\bm x}_i,{\bm x}_j,i\varepsilon_n)
\delta\check{\Sigma}({\bm x}_j,i\omega_m)
\check{G}({\bm x}_j,{\bm x}_i,i\varepsilon_n-i\omega_m)
\delta\check{\Sigma}({\bm x}_i,-i\omega_m)
\right],
\label{eq:trace}
\end{align}
where $\varepsilon_n = (2n+1)\pi T_{\rm c}$ and $\omega_m\equiv 2m \pi T$ ($n,m\in\mathbb{Z}$) are the Matsubara frequencies of fermions and bosons, respectively. Let us consider the situation in which the superconducting order modulates along the $x$-axis and has uniformity in the $y$-$z$ plane. Then, the Green's function and sefl-energy fluctuation can be expanded as 
\begin{gather}
\check{G}({\bm x}_i,{\bm x}_j,i\varepsilon_n) 
= \frac{1}{L_yL_z}\sum_{{\bm k}_{\parallel}}
e^{ik_y(y_i-y_j)+ik_z(z_i-z_j)}\check{G}_{ij}(k_{\parallel}),
\label{eq:Gf} \\
\delta\Sigma({\bm x},i\omega_m)=\frac{1}{L_yL_z}\sum_{{\bm q}_{\parallel}}e^{iq_yy+iq_zz}\delta\Sigma(x,Q_{\parallel}).
\end{gather}
Here we indotruce the abbreviations, $G_{ij}(k_{\parallel})\equiv G(x_i,x_j,{\bm k}_{\parallel},i\varepsilon_n)$, $k_{\parallel}\equiv ({\bm k}_{\parallel},i\varepsilon_n)$, and $Q_{\parallel}\equiv ({\bm q}_{\parallel},i\omega_m)$. By substituting this into Eq.~\eqref{eq:trace}, ${\rm Tr}\left( \check{\bf G}_{\rm eq}\delta \check{\bm{\Sigma}}\right)^2$ reads
\begin{align}
{\rm Tr}\left( \check{G}_{\rm eq}\delta \check{\Sigma}\right)^2
= \sum _{k_{\parallel},Q_{\parallel}}\int dx_i\int dx_j {\rm tr}\left[
\check{G}_{ij}(k_{\parallel})
\delta \check{\Sigma}\left(x_j,Q_{\parallel}\right)
\check{G}_{ji}(k_{\parallel}-Q_{\parallel})
\delta \check{\Sigma}\left(x_i,-Q_{\parallel}\right)
\right],
\label{eq:GSGS}
\end{align}
where $\sum_{k_{\parallel}}\equiv T\sum_m \frac{1}{L_yL_z}\sum_{{\bm k}_{\parallel}}$ and  $\sum_{Q_{\parallel}}\equiv T\sum_m \frac{1}{L_yL_z}\sum_{{\bm q}_{\parallel}}$.

The self-energy fluctuations can be block-diagonalized in terms of the $S_z=\pm \frac{1}{2}$ eigenstates as 
\begin{align}
\delta \check{\Sigma}(x,Q_{\parallel}) 
=& \begin{pmatrix}
ie\phi(x,Q_{\parallel}) & \delta\Delta(x,Q_{\parallel})i\sigma_y \\ 
-i\sigma_y\delta\Delta^{\ast}(x,-Q_{\parallel}) & -ie\phi(x,Q_{\parallel})
\end{pmatrix}
= \delta \hat{\Sigma}^{(+)}(x,Q_{\parallel}) \oplus 
\delta \hat{\Sigma}^{(-)}(x,Q_{\parallel}) .
\end{align}
Here, $\hat{\Sigma}^{(+)}$ and $\hat{\Sigma}^{(-)}$ are the self-energy fluctuations in the $S_z=\frac{1}{2}$ and $-\frac{1}{2}$ sectors respectively. 
Let us recast the $2\times 2$ matrix form of the self-energy fluctuations in the particle-hole space into
\begin{align}
\delta \hat{\Sigma}^{(+)}(x_i,Q_{\parallel})=  
ie\phi(x_i,Q_{\parallel}) \hat{\tau}_z + \frac{1}{2}\hat{\tau}_x\delta\Delta^+(x_i,Q_{\parallel}) 
+ \frac{1}{2}i\hat{\tau}_y\delta\Delta^-(x_i,Q_{\parallel})
\equiv\sum_{a=\phi,+,-}\Sigma_{a,i}(Q_{\parallel})\hat{\Lambda}_a,
\label{eq:delta_sigma_p} 
\end{align}
and $\delta \hat{\Sigma}^{(-)}=\tau_z\delta \hat{\Sigma}^{(+)}\tau_z$. Here, $\delta \Delta^{+}\equiv \delta \Delta + \delta \Delta^{\ast}$ and $\delta \Delta^{-}\equiv \delta \Delta - \delta \Delta^{\ast}$ describe the fluctuations of the real part (amplitude) and imaginary part (phase) of the gap function, respectively.
We have also introduced the vertex operators as 
$(\hat{\Lambda}_{+}, \hat{\Lambda}_{-}, \hat{\Lambda}_{\phi}) \equiv \hat{\tau}_x/2, i\hat{\tau}_y/2, 
ie\hat{\tau}_z)$.
As ${\rm Tr}( \check{\bf G}\delta \check{\bm{\Sigma}})^2$ does not have the mixing term of the $S_z=\pm \frac{1}{2}$ eignestates, Eq.~\eqref{eq:GSGS} reads
\begin{align}
{\rm Tr}\left( \check{G}\delta \check{\Sigma}\right)^2
=
2\sum_{a,b}\int dx_i\int dx_j\sum _{Q_{\parallel}} \Sigma_{a,i}(Q_{\parallel})
\chi^{ab}_{ij}(Q_{\parallel})\Sigma_{b,j}(-Q_{\parallel}),
\end{align}
where $x_i\in [0,LN_{\rm cell}]$ and ${\Sigma}_a$ denote the self-energy fluctuations, 
$\Sigma_{a=\phi,j}(Q_{\parallel})=\phi(x_j,Q_{\parallel})$ and $\Sigma_{a=\pm,j}(Q_{\parallel})=\delta\Delta^{\pm}(x_j,Q_{\parallel})$. We have introduced 
$\delta\Delta^{\pm}(x,Q_{\parallel}) \equiv \delta\Delta(x,Q_{\parallel}) \pm \delta\Delta^{\ast}(x,-Q_{\parallel})$, where 
$\delta\Delta^{+}$ and $\delta\Delta^{-}$ correspond to the fluctuations of the superconducting amplitude and phase, respectively. 
The polarization function $\chi_{ij}^{ab}$ ($a,b=\phi,+,-$) is defined as 
\begin{align}
\chi^{ab}_{ij}(Q_{\parallel}) = \frac{1}{2}\sum_{\sigma}\sum_{k_{\parallel}}{\rm tr}_2\left[
\hat{G}^{(\sigma)}_{ji}(k_{\parallel})\hat{\Lambda}_a
\hat{G}^{(\sigma)}_{ij}(k_{\parallel}-Q_{\parallel})\hat{\Lambda}_b
\right] .
\label{eq:chi}
\end{align}
To this end, the effective action for the fluctuations of the bosonic fields, 
$\bm{\Sigma}_i(Q_{\parallel})\equiv 
[\Sigma_{+,i}(Q_{\parallel}),\Sigma_{-,i}(Q_{\parallel}),\Sigma_{\phi,i}(Q_{\parallel})]^{\rm t}
= [
\delta\Delta^{+}(x_i,Q_{\parallel}),
\delta\Delta^{-}(x_i,Q_{\parallel}),
\phi(x_i,Q_{\parallel})
]^{\rm t}
$, 
reduces to Eq.~\eqref{eq:Sf2} as
\begin{align}
\mathcal{S}_{\rm fluc}[\bm{\Sigma}]
=\frac{1}{2}\sum_{Q_{\parallel}}\int dx_i \int dx_j
\bm{\Sigma}^{\rm t}_i(Q_{\parallel})
\mathcal{D}^{-1}_{ij}(Q_{\parallel}) 
\bm{\Sigma}_j(-Q_{\parallel}), 
\label{eq:Sfluct}
\end{align}
where $a^{\rm t}$ denotes the transpose of a matrix $a$ and the $3\times 3$ matrix of the fluctuation propagators is defined as 
\beq
\mathcal{D}_{ij}(Q_{\parallel}) 
\equiv 
\begin{pmatrix}
\displaystyle{\frac{1}{2V}\delta_{ij}+\chi^{++}_{ij}(Q_{\parallel})} & \chi^{+-}_{ij}(Q_{\parallel}) & \chi^{+\phi}_{ij}(Q_{\parallel}) \\
\chi^{-+}_{ij}(Q_{\parallel}) & \displaystyle{-\frac{1}{2V}\delta_{ij}+\chi^{--}_{ij}(Q_{\parallel})} & \chi^{-\phi}_{ij}(Q_{\parallel}) \\
\chi^{\phi+}_{ij}(Q_{\parallel}) & \chi^{\phi-}_{ij}(Q_{\parallel}) & 
\displaystyle{\delta_{ij}\frac{{\bm q}^2_{\parallel}-{\partial^2_{x_j}}}{4\pi}+\chi^{\phi\phi}_{ij}(Q_{\parallel})}
\end{pmatrix}^{-1}
\eeq
In numerical calculations, we consider the tight-biding model introduced in Sec.~\ref{sec:equilibrium}. Let us discretize a half period of the FFLO state, $L/2$, to $N$th grids with the lattice constant $a$. Then, the operator ${\partial^2_{x_i}}$ in Eq.~\eqref{eq:Sfluct} stands for $(\delta_{i,j+1}+\delta_{i+1,j}-2\delta_{ij})/a^2$ and the integral over $x_i\in[0,LN_{\rm cell})$, is replaced to $\frac{L}{2N}\sum_{x_i=1}^{2NN_{\rm cell}}$.

\end{widetext}

\section{Bare correlation functions}
\label{sec:chi}

In this Appendix, we summarize the expression of the bare correlation functions in terms of the eigenstates of the BdG Hamiltonian. Here we consider the BdG equation in the presence of the periodic potential described in Eq.~\eqref{eq:delta_periodic2}. Let $E_{\alpha,{\bm K}}$ and ${\bm w}_{\alpha,{\bm K}}$ be the eigenvalues and the eigenfunctions of the BdG Hamiltonian at zero field ($B=0$). Then, they are given by diagonalizing the BdG equation in the $S_z=+1$ sector,
\beq
\sum_j \mathcal{H}^{(0)}_{ij}({\bm K}) {\bm w}_{\alpha,{\bm K}}(x_j)
=E_{\alpha,{\bm K}}{\bm w}_{\alpha,{\bm K}}(x_j),
\eeq
where $\alpha$ labels the eigenstates for ${\bm K}\equiv (K_x,{\bm k}_{\parallel})$. The eigenfunction obeys the normalization condition in Eq.~\eqref{eq:normw}. The BdG Hamiltonian in the $S_z=+1$ sector at zero fields is obtained from Eq.~\eqref{eq:Heq} as $\mathcal{H}^{{(0)}}_{ij}({\bm K}) = e^{-i{\bm K}\cdot{\bm x}_i}\mathcal{H}^{{(0)}}_{ij} e^{i{\bm K}\cdot{\bm x}_j}$, where
\beq
\mathcal{H}^{{(0)}}_{ij}= \begin{pmatrix} 
\xi^{(0)}_{ij} & \Delta({\bm x}_i) \\ 
\Delta({\bm x}_i) & -\xi^{(0)}_{ij}
\end{pmatrix}.
\eeq
Then, the Green's functions in Eqs.~\eqref{eq:G+} and \eqref{eq:G-} can be rewritten in terms of the Bloch wave number $K_x$ as
\begin{align}
{G}^{\sigma}({\bm x}_i,{\bm x}_j,i\varepsilon_n) = \sum_{\bm K}e^{iK_x}e^{i{\bm k}_{\parallel}\cdot{\bm x}_{\parallel}}\tilde{G}^{\sigma}(x_i,x_j,{\bm K},i\varepsilon_n) ,
\label{eq:Gbloch}
\end{align}
where $ \sum_{\bm K}\equiv \frac{1}{2N_{\rm cell}}\sum_{K_x} \frac{1}{L_yL_z}\sum_{k_y,k_z}$. 
The Green's function for the Bloch wave number $K_x$ is given by 
\begin{align}
\tilde{G}^+_{ij}(K) 
=&\sum_{\alpha}{}^{\prime} \bigg[
\frac{{\bm w}_{\alpha,{\bm K}}(x_i)
{\bm w}^{\dag}_{\alpha,{\bm K}}(x_j)}{i\varepsilon_n-E^+_{\alpha,{\bm K}}}\nn \\
&+\frac{(i\tau_y{\bm w}^{(0)\ast}_{\alpha,-{\bm K}}(x_i))
(i\tau_y{\bm w}_{\alpha,-{\bm K}}(x_j))^{\rm t}}{i\varepsilon_n+{E^-_{\alpha,-{\bm K}}}}
\bigg], \label{eq:Gbloch+}\\
\tilde{G}^-_{ij}(K) 
=&\sum_{\alpha}{}^{\prime}\tau_z \bigg[
\frac{{\bm w}_{\alpha,{\bm K}}(x_i)
{\bm w}^{\dag}_{\alpha,{\bm K}}(x_j)}{i\varepsilon_n-E^-_{\alpha,{\bm K}}}\nn \\
&+\frac{(i\tau_y{\bm w}^{\ast}_{\alpha,-{\bm K}}(x_i))
(i\tau_y{\bm w}_{\alpha,-{\bm K}}(x_j))^{\rm t}}{i\varepsilon_n+{E^+_{\alpha,-{\bm K}}}}
\bigg]\tau_z,\label{eq:Gbloch-}
\end{align}
where $E^{\pm}_{\alpha,{\bm K}}= E_{\alpha,{\bm K}}\pm \mu_0 B$ is the quasiparticle energy under an applied magnetic field $B$.
We note that $\sum^{\prime}_{\alpha}$ implies the sum over $\alpha$ that satisfies $E_{\alpha,{\bm K}}>0$. From Eq.~\eqref{eq:w_bc}, the Green's functions obey the periodic boundary condition
\begin{align}
&\tilde{G}^{\pm} (x_i+L/2,x_j,K) 
= U^{\dag}_0\tilde{G}^{\pm}(x_i,x_j,K), 
\label{eq:Gperiod1}\\
&\tilde{G}^{\pm} (x_i,x_j+L/2,K)
= \tilde{G}^{\pm} (x_i,x_j,K)U_0.
\label{eq:Gperiod2}
\end{align}

The expression of the Green's functions can be further simplified to
\begin{gather}
\tilde{G}^{+}_{ij}(K) 
= {g}^{+}_{ij}(K) , \quad
\tilde{G}^{-}_{ij}(K) 
= \tau_z{g}^{-}_{ij}(K) \tau_z, 
\label{eq:G+-}
\end{gather}
with
\beq
{g}^{\sigma}_{ij}({\bm K},i\varepsilon_n)  
= \sum_{s=\pm 1}\sum_{\alpha}{}^{\prime} 
\frac{{\bm W}_{\alpha,s{\bm K}}(x_i)
{\bm W}^{\dag}_{\alpha,s{\bm K}}(x_j)}{i\varepsilon_n-sE^{(s\sigma)}_{\alpha,s{\bm K}}}.
\eeq
Here we set ${\bm W}_{\alpha,s{\bm K}}={\bm w}_{\alpha,{\bm K}}$ for $s=+$ and
${\bm W}_{\alpha,s{\bm K}}=i\tau_y{\bm w}^{\ast}_{\alpha,-{\bm K}}$ for $s=-$, where $s=\pm 1$ is the label associated with the particle-hole symmetry.

The effective action in Eq.~\eqref{eq:Sfluct} contains the integral over $x_i\in[0,LN_{\rm cell}]$. In the FFLO states, however, by using the antiperiodicity of the FFLO order in Eq.~\eqref{eq:delta_periodic2} and the Bloch wave functions in Eq.~\eqref{eq:bloch4}, the integral can be reduced to $x_i\in[0,L/2]$, the supercell of the FFLO order. To reduce the effective action in Eq.~\eqref{eq:Sfluct} to the supercell of the FFLO period $L/2$, let us first expand the bosonic fields in terms of the Bloch wave numbers $Q_x= 2\pi m/LN_{\rm cell}$ as 
\begin{gather}
\delta\Delta^{\pm}(x,Q_{\bm \parallel}) = \frac{1}{2N_{\rm cell}} \sum_{Q_x} e^{iQ_xx}\delta\tilde{\Delta}^{\pm}(x,Q_x,Q_{\bm \parallel}), 
\\
\phi(x,Q_{\bm \parallel}) = \frac{1}{2N_{\rm cell}}\sum_{Q_x} e^{iQ_xx}\tilde{\phi}(x,Q_x,Q_{\bm \parallel}),
\label{eq:B_bloch}
\end{gather}
where we set $\delta\tilde{\Delta}^{\pm}(x,Q_x,Q_{\bm \parallel}) \equiv \delta\tilde{\Delta}(x,Q_x,Q_{\bm \parallel}) \pm \delta\tilde{\Delta}^{\ast}(x,-Q_x,-Q_{\bm \parallel})$. The bosonic fields obey the periodic boundary conditions 
\begin{gather}
\delta\tilde{\Delta}^{\pm}(x+L/2,Q_x,Q_{\bm \parallel}) = e^{i\chi_0}\delta\tilde{\Delta}^{\pm}(x,Q_x,Q_{\bm \parallel}) ,\\ 
\tilde{\phi}^{\pm}(x+L/2,Q_x,Q_{\bm \parallel}) = \tilde{\phi}^{\pm}(x,Q_x,Q_{\bm \parallel}).
\end{gather}
By substituting this into Eq.~\eqref{eq:Sfluct} and using the boundary conditions in Eqs.~\eqref{eq:Gperiod1} and \eqref{eq:Gperiod2}, the fluctuation action is reduced to the integral over the FFLO supercell ${x}_i\in[0,L/2)$ as 
\begin{align}
\mathcal{S}_{\rm fluc}[\tilde{\bm{\Sigma}}]
=&\frac{1}{4N_{\rm cell}}\sum_{Q_{\parallel}}\int^{L/2}_0 dx_i \int^{L/2}_0 dx_j\nn \\
&\times
\bm{\Sigma}^{\rm t}_i(Q_{\parallel})
\mathcal{D}^{-1}_{ij}(Q_{\parallel}) 
\tilde{\bm{\Sigma}}_j(-Q_{\parallel}), 
\label{eq:Sfluct2}
\end{align}
where $\tilde{\bm{\Sigma}}$ composists of the fluctuations of the bosonic fields defined in the FFLO supercell, $\tilde{\bm{\Sigma}}^{\rm t}_i(Q)=[\delta\tilde{\Delta}^+(x_i,Q),\delta\tilde{\Delta}^+(x_i,Q),\tilde{\phi}(x_i,Q)]$.

The correlation functions are defined in Eq.~\eqref{eq:chi}. They can be expressed in terms of $\hat{\tilde{G}}^{\sigma}_{ji}(K)$ as 
\begin{align}
\tilde{\chi}^{ab}_{ij}(Q) 
\equiv&\sum_{\sigma,{K}}
\frac{1}{2}{\rm tr}_2\left[
\hat{\tilde{G}}^{\sigma}_{ji}(K)\hat{\Lambda}_a^{\sigma}
\hat{\tilde{G}}^{\sigma}_{ij}(K-Q)\hat{\Lambda}_b^{\sigma}
\right],
\label{eq:chiab}
\end{align}
where $Q\equiv (Q_x,{\bm q}_{\parallel},i\omega_m)$. Substituting Eqs.~\eqref{eq:Gbloch+} and \eqref{eq:Gbloch-} into Eq.~\eqref{eq:chiab} and performing the Matsubara sum over $\varepsilon_n$, the bare correlation function reads
\begin{align}
\tilde{\chi}^{ab}_{ij}(Q)
&=\sum_{s_1,s_2} \sum_{\alpha,\beta}{}^{\prime}\sum_{\bm K}\nn \\
&\times 
\left\{\frac{1}{2}\sum_{\sigma}
\frac{f(s_1E^{(s_1\sigma)}_{\alpha,s_1{\bm K}})-f(s_2E^{(s_2\sigma)}_{\beta,s_2({\bm K}-{\bm Q})})}
{-i\omega_m + s_1E^{(s_1\sigma)}_{\alpha,s_1{\bm K}} - s_2E^{(s_2\sigma)}_{\beta,s_2({\bm K}-{\bm Q})}}
\right\}\nn \\
& \times \left(
{\bm W}^{\dag}_{\alpha,s_1{\bm K}}(x_i)\hat{\Lambda}_a{\bm W}_{\beta,s_2({\bm K}-{\bm Q})}(x_i)
\right)\nn \\
& \times \left(
{\bm W}^{\dag}_{\beta,s_2({\bm K}-{\bm Q})}(x_j)\hat{\Lambda}_b{\bm W}_{\alpha,s_1{\bm K}}(x_j)
\right),
\end{align}
where $f(x)=1/(e^{x/T}+1) $ is the Fermi distribution function at temperature $T$. The expression of $\tilde{\chi}^{ab}$ includes the summation over the band ($\alpha,\beta$), the spin ($\sigma$), and the particle-hole symmetry ($s_1,s_2$) in addition to the momentum ${\bm K}$.

Lastly, we recast the expression of the bare correlation functions into a form that is suitable for numerical calculations. Let $N$ be the number of the eigenstates with positive eigenvalues. Then, we introduce the $N\times N$ matrices, $\mathbb{W}$ and $\bar{\mathbb{W}}$, as 
\begin{align}
&[\mathbb{W}_a(x_i)]_{\alpha\beta} 
\equiv {\bm W}^{\dag}_{\alpha,s_1{\bm K}}(x_i)\hat{\Lambda}_a{\bm W}_{\beta,s_2({\bm K}-{\bm Q})}(x_i), \\
&[\bar{\mathbb{W}}_b(x_j)]_{\alpha\beta}
\equiv {\bm W}^{\rm t}_{\alpha,s_1{\bm K}}(x_j)\hat{\Lambda}^{\rm t}_b{\bm W}^{\ast}_{\beta,s_2({\bm K}-{\bm Q})}(x_j).
\end{align}
Similarly, let $\mathbb{M}$ be the $N\times N$ matrix defined by
\begin{align}
[\mathbb{M}]_{\alpha\beta}
\equiv \frac{1}{2}\sum_{\sigma=\pm}
\frac{f(s_1E^{(s_1\sigma)}_{\alpha,s_1{\bm K}})-f(s_2E^{(s_2\sigma)}_{\beta,s_2({\bm K}-{\bm Q})})}
{-i\omega_m + s_1E^{(s_1\sigma)}_{\alpha,s_1{\bm K}} - s_2E^{(s_2\sigma)}_{\beta,s_2({\bm K}-{\bm Q})}}.
\end{align}
By using these, the expression of the bare correlation function, $\tilde{\chi}^{ab}_{ij}(Q)$, can be simplified to the form that is suitable for numerical calculations,
\begin{align}
\tilde{\chi}^{ab}_{ij}(Q)
=&\sum_{s_1,s_2}\sum_{\bm K} {\rm sum}\left[
\mathbb{M}\odot\mathbb{W}_a(x_i)\odot\bar{\mathbb{W}}_b(x_j)
\right],
\end{align}
where ${\rm sum}(\mathbb{A})$ implies the sum of all the elements in an entire array of $\mathbb{A}$ and $\mathbb{A}\odot\mathbb{B}$ denotes the Hadamard product of arbitrary matrices, $(\mathbb{A}\odot\mathbb{B})_{ij} =\mathbb{A}_{ij}\mathbb{B}_{ij}$. 

\bibliography{fflo}

\begin{thebibliography}{82}%
\makeatletter
\providecommand \@ifxundefined [1]{%
 \@ifx{#1\undefined}
}%
\providecommand \@ifnum [1]{%
 \ifnum #1\expandafter \@firstoftwo
 \else \expandafter \@secondoftwo
 \fi
}%
\providecommand \@ifx [1]{%
 \ifx #1\expandafter \@firstoftwo
 \else \expandafter \@secondoftwo
 \fi
}%
\providecommand \natexlab [1]{#1}%
\providecommand \enquote  [1]{``#1''}%
\providecommand \bibnamefont  [1]{#1}%
\providecommand \bibfnamefont [1]{#1}%
\providecommand \citenamefont [1]{#1}%
\providecommand \href@noop [0]{\@secondoftwo}%
\providecommand \href [0]{\begingroup \@sanitize@url \@href}%
\providecommand \@href[1]{\@@startlink{#1}\@@href}%
\providecommand \@@href[1]{\endgroup#1\@@endlink}%
\providecommand \@sanitize@url [0]{\catcode `\\12\catcode `\$12\catcode
  `\&12\catcode `\#12\catcode `\^12\catcode `\_12\catcode `\%12\relax}%
\providecommand \@@startlink[1]{}%
\providecommand \@@endlink[0]{}%
\providecommand \url  [0]{\begingroup\@sanitize@url \@url }%
\providecommand \@url [1]{\endgroup\@href {#1}{\urlprefix }}%
\providecommand \urlprefix  [0]{URL }%
\providecommand \Eprint [0]{\href }%
\providecommand \doibase [0]{https://doi.org/}%
\providecommand \selectlanguage [0]{\@gobble}%
\providecommand \bibinfo  [0]{\@secondoftwo}%
\providecommand \bibfield  [0]{\@secondoftwo}%
\providecommand \translation [1]{[#1]}%
\providecommand \BibitemOpen [0]{}%
\providecommand \bibitemStop [0]{}%
\providecommand \bibitemNoStop [0]{.\EOS\space}%
\providecommand \EOS [0]{\spacefactor3000\relax}%
\providecommand \BibitemShut  [1]{\csname bibitem#1\endcsname}%
\let\auto@bib@innerbib\@empty
\bibitem [{\citenamefont {Fulde}\ and\ \citenamefont {Ferrell}(1964)}]{ff}%
  \BibitemOpen
  \bibfield  {author} {\bibinfo {author} {\bibfnamefont {P.}~\bibnamefont
  {Fulde}}\ and\ \bibinfo {author} {\bibfnamefont {R.~A.}\ \bibnamefont
  {Ferrell}},\ }\bibfield  {title} {\bibinfo {title} {{Superconductivity in a
  Strong Spin-Exchange Field}},\ }\href
  {https://doi.org/10.1103/PhysRev.135.A550} {\bibfield  {journal} {\bibinfo
  {journal} {Phys. Rev.}\ }\textbf {\bibinfo {volume} {135}},\ \bibinfo {pages}
  {A550} (\bibinfo {year} {1964})}\BibitemShut {NoStop}%
\bibitem [{\citenamefont {Larkin}\ and\ \citenamefont
  {Ovchinnikov}(1964)}]{lo}%
  \BibitemOpen
  \bibfield  {author} {\bibinfo {author} {\bibfnamefont {A.~I.}\ \bibnamefont
  {Larkin}}\ and\ \bibinfo {author} {\bibfnamefont {Y.~N.}\ \bibnamefont
  {Ovchinnikov}},\ }\bibfield  {title} {\bibinfo {title} {{Nonuniform state of
  superconductors}},\ }\href@noop {} {\bibfield  {journal} {\bibinfo  {journal}
  {Zh. Eksp. Teor. Fiz.}\ }\textbf {\bibinfo {volume} {47}},\ \bibinfo {pages}
  {1136} (\bibinfo {year} {1964})},\ \bibinfo {note} {[Sov. Phys. JETP {\bf
  20}, 762 (1965)]}\BibitemShut {NoStop}%
\bibitem [{\citenamefont {Matsuda}\ and\ \citenamefont
  {Shimahara}(2007)}]{mat07}%
  \BibitemOpen
  \bibfield  {author} {\bibinfo {author} {\bibfnamefont {Y.}~\bibnamefont
  {Matsuda}}\ and\ \bibinfo {author} {\bibfnamefont {H.}~\bibnamefont
  {Shimahara}},\ }\bibfield  {title} {\bibinfo {title}
  {{Fulde-Ferrell-Larkin-Ovchinnikov State in Heavy Fermion Superconductors}},\
  }\href {https://doi.org/10.1143/JPSJ.76.051005} {\bibfield  {journal}
  {\bibinfo  {journal} {J. Phys. Soc. Jpn.}\ }\textbf {\bibinfo {volume}
  {76}},\ \bibinfo {pages} {051005} (\bibinfo {year} {2007})}\BibitemShut
  {NoStop}%
\bibitem [{\citenamefont {Beyer}\ and\ \citenamefont {Wosnitza}(2013)}]{bey13}%
  \BibitemOpen
  \bibfield  {author} {\bibinfo {author} {\bibfnamefont {R.}~\bibnamefont
  {Beyer}}\ and\ \bibinfo {author} {\bibfnamefont {J.}~\bibnamefont
  {Wosnitza}},\ }\bibfield  {title} {\bibinfo {title} {{Emerging evidence for
  FFLO states in layered organic superconductors (Review Article)}},\ }\href
  {https://doi.org/10.1063/1.4794996} {\bibfield  {journal} {\bibinfo
  {journal} {Low Temp. Phys.}\ }\textbf {\bibinfo {volume} {39}},\ \bibinfo
  {pages} {225} (\bibinfo {year} {2013})}\BibitemShut {NoStop}%
\bibitem [{\citenamefont {Wosnitza}(2018)}]{wos18}%
  \BibitemOpen
  \bibfield  {author} {\bibinfo {author} {\bibfnamefont {J.}~\bibnamefont
  {Wosnitza}},\ }\bibfield  {title} {\bibinfo {title} {{FFLO States in Layered
  Organic Superconductors}},\ }\href
  {https://doi.org/https://doi.org/10.1002/andp.201700282} {\bibfield
  {journal} {\bibinfo  {journal} {Ann. Phys. (Berlin)}\ }\textbf {\bibinfo
  {volume} {530}},\ \bibinfo {pages} {1700282} (\bibinfo {year}
  {2018})}\BibitemShut {NoStop}%
\bibitem [{\citenamefont {Mizushima}\ \emph {et~al.}(2005)\citenamefont
  {Mizushima}, \citenamefont {Machida},\ and\ \citenamefont {Ichioka}}]{miz05}%
  \BibitemOpen
  \bibfield  {author} {\bibinfo {author} {\bibfnamefont {T.}~\bibnamefont
  {Mizushima}}, \bibinfo {author} {\bibfnamefont {K.}~\bibnamefont {Machida}},\
  and\ \bibinfo {author} {\bibfnamefont {M.}~\bibnamefont {Ichioka}},\
  }\bibfield  {title} {\bibinfo {title} {{Direct Imaging of Spatially Modulated
  Superfluid Phases in Atomic Fermion Systems}},\ }\href
  {https://doi.org/10.1103/PhysRevLett.94.060404} {\bibfield  {journal}
  {\bibinfo  {journal} {Phys. Rev. Lett.}\ }\textbf {\bibinfo {volume} {94}},\
  \bibinfo {pages} {060404} (\bibinfo {year} {2005})}\BibitemShut {NoStop}%
\bibitem [{\citenamefont {Kinnunen}\ \emph {et~al.}(2018)\citenamefont
  {Kinnunen}, \citenamefont {Baarsma}, \citenamefont {Martikainen},\ and\
  \citenamefont {T\"{o}rm\"{a}}}]{kin18}%
  \BibitemOpen
  \bibfield  {author} {\bibinfo {author} {\bibfnamefont {J.~J.}\ \bibnamefont
  {Kinnunen}}, \bibinfo {author} {\bibfnamefont {J.~E.}\ \bibnamefont
  {Baarsma}}, \bibinfo {author} {\bibfnamefont {J.-P.}\ \bibnamefont
  {Martikainen}},\ and\ \bibinfo {author} {\bibfnamefont {P.}~\bibnamefont
  {T\"{o}rm\"{a}}},\ }\bibfield  {title} {\bibinfo {title} {{The
  Fulde-Ferrell-Larkin-Ovchinnikov state for ultracold fermions in lattice and
  harmonic potentials: a review}},\ }\href
  {https://doi.org/10.1088/1361-6633/aaa4ad} {\bibfield  {journal} {\bibinfo
  {journal} {Rep. Prog. Phys.}\ }\textbf {\bibinfo {volume} {81}},\ \bibinfo
  {pages} {046401} (\bibinfo {year} {2018})}\BibitemShut {NoStop}%
\bibitem [{\citenamefont {Sedrakian}\ and\ \citenamefont
  {Clark}(2019)}]{sed19}%
  \BibitemOpen
  \bibfield  {author} {\bibinfo {author} {\bibfnamefont {A.}~\bibnamefont
  {Sedrakian}}\ and\ \bibinfo {author} {\bibfnamefont {J.~W.}\ \bibnamefont
  {Clark}},\ }\bibfield  {title} {\bibinfo {title} {{Superfluidity in nuclear
  systems and neutron stars}},\ }\href
  {https://doi.org/10.1140/epja/i2019-12863-6} {\bibfield  {journal} {\bibinfo
  {journal} {Eur. Phys. J. A}\ }\textbf {\bibinfo {volume} {55}},\ \bibinfo
  {pages} {167} (\bibinfo {year} {2019})}\BibitemShut {NoStop}%
\bibitem [{\citenamefont {Casalbuoni}\ and\ \citenamefont
  {Nardulli}(2004)}]{cas04}%
  \BibitemOpen
  \bibfield  {author} {\bibinfo {author} {\bibfnamefont {R.}~\bibnamefont
  {Casalbuoni}}\ and\ \bibinfo {author} {\bibfnamefont {G.}~\bibnamefont
  {Nardulli}},\ }\bibfield  {title} {\bibinfo {title} {{Inhomogeneous
  superconductivity in condensed matter and QCD}},\ }\href
  {https://doi.org/10.1103/RevModPhys.76.263} {\bibfield  {journal} {\bibinfo
  {journal} {Rev. Mod. Phys.}\ }\textbf {\bibinfo {volume} {76}},\ \bibinfo
  {pages} {263} (\bibinfo {year} {2004})}\BibitemShut {NoStop}%
\bibitem [{\citenamefont {Kinjo}\ \emph {et~al.}(2022)\citenamefont {Kinjo},
  \citenamefont {Manago}, \citenamefont {Kitagawa}, \citenamefont {Mao},
  \citenamefont {Yonezawa}, \citenamefont {Maeno},\ and\ \citenamefont
  {Ishida}}]{kin22}%
  \BibitemOpen
  \bibfield  {author} {\bibinfo {author} {\bibfnamefont {K.}~\bibnamefont
  {Kinjo}}, \bibinfo {author} {\bibfnamefont {M.}~\bibnamefont {Manago}},
  \bibinfo {author} {\bibfnamefont {S.}~\bibnamefont {Kitagawa}}, \bibinfo
  {author} {\bibfnamefont {Z.~Q.}\ \bibnamefont {Mao}}, \bibinfo {author}
  {\bibfnamefont {S.}~\bibnamefont {Yonezawa}}, \bibinfo {author}
  {\bibfnamefont {Y.}~\bibnamefont {Maeno}},\ and\ \bibinfo {author}
  {\bibfnamefont {K.}~\bibnamefont {Ishida}},\ }\bibfield  {title} {\bibinfo
  {title} {{Superconducting spin smecticity evidencing the
  Fulde-Ferrell-Larkin-Ovchinnikov state in Sr$_2$RuO$_4$}},\ }\href
  {https://doi.org/10.1126/science.abb0332} {\bibfield  {journal} {\bibinfo
  {journal} {Science}\ }\textbf {\bibinfo {volume} {376}},\ \bibinfo {pages}
  {397} (\bibinfo {year} {2022})}\BibitemShut {NoStop}%
\bibitem [{\citenamefont {Cho}\ \emph {et~al.}(2017)\citenamefont {Cho},
  \citenamefont {Yang}, \citenamefont {Yuan}, \citenamefont {Shen},
  \citenamefont {Wolf},\ and\ \citenamefont {Lortz}}]{cho17}%
  \BibitemOpen
  \bibfield  {author} {\bibinfo {author} {\bibfnamefont {C.-w.}\ \bibnamefont
  {Cho}}, \bibinfo {author} {\bibfnamefont {J.~H.}\ \bibnamefont {Yang}},
  \bibinfo {author} {\bibfnamefont {N.~F.~Q.}\ \bibnamefont {Yuan}}, \bibinfo
  {author} {\bibfnamefont {J.}~\bibnamefont {Shen}}, \bibinfo {author}
  {\bibfnamefont {T.}~\bibnamefont {Wolf}},\ and\ \bibinfo {author}
  {\bibfnamefont {R.}~\bibnamefont {Lortz}},\ }\bibfield  {title} {\bibinfo
  {title} {{Thermodynamic Evidence for the Fulde-Ferrell-Larkin-Ovchinnikov
  State in the ${\mathrm{KFe}}_{2}{\mathrm{As}}_{2}$ Superconductor}},\ }\href
  {https://doi.org/10.1103/PhysRevLett.119.217002} {\bibfield  {journal}
  {\bibinfo  {journal} {Phys. Rev. Lett.}\ }\textbf {\bibinfo {volume} {119}},\
  \bibinfo {pages} {217002} (\bibinfo {year} {2017})}\BibitemShut {NoStop}%
\bibitem [{\citenamefont {Kasahara}\ \emph {et~al.}(2020)\citenamefont
  {Kasahara}, \citenamefont {Sato}, \citenamefont {Licciardello}, \citenamefont
  {\ifmmode~\check{C}\else \v{C}\fi{}ulo}, \citenamefont
  {Arsenijevi\ifmmode~\acute{c}\else \'{c}\fi{}}, \citenamefont {Ottenbros},
  \citenamefont {Tominaga}, \citenamefont {B\"oker}, \citenamefont {Eremin},
  \citenamefont {Shibauchi}, \citenamefont {Wosnitza}, \citenamefont {Hussey},\
  and\ \citenamefont {Matsuda}}]{kas20}%
  \BibitemOpen
  \bibfield  {author} {\bibinfo {author} {\bibfnamefont {S.}~\bibnamefont
  {Kasahara}}, \bibinfo {author} {\bibfnamefont {Y.}~\bibnamefont {Sato}},
  \bibinfo {author} {\bibfnamefont {S.}~\bibnamefont {Licciardello}}, \bibinfo
  {author} {\bibfnamefont {M.}~\bibnamefont {\ifmmode~\check{C}\else
  \v{C}\fi{}ulo}}, \bibinfo {author} {\bibfnamefont {S.}~\bibnamefont
  {Arsenijevi\ifmmode~\acute{c}\else \'{c}\fi{}}}, \bibinfo {author}
  {\bibfnamefont {T.}~\bibnamefont {Ottenbros}}, \bibinfo {author}
  {\bibfnamefont {T.}~\bibnamefont {Tominaga}}, \bibinfo {author}
  {\bibfnamefont {J.}~\bibnamefont {B\"oker}}, \bibinfo {author} {\bibfnamefont
  {I.}~\bibnamefont {Eremin}}, \bibinfo {author} {\bibfnamefont
  {T.}~\bibnamefont {Shibauchi}}, \bibinfo {author} {\bibfnamefont
  {J.}~\bibnamefont {Wosnitza}}, \bibinfo {author} {\bibfnamefont {N.~E.}\
  \bibnamefont {Hussey}},\ and\ \bibinfo {author} {\bibfnamefont
  {Y.}~\bibnamefont {Matsuda}},\ }\bibfield  {title} {\bibinfo {title}
  {{Evidence for an Fulde-Ferrell-Larkin-Ovchinnikov State with Segmented
  Vortices in the BCS-BEC-Crossover Superconductor FeSe}},\ }\href
  {https://doi.org/10.1103/PhysRevLett.124.107001} {\bibfield  {journal}
  {\bibinfo  {journal} {Phys. Rev. Lett.}\ }\textbf {\bibinfo {volume} {124}},\
  \bibinfo {pages} {107001} (\bibinfo {year} {2020})}\BibitemShut {NoStop}%
\bibitem [{\citenamefont {Kasahara}\ \emph {et~al.}(2021)\citenamefont
  {Kasahara}, \citenamefont {Suzuki}, \citenamefont {Machida}, \citenamefont
  {Sato}, \citenamefont {Ukai}, \citenamefont {Murayama}, \citenamefont
  {Suetsugu}, \citenamefont {Kasahara}, \citenamefont {Shibauchi},
  \citenamefont {Hanaguri},\ and\ \citenamefont {Matsuda}}]{kas21}%
  \BibitemOpen
  \bibfield  {author} {\bibinfo {author} {\bibfnamefont {S.}~\bibnamefont
  {Kasahara}}, \bibinfo {author} {\bibfnamefont {H.}~\bibnamefont {Suzuki}},
  \bibinfo {author} {\bibfnamefont {T.}~\bibnamefont {Machida}}, \bibinfo
  {author} {\bibfnamefont {Y.}~\bibnamefont {Sato}}, \bibinfo {author}
  {\bibfnamefont {Y.}~\bibnamefont {Ukai}}, \bibinfo {author} {\bibfnamefont
  {H.}~\bibnamefont {Murayama}}, \bibinfo {author} {\bibfnamefont
  {S.}~\bibnamefont {Suetsugu}}, \bibinfo {author} {\bibfnamefont
  {Y.}~\bibnamefont {Kasahara}}, \bibinfo {author} {\bibfnamefont
  {T.}~\bibnamefont {Shibauchi}}, \bibinfo {author} {\bibfnamefont
  {T.}~\bibnamefont {Hanaguri}},\ and\ \bibinfo {author} {\bibfnamefont
  {Y.}~\bibnamefont {Matsuda}},\ }\bibfield  {title} {\bibinfo {title}
  {{Quasiparticle Nodal Plane in the Fulde-Ferrell-Larkin-Ovchinnikov State of
  FeSe}},\ }\href {https://doi.org/10.1103/PhysRevLett.127.257001} {\bibfield
  {journal} {\bibinfo  {journal} {Phys. Rev. Lett.}\ }\textbf {\bibinfo
  {volume} {127}},\ \bibinfo {pages} {257001} (\bibinfo {year}
  {2021})}\BibitemShut {NoStop}%
\bibitem [{\citenamefont {Shimano}\ and\ \citenamefont {Tsuji}(2020)}]{shi20}%
  \BibitemOpen
  \bibfield  {author} {\bibinfo {author} {\bibfnamefont {R.}~\bibnamefont
  {Shimano}}\ and\ \bibinfo {author} {\bibfnamefont {N.}~\bibnamefont
  {Tsuji}},\ }\bibfield  {title} {\bibinfo {title} {{Higgs Mode in
  Superconductors}},\ }\href
  {https://doi.org/10.1146/annurev-conmatphys-031119-050813} {\bibfield
  {journal} {\bibinfo  {journal} {Annu. Rev. Condens. Matter Phys.}\ }\textbf
  {\bibinfo {volume} {11}},\ \bibinfo {pages} {103} (\bibinfo {year}
  {2020})}\BibitemShut {NoStop}%
\bibitem [{\citenamefont {Lin}\ \emph {et~al.}(2020)\citenamefont {Lin},
  \citenamefont {Kim}, \citenamefont {Bauer}, \citenamefont {Ronning},
  \citenamefont {Thompson},\ and\ \citenamefont {Movshovich}}]{lin20}%
  \BibitemOpen
  \bibfield  {author} {\bibinfo {author} {\bibfnamefont {S.-Z.}\ \bibnamefont
  {Lin}}, \bibinfo {author} {\bibfnamefont {D.~Y.}\ \bibnamefont {Kim}},
  \bibinfo {author} {\bibfnamefont {E.~D.}\ \bibnamefont {Bauer}}, \bibinfo
  {author} {\bibfnamefont {F.}~\bibnamefont {Ronning}}, \bibinfo {author}
  {\bibfnamefont {J.~D.}\ \bibnamefont {Thompson}},\ and\ \bibinfo {author}
  {\bibfnamefont {R.}~\bibnamefont {Movshovich}},\ }\bibfield  {title}
  {\bibinfo {title} {{Interplay of the Spin Density Wave and a Possible
  Fulde-Ferrell-Larkin-Ovchinnikov State in ${\mathrm{CeCoIn}}_{5}$ in Rotating
  Magnetic Field}},\ }\href {https://doi.org/10.1103/PhysRevLett.124.217001}
  {\bibfield  {journal} {\bibinfo  {journal} {Phys. Rev. Lett.}\ }\textbf
  {\bibinfo {volume} {124}},\ \bibinfo {pages} {217001} (\bibinfo {year}
  {2020})}\BibitemShut {NoStop}%
\bibitem [{\citenamefont {Kittaka}\ \emph {et~al.}(2023)\citenamefont
  {Kittaka}, \citenamefont {Kono}, \citenamefont {Tsunashima}, \citenamefont
  {Kimoto}, \citenamefont {Yokoyama}, \citenamefont {Shimizu}, \citenamefont
  {Sakakibara}, \citenamefont {Yamashita},\ and\ \citenamefont
  {Machida}}]{kit23}%
  \BibitemOpen
  \bibfield  {author} {\bibinfo {author} {\bibfnamefont {S.}~\bibnamefont
  {Kittaka}}, \bibinfo {author} {\bibfnamefont {Y.}~\bibnamefont {Kono}},
  \bibinfo {author} {\bibfnamefont {K.}~\bibnamefont {Tsunashima}}, \bibinfo
  {author} {\bibfnamefont {D.}~\bibnamefont {Kimoto}}, \bibinfo {author}
  {\bibfnamefont {M.}~\bibnamefont {Yokoyama}}, \bibinfo {author}
  {\bibfnamefont {Y.}~\bibnamefont {Shimizu}}, \bibinfo {author} {\bibfnamefont
  {T.}~\bibnamefont {Sakakibara}}, \bibinfo {author} {\bibfnamefont
  {M.}~\bibnamefont {Yamashita}},\ and\ \bibinfo {author} {\bibfnamefont
  {K.}~\bibnamefont {Machida}},\ }\bibfield  {title} {\bibinfo {title}
  {{Modulation vector of the Fulde-Ferrell-Larkin-Ovchinnikov state in
  ${\mathrm{CeCoIn}}_{5}$ revealed by high-resolution magnetostriction
  measurements}},\ }\href {https://doi.org/10.1103/PhysRevB.107.L220505}
  {\bibfield  {journal} {\bibinfo  {journal} {Phys. Rev. B}\ }\textbf {\bibinfo
  {volume} {107}},\ \bibinfo {pages} {L220505} (\bibinfo {year}
  {2023})}\BibitemShut {NoStop}%
\bibitem [{\citenamefont {Kitagawa}\ \emph {et~al.}(2018)\citenamefont
  {Kitagawa}, \citenamefont {Nakamine}, \citenamefont {Ishida}, \citenamefont
  {Jeevan}, \citenamefont {Geibel},\ and\ \citenamefont {Steglich}}]{kit18}%
  \BibitemOpen
  \bibfield  {author} {\bibinfo {author} {\bibfnamefont {S.}~\bibnamefont
  {Kitagawa}}, \bibinfo {author} {\bibfnamefont {G.}~\bibnamefont {Nakamine}},
  \bibinfo {author} {\bibfnamefont {K.}~\bibnamefont {Ishida}}, \bibinfo
  {author} {\bibfnamefont {H.~S.}\ \bibnamefont {Jeevan}}, \bibinfo {author}
  {\bibfnamefont {C.}~\bibnamefont {Geibel}},\ and\ \bibinfo {author}
  {\bibfnamefont {F.}~\bibnamefont {Steglich}},\ }\bibfield  {title} {\bibinfo
  {title} {{Evidence for the Presence of the Fulde-Ferrell-Larkin-Ovchinnikov
  State in ${\mathrm{CeCu}}_{2}{\mathrm{Si}}_{2}$ Revealed Using
  $^{63}\mathrm{Cu}$ NMR}},\ }\href
  {https://doi.org/10.1103/PhysRevLett.121.157004} {\bibfield  {journal}
  {\bibinfo  {journal} {Phys. Rev. Lett.}\ }\textbf {\bibinfo {volume} {121}},\
  \bibinfo {pages} {157004} (\bibinfo {year} {2018})}\BibitemShut {NoStop}%
\bibitem [{\citenamefont {Cho}\ \emph {et~al.}(2021)\citenamefont {Cho},
  \citenamefont {Lyu}, \citenamefont {Ng}, \citenamefont {He}, \citenamefont
  {Lo}, \citenamefont {Chareev}, \citenamefont {Abdel-Baset}, \citenamefont
  {Abdel-Hafiez},\ and\ \citenamefont {Lortz}}]{cho21}%
  \BibitemOpen
  \bibfield  {author} {\bibinfo {author} {\bibfnamefont {C.-w.}\ \bibnamefont
  {Cho}}, \bibinfo {author} {\bibfnamefont {J.}~\bibnamefont {Lyu}}, \bibinfo
  {author} {\bibfnamefont {C.~Y.}\ \bibnamefont {Ng}}, \bibinfo {author}
  {\bibfnamefont {J.~J.}\ \bibnamefont {He}}, \bibinfo {author} {\bibfnamefont
  {K.~T.}\ \bibnamefont {Lo}}, \bibinfo {author} {\bibfnamefont
  {D.}~\bibnamefont {Chareev}}, \bibinfo {author} {\bibfnamefont {T.~A.}\
  \bibnamefont {Abdel-Baset}}, \bibinfo {author} {\bibfnamefont
  {M.}~\bibnamefont {Abdel-Hafiez}},\ and\ \bibinfo {author} {\bibfnamefont
  {R.}~\bibnamefont {Lortz}},\ }\bibfield  {title} {\bibinfo {title} {{Evidence
  for the Fulde-Ferrell-Larkin-Ovchinnikov state in bulk NbS$_2$}},\ }\href
  {https://doi.org/10.1038/s41467-021-23976-2} {\bibfield  {journal} {\bibinfo
  {journal} {Nat. Commun.}\ }\textbf {\bibinfo {volume} {12}},\ \bibinfo
  {pages} {3676} (\bibinfo {year} {2021})}\BibitemShut {NoStop}%
\bibitem [{\citenamefont {Sugiura}\ \emph {et~al.}(2019)\citenamefont
  {Sugiura}, \citenamefont {Isono}, \citenamefont {Terashima}, \citenamefont
  {Yasuzuka}, \citenamefont {Schlueter},\ and\ \citenamefont {Uji}}]{sug19}%
  \BibitemOpen
  \bibfield  {author} {\bibinfo {author} {\bibfnamefont {S.}~\bibnamefont
  {Sugiura}}, \bibinfo {author} {\bibfnamefont {T.}~\bibnamefont {Isono}},
  \bibinfo {author} {\bibfnamefont {T.}~\bibnamefont {Terashima}}, \bibinfo
  {author} {\bibfnamefont {S.}~\bibnamefont {Yasuzuka}}, \bibinfo {author}
  {\bibfnamefont {J.~A.}\ \bibnamefont {Schlueter}},\ and\ \bibinfo {author}
  {\bibfnamefont {S.}~\bibnamefont {Uji}},\ }\bibfield  {title} {\bibinfo
  {title} {{Fulde-Ferrell-Larkin-Ovchinnikov and vortex phases in a layered
  organic superconductor}},\ }\href {https://doi.org/10.1038/s41535-019-0147-2}
  {\bibfield  {journal} {\bibinfo  {journal} {npj Quantum Mater.}\ }\textbf
  {\bibinfo {volume} {4}},\ \bibinfo {pages} {7} (\bibinfo {year}
  {2019})}\BibitemShut {NoStop}%
\bibitem [{\citenamefont {Imajo}\ \emph {et~al.}(2022)\citenamefont {Imajo},
  \citenamefont {Nomura}, \citenamefont {Kohama},\ and\ \citenamefont
  {Kindo}}]{ima22}%
  \BibitemOpen
  \bibfield  {author} {\bibinfo {author} {\bibfnamefont {S.}~\bibnamefont
  {Imajo}}, \bibinfo {author} {\bibfnamefont {T.}~\bibnamefont {Nomura}},
  \bibinfo {author} {\bibfnamefont {Y.}~\bibnamefont {Kohama}},\ and\ \bibinfo
  {author} {\bibfnamefont {K.}~\bibnamefont {Kindo}},\ }\bibfield  {title}
  {\bibinfo {title} {{Emergent anisotropy in the
  Fulde-Ferrell-Larkin-Ovchinnikov state}},\ }\href
  {https://doi.org/10.1038/s41467-022-33354-1} {\bibfield  {journal} {\bibinfo
  {journal} {Nat. Commun.}\ }\textbf {\bibinfo {volume} {13}},\ \bibinfo
  {pages} {5590} (\bibinfo {year} {2022})}\BibitemShut {NoStop}%
\bibitem [{\citenamefont {Kotte}\ \emph {et~al.}(2022)\citenamefont {Kotte},
  \citenamefont {K\"uhne}, \citenamefont {Schlueter}, \citenamefont
  {Zwicknagl},\ and\ \citenamefont {Wosnitza}}]{kot22}%
  \BibitemOpen
  \bibfield  {author} {\bibinfo {author} {\bibfnamefont {T.}~\bibnamefont
  {Kotte}}, \bibinfo {author} {\bibfnamefont {H.}~\bibnamefont {K\"uhne}},
  \bibinfo {author} {\bibfnamefont {J.~A.}\ \bibnamefont {Schlueter}}, \bibinfo
  {author} {\bibfnamefont {G.}~\bibnamefont {Zwicknagl}},\ and\ \bibinfo
  {author} {\bibfnamefont {J.}~\bibnamefont {Wosnitza}},\ }\bibfield  {title}
  {\bibinfo {title} {Orbital-induced crossover of the
  fulde-ferrell-larkin-ovchinnikov phase into abrikosov-like states},\ }\href
  {https://doi.org/10.1103/PhysRevB.106.L060503} {\bibfield  {journal}
  {\bibinfo  {journal} {Phys. Rev. B}\ }\textbf {\bibinfo {volume} {106}},\
  \bibinfo {pages} {L060503} (\bibinfo {year} {2022})}\BibitemShut {NoStop}%
\bibitem [{\citenamefont {Molatta}\ \emph {et~al.}(2024)\citenamefont
  {Molatta}, \citenamefont {Kotte}, \citenamefont {Opherden}, \citenamefont
  {Koutroulakis}, \citenamefont {Schlueter}, \citenamefont {Zwicknagl},
  \citenamefont {Brown}, \citenamefont {Wosnitza},\ and\ \citenamefont
  {K\"uhne}}]{mol24}%
  \BibitemOpen
  \bibfield  {author} {\bibinfo {author} {\bibfnamefont {S.}~\bibnamefont
  {Molatta}}, \bibinfo {author} {\bibfnamefont {T.}~\bibnamefont {Kotte}},
  \bibinfo {author} {\bibfnamefont {D.}~\bibnamefont {Opherden}}, \bibinfo
  {author} {\bibfnamefont {G.}~\bibnamefont {Koutroulakis}}, \bibinfo {author}
  {\bibfnamefont {J.~A.}\ \bibnamefont {Schlueter}}, \bibinfo {author}
  {\bibfnamefont {G.}~\bibnamefont {Zwicknagl}}, \bibinfo {author}
  {\bibfnamefont {S.~E.}\ \bibnamefont {Brown}}, \bibinfo {author}
  {\bibfnamefont {J.}~\bibnamefont {Wosnitza}},\ and\ \bibinfo {author}
  {\bibfnamefont {H.}~\bibnamefont {K\"uhne}},\ }\bibfield  {title} {\bibinfo
  {title} {{Order-parameter evolution in the Fulde-Ferrell-Larkin-Ovchinnikov
  phase}},\ }\href {https://doi.org/10.1103/PhysRevB.109.L020504} {\bibfield
  {journal} {\bibinfo  {journal} {Phys. Rev. B}\ }\textbf {\bibinfo {volume}
  {109}},\ \bibinfo {pages} {L020504} (\bibinfo {year} {2024})}\BibitemShut
  {NoStop}%
\bibitem [{\citenamefont {Devarakonda}\ \emph {et~al.}(2021)\citenamefont
  {Devarakonda}, \citenamefont {Suzuki}, \citenamefont {Fang}, \citenamefont
  {Zhu}, \citenamefont {Graf}, \citenamefont {Kriener}, \citenamefont {Fu},
  \citenamefont {Kaxiras},\ and\ \citenamefont {Checkelsky}}]{dev21}%
  \BibitemOpen
  \bibfield  {author} {\bibinfo {author} {\bibfnamefont {A.}~\bibnamefont
  {Devarakonda}}, \bibinfo {author} {\bibfnamefont {T.}~\bibnamefont {Suzuki}},
  \bibinfo {author} {\bibfnamefont {S.}~\bibnamefont {Fang}}, \bibinfo {author}
  {\bibfnamefont {J.}~\bibnamefont {Zhu}}, \bibinfo {author} {\bibfnamefont
  {D.}~\bibnamefont {Graf}}, \bibinfo {author} {\bibfnamefont {M.}~\bibnamefont
  {Kriener}}, \bibinfo {author} {\bibfnamefont {L.}~\bibnamefont {Fu}},
  \bibinfo {author} {\bibfnamefont {E.}~\bibnamefont {Kaxiras}},\ and\ \bibinfo
  {author} {\bibfnamefont {J.~G.}\ \bibnamefont {Checkelsky}},\ }\bibfield
  {title} {\bibinfo {title} {{Signatures of bosonic Landau levels in a
  finite-momentum superconductor}},\ }\href
  {https://doi.org/10.1038/s41586-021-03915-3} {\bibfield  {journal} {\bibinfo
  {journal} {Nature}\ }\textbf {\bibinfo {volume} {599}},\ \bibinfo {pages}
  {51} (\bibinfo {year} {2021})}\BibitemShut {NoStop}%
\bibitem [{\citenamefont {Sohn}\ \emph {et~al.}(2018)\citenamefont {Sohn},
  \citenamefont {Xi}, \citenamefont {He}, \citenamefont {Jiang}, \citenamefont
  {Wang}, \citenamefont {Kang}, \citenamefont {Park}, \citenamefont {Berger},
  \citenamefont {Forr\'{o}}, \citenamefont {Law}, \citenamefont {Shan},\ and\
  \citenamefont {Mak}}]{soh18}%
  \BibitemOpen
  \bibfield  {author} {\bibinfo {author} {\bibfnamefont {E.}~\bibnamefont
  {Sohn}}, \bibinfo {author} {\bibfnamefont {X.}~\bibnamefont {Xi}}, \bibinfo
  {author} {\bibfnamefont {W.-Y.}\ \bibnamefont {He}}, \bibinfo {author}
  {\bibfnamefont {S.}~\bibnamefont {Jiang}}, \bibinfo {author} {\bibfnamefont
  {Z.}~\bibnamefont {Wang}}, \bibinfo {author} {\bibfnamefont {K.}~\bibnamefont
  {Kang}}, \bibinfo {author} {\bibfnamefont {J.-H.}\ \bibnamefont {Park}},
  \bibinfo {author} {\bibfnamefont {H.}~\bibnamefont {Berger}}, \bibinfo
  {author} {\bibfnamefont {L.}~\bibnamefont {Forr\'{o}}}, \bibinfo {author}
  {\bibfnamefont {K.~T.}\ \bibnamefont {Law}}, \bibinfo {author} {\bibfnamefont
  {J.}~\bibnamefont {Shan}},\ and\ \bibinfo {author} {\bibfnamefont {K.~F.}\
  \bibnamefont {Mak}},\ }\bibfield  {title} {\bibinfo {title} {{An unusual
  continuous paramagnetic-limited superconducting phase transition in 2D
  NbSe$_2$}},\ }\href {https://doi.org/10.1038/s41563-018-0061-1} {\bibfield
  {journal} {\bibinfo  {journal} {Nat. Mater.}\ }\textbf {\bibinfo {volume}
  {17}},\ \bibinfo {pages} {504} (\bibinfo {year} {2018})}\BibitemShut
  {NoStop}%
\bibitem [{\citenamefont {Wan}\ \emph {et~al.}(2023)\citenamefont {Wan},
  \citenamefont {Zheliuk}, \citenamefont {Yuan}, \citenamefont {Peng},
  \citenamefont {Zhang}, \citenamefont {Liang}, \citenamefont {Zeitler},
  \citenamefont {Wiedmann}, \citenamefont {Hussey}, \citenamefont {Palstra},\
  and\ \citenamefont {Ye}}]{wan23}%
  \BibitemOpen
  \bibfield  {author} {\bibinfo {author} {\bibfnamefont {P.}~\bibnamefont
  {Wan}}, \bibinfo {author} {\bibfnamefont {O.}~\bibnamefont {Zheliuk}},
  \bibinfo {author} {\bibfnamefont {N.~F.~Q.}\ \bibnamefont {Yuan}}, \bibinfo
  {author} {\bibfnamefont {X.}~\bibnamefont {Peng}}, \bibinfo {author}
  {\bibfnamefont {L.}~\bibnamefont {Zhang}}, \bibinfo {author} {\bibfnamefont
  {M.}~\bibnamefont {Liang}}, \bibinfo {author} {\bibfnamefont
  {U.}~\bibnamefont {Zeitler}}, \bibinfo {author} {\bibfnamefont
  {S.}~\bibnamefont {Wiedmann}}, \bibinfo {author} {\bibfnamefont {N.~E.}\
  \bibnamefont {Hussey}}, \bibinfo {author} {\bibfnamefont {T.~T.~M.}\
  \bibnamefont {Palstra}},\ and\ \bibinfo {author} {\bibfnamefont
  {J.}~\bibnamefont {Ye}},\ }\bibfield  {title} {\bibinfo {title} {{Orbital
  Fulde-Ferrell-Larkin-Ovchinnikov state in an Ising superconductor}},\ }\href
  {https://doi.org/10.1038/s41586-023-05967-z} {\bibfield  {journal} {\bibinfo
  {journal} {Nature}\ }\textbf {\bibinfo {volume} {619}},\ \bibinfo {pages}
  {46} (\bibinfo {year} {2023})}\BibitemShut {NoStop}%
\bibitem [{\citenamefont {Zhao}\ \emph {et~al.}(2023)\citenamefont {Zhao},
  \citenamefont {Debbeler}, \citenamefont {K\"{u}hne}, \citenamefont {Fecher},
  \citenamefont {Gross},\ and\ \citenamefont {Smet}}]{zha23}%
  \BibitemOpen
  \bibfield  {author} {\bibinfo {author} {\bibfnamefont {D.}~\bibnamefont
  {Zhao}}, \bibinfo {author} {\bibfnamefont {L.}~\bibnamefont {Debbeler}},
  \bibinfo {author} {\bibfnamefont {M.}~\bibnamefont {K\"{u}hne}}, \bibinfo
  {author} {\bibfnamefont {S.}~\bibnamefont {Fecher}}, \bibinfo {author}
  {\bibfnamefont {N.}~\bibnamefont {Gross}},\ and\ \bibinfo {author}
  {\bibfnamefont {J.}~\bibnamefont {Smet}},\ }\bibfield  {title} {\bibinfo
  {title} {{Evidence of finite-momentum pairing in a centrosymmetric
  bilayer}},\ }\href {https://doi.org/10.1038/s41567-023-02202-4} {\bibfield
  {journal} {\bibinfo  {journal} {Nat. Phys.}\ }\textbf {\bibinfo {volume}
  {19}},\ \bibinfo {pages} {1599} (\bibinfo {year} {2023})}\BibitemShut
  {NoStop}%
\bibitem [{\citenamefont {Machida}\ and\ \citenamefont
  {Nakanishi}(1984)}]{mac84}%
  \BibitemOpen
  \bibfield  {author} {\bibinfo {author} {\bibfnamefont {K.}~\bibnamefont
  {Machida}}\ and\ \bibinfo {author} {\bibfnamefont {H.}~\bibnamefont
  {Nakanishi}},\ }\bibfield  {title} {\bibinfo {title} {{Superconductivity
  under a ferromagnetic molecular field}},\ }\href
  {https://doi.org/10.1103/PhysRevB.30.122} {\bibfield  {journal} {\bibinfo
  {journal} {Phys. Rev. B}\ }\textbf {\bibinfo {volume} {30}},\ \bibinfo
  {pages} {122} (\bibinfo {year} {1984})}\BibitemShut {NoStop}%
\bibitem [{\citenamefont {Vorontsov}\ \emph {et~al.}(2005)\citenamefont
  {Vorontsov}, \citenamefont {Sauls},\ and\ \citenamefont {Graf}}]{vor05}%
  \BibitemOpen
  \bibfield  {author} {\bibinfo {author} {\bibfnamefont {A.~B.}\ \bibnamefont
  {Vorontsov}}, \bibinfo {author} {\bibfnamefont {J.~A.}\ \bibnamefont
  {Sauls}},\ and\ \bibinfo {author} {\bibfnamefont {M.~J.}\ \bibnamefont
  {Graf}},\ }\bibfield  {title} {\bibinfo {title} {{Phase diagram and
  spectroscopy of Fulde-Ferrell-Larkin-Ovchinnikov states of two-dimensional
  $d$-wave superconductors}},\ }\href
  {https://doi.org/10.1103/PhysRevB.72.184501} {\bibfield  {journal} {\bibinfo
  {journal} {Phys. Rev. B}\ }\textbf {\bibinfo {volume} {72}},\ \bibinfo
  {pages} {184501} (\bibinfo {year} {2005})}\BibitemShut {NoStop}%
\bibitem [{\citenamefont {Ichioka}\ \emph {et~al.}(2007)\citenamefont
  {Ichioka}, \citenamefont {Adachi}, \citenamefont {Mizushima},\ and\
  \citenamefont {Machida}}]{ich07}%
  \BibitemOpen
  \bibfield  {author} {\bibinfo {author} {\bibfnamefont {M.}~\bibnamefont
  {Ichioka}}, \bibinfo {author} {\bibfnamefont {H.}~\bibnamefont {Adachi}},
  \bibinfo {author} {\bibfnamefont {T.}~\bibnamefont {Mizushima}},\ and\
  \bibinfo {author} {\bibfnamefont {K.}~\bibnamefont {Machida}},\ }\bibfield
  {title} {\bibinfo {title} {{Vortex state in a
  Fulde-Ferrell-Larkin-Ovchinnikov superconductor based on quasiclassical
  theory}},\ }\href {https://doi.org/10.1103/PhysRevB.76.014503} {\bibfield
  {journal} {\bibinfo  {journal} {Phys. Rev. B}\ }\textbf {\bibinfo {volume}
  {76}},\ \bibinfo {pages} {014503} (\bibinfo {year} {2007})}\BibitemShut
  {NoStop}%
\bibitem [{\citenamefont {M.~Suzuki}\ \emph {et~al.}(2011)\citenamefont
  {M.~Suzuki}, \citenamefont {Tsutsumi}, \citenamefont {Nakai}, \citenamefont
  {Ichioka},\ and\ \citenamefont {Machida}}]{suz11}%
  \BibitemOpen
  \bibfield  {author} {\bibinfo {author} {\bibfnamefont {K.}~\bibnamefont
  {M.~Suzuki}}, \bibinfo {author} {\bibfnamefont {Y.}~\bibnamefont {Tsutsumi}},
  \bibinfo {author} {\bibfnamefont {N.}~\bibnamefont {Nakai}}, \bibinfo
  {author} {\bibfnamefont {M.}~\bibnamefont {Ichioka}},\ and\ \bibinfo {author}
  {\bibfnamefont {K.}~\bibnamefont {Machida}},\ }\bibfield  {title} {\bibinfo
  {title} {{Field Evolution of the Fulde-Ferrell-Larkin-Ovchinnikov State in a
  Superconductor with Strong Pauli Effects}},\ }\href
  {https://doi.org/10.1143/JPSJ.80.123706} {\bibfield  {journal} {\bibinfo
  {journal} {J. Phys. Soc. Jpn.}\ }\textbf {\bibinfo {volume} {80}},\ \bibinfo
  {pages} {123706} (\bibinfo {year} {2011})}\BibitemShut {NoStop}%
\bibitem [{\citenamefont {Rosemeyer}\ and\ \citenamefont
  {Vorontsov}(2016)}]{ros16}%
  \BibitemOpen
  \bibfield  {author} {\bibinfo {author} {\bibfnamefont {B.~M.}\ \bibnamefont
  {Rosemeyer}}\ and\ \bibinfo {author} {\bibfnamefont {A.~B.}\ \bibnamefont
  {Vorontsov}},\ }\bibfield  {title} {\bibinfo {title} {{Spin susceptibility of
  Andreev bound states}},\ }\href {https://doi.org/10.1103/PhysRevB.94.144501}
  {\bibfield  {journal} {\bibinfo  {journal} {Phys. Rev. B}\ }\textbf {\bibinfo
  {volume} {94}},\ \bibinfo {pages} {144501} (\bibinfo {year}
  {2016})}\BibitemShut {NoStop}%
\bibitem [{\citenamefont {Mizushima}\ and\ \citenamefont
  {Machida}(2018)}]{miz18}%
  \BibitemOpen
  \bibfield  {author} {\bibinfo {author} {\bibfnamefont {T.}~\bibnamefont
  {Mizushima}}\ and\ \bibinfo {author} {\bibfnamefont {K.}~\bibnamefont
  {Machida}},\ }\bibfield  {title} {\bibinfo {title} {{Multifaceted properties
  of Andreev bound states: interplay of symmetry and topology}},\ }\href
  {https://doi.org/10.1098/rsta.2015.0355} {\bibfield  {journal} {\bibinfo
  {journal} {Phil. Trans. R. Soc. A.}\ }\textbf {\bibinfo {volume} {376}},\
  \bibinfo {pages} {20150355} (\bibinfo {year} {2018})}\BibitemShut {NoStop}%
\bibitem [{\citenamefont {Suzuki}\ \emph {et~al.}(2020)\citenamefont {Suzuki},
  \citenamefont {Machida}, \citenamefont {Tsutsumi},\ and\ \citenamefont
  {Ichioka}}]{suz20}%
  \BibitemOpen
  \bibfield  {author} {\bibinfo {author} {\bibfnamefont {K.~M.}\ \bibnamefont
  {Suzuki}}, \bibinfo {author} {\bibfnamefont {K.}~\bibnamefont {Machida}},
  \bibinfo {author} {\bibfnamefont {Y.}~\bibnamefont {Tsutsumi}},\ and\
  \bibinfo {author} {\bibfnamefont {M.}~\bibnamefont {Ichioka}},\ }\bibfield
  {title} {\bibinfo {title} {{Microscopic Eilenberger theory of
  Fulde-Ferrell-Larkin-Ovchinnikov states in the presence of vortices}},\
  }\href {https://doi.org/10.1103/PhysRevB.101.214516} {\bibfield  {journal}
  {\bibinfo  {journal} {Phys. Rev. B}\ }\textbf {\bibinfo {volume} {101}},\
  \bibinfo {pages} {214516} (\bibinfo {year} {2020})}\BibitemShut {NoStop}%
\bibitem [{\citenamefont {Leggett}(1966)}]{leg66}%
  \BibitemOpen
  \bibfield  {author} {\bibinfo {author} {\bibfnamefont {A.~J.}\ \bibnamefont
  {Leggett}},\ }\bibfield  {title} {\bibinfo {title} {{Number-Phase
  Fluctuations in Two-Band Superconductors}},\ }\href
  {https://doi.org/10.1143/PTP.36.901} {\bibfield  {journal} {\bibinfo
  {journal} {Prog. Theor. Phys.}\ }\textbf {\bibinfo {volume} {36}},\ \bibinfo
  {pages} {901} (\bibinfo {year} {1966})}\BibitemShut {NoStop}%
\bibitem [{\citenamefont {Maki}(1974)}]{mak74}%
  \BibitemOpen
  \bibfield  {author} {\bibinfo {author} {\bibfnamefont {K.}~\bibnamefont
  {Maki}},\ }\bibfield  {title} {\bibinfo {title} {{Propagation of zero sound
  in the Balian-Werthamer state}},\ }\href {https://doi.org/10.1007/BF00654896}
  {\bibfield  {journal} {\bibinfo  {journal} {J. Low Temp. Phys.}\ }\textbf
  {\bibinfo {volume} {16}},\ \bibinfo {pages} {465} (\bibinfo {year}
  {1974})}\BibitemShut {NoStop}%
\bibitem [{\citenamefont {Nagai}(1975)}]{nag75}%
  \BibitemOpen
  \bibfield  {author} {\bibinfo {author} {\bibfnamefont {K.}~\bibnamefont
  {Nagai}},\ }\bibfield  {title} {\bibinfo {title} {{Collective Excitations
  from the Balian-Werthamer State}},\ }\href {https://doi.org/10.1143/PTP.54.1}
  {\bibfield  {journal} {\bibinfo  {journal} {Prog. Theor. Phys.}\ }\textbf
  {\bibinfo {volume} {54}},\ \bibinfo {pages} {1} (\bibinfo {year}
  {1975})}\BibitemShut {NoStop}%
\bibitem [{\citenamefont {Tewordt}\ \emph {et~al.}(1975)\citenamefont
  {Tewordt}, \citenamefont {Fay}, \citenamefont {D{\o}rre},\ and\ \citenamefont
  {Einzel}}]{tew75}%
  \BibitemOpen
  \bibfield  {author} {\bibinfo {author} {\bibfnamefont {L.}~\bibnamefont
  {Tewordt}}, \bibinfo {author} {\bibfnamefont {D.}~\bibnamefont {Fay}},
  \bibinfo {author} {\bibfnamefont {P.}~\bibnamefont {D{\o}rre}},\ and\
  \bibinfo {author} {\bibfnamefont {D.}~\bibnamefont {Einzel}},\ }\bibfield
  {title} {\bibinfo {title} {{Self-consistent calculation of the longitudinal
  NMR for the Balian-Werthamer and Anderson-Brinkman-Morel states of superfluid
  3He}},\ }\href {https://doi.org/10.1007/BF01141616} {\bibfield  {journal}
  {\bibinfo  {journal} {J. Low Temp. Phys.}\ }\textbf {\bibinfo {volume}
  {21}},\ \bibinfo {pages} {645} (\bibinfo {year} {1975})}\BibitemShut
  {NoStop}%
\bibitem [{\citenamefont {Maki}(1976)}]{mak76}%
  \BibitemOpen
  \bibfield  {author} {\bibinfo {author} {\bibfnamefont {K.}~\bibnamefont
  {Maki}},\ }\bibfield  {title} {\bibinfo {title} {{Collective modes and spin
  waves in superfluid $^3$He-B}},\ }\href {https://doi.org/10.1007/BF00657178}
  {\bibfield  {journal} {\bibinfo  {journal} {J. Low Temp. Phys.}\ }\textbf
  {\bibinfo {volume} {24}},\ \bibinfo {pages} {755} (\bibinfo {year}
  {1976})}\BibitemShut {NoStop}%
\bibitem [{\citenamefont {W\"olfle}(1976)}]{wol76}%
  \BibitemOpen
  \bibfield  {author} {\bibinfo {author} {\bibfnamefont {P.}~\bibnamefont
  {W\"olfle}},\ }\bibfield  {title} {\bibinfo {title} {{Order-Parameter
  Collective Modes in $^{3}\mathrm{He}\ensuremath{-}A$}},\ }\href
  {https://doi.org/10.1103/PhysRevLett.37.1279} {\bibfield  {journal} {\bibinfo
   {journal} {Phys. Rev. Lett.}\ }\textbf {\bibinfo {volume} {37}},\ \bibinfo
  {pages} {1279} (\bibinfo {year} {1976})}\BibitemShut {NoStop}%
\bibitem [{\citenamefont {Tewordt}\ and\ \citenamefont
  {Schopohl}(1979)}]{tew79}%
  \BibitemOpen
  \bibfield  {author} {\bibinfo {author} {\bibfnamefont {L.}~\bibnamefont
  {Tewordt}}\ and\ \bibinfo {author} {\bibfnamefont {N.}~\bibnamefont
  {Schopohl}},\ }\bibfield  {title} {\bibinfo {title} {{Order-parameter
  collective modes in $^3$He-A and their effect on nuclear magnetic resonance
  (NMR) and ultrasonic attenuation}},\ }\href
  {https://doi.org/10.1007/BF00114937} {\bibfield  {journal} {\bibinfo
  {journal} {J. Low Temp. Phys.}\ }\textbf {\bibinfo {volume} {34}},\ \bibinfo
  {pages} {489} (\bibinfo {year} {1979})}\BibitemShut {NoStop}%
\bibitem [{\citenamefont {Giannetta}\ \emph {et~al.}(1980)\citenamefont
  {Giannetta}, \citenamefont {Ahonen}, \citenamefont {Polturak}, \citenamefont
  {Saunders}, \citenamefont {Zeise}, \citenamefont {Richardson},\ and\
  \citenamefont {Lee}}]{gia80}%
  \BibitemOpen
  \bibfield  {author} {\bibinfo {author} {\bibfnamefont {R.~W.}\ \bibnamefont
  {Giannetta}}, \bibinfo {author} {\bibfnamefont {A.}~\bibnamefont {Ahonen}},
  \bibinfo {author} {\bibfnamefont {E.}~\bibnamefont {Polturak}}, \bibinfo
  {author} {\bibfnamefont {J.}~\bibnamefont {Saunders}}, \bibinfo {author}
  {\bibfnamefont {E.~K.}\ \bibnamefont {Zeise}}, \bibinfo {author}
  {\bibfnamefont {R.~C.}\ \bibnamefont {Richardson}},\ and\ \bibinfo {author}
  {\bibfnamefont {D.~M.}\ \bibnamefont {Lee}},\ }\bibfield  {title} {\bibinfo
  {title} {{Observation of a {N}ew {S}ound-{A}ttenuation {P}eak in {S}uperfluid
  $^{3}\mathrm{He}$-${B}$}},\ }\href
  {https://doi.org/10.1103/PhysRevLett.45.262} {\bibfield  {journal} {\bibinfo
  {journal} {Phys. Rev. Lett.}\ }\textbf {\bibinfo {volume} {45}},\ \bibinfo
  {pages} {262} (\bibinfo {year} {1980})}\BibitemShut {NoStop}%
\bibitem [{\citenamefont {Mast}\ \emph {et~al.}(1980)\citenamefont {Mast},
  \citenamefont {Sarma}, \citenamefont {Owers-Bradley}, \citenamefont {Calder},
  \citenamefont {Ketterson},\ and\ \citenamefont {Halperin}}]{mas80}%
  \BibitemOpen
  \bibfield  {author} {\bibinfo {author} {\bibfnamefont {D.~B.}\ \bibnamefont
  {Mast}}, \bibinfo {author} {\bibfnamefont {B.~K.}\ \bibnamefont {Sarma}},
  \bibinfo {author} {\bibfnamefont {J.~R.}\ \bibnamefont {Owers-Bradley}},
  \bibinfo {author} {\bibfnamefont {I.~D.}\ \bibnamefont {Calder}}, \bibinfo
  {author} {\bibfnamefont {J.~B.}\ \bibnamefont {Ketterson}},\ and\ \bibinfo
  {author} {\bibfnamefont {W.~P.}\ \bibnamefont {Halperin}},\ }\bibfield
  {title} {\bibinfo {title} {{Measurements of {H}igh-{F}requency {S}ound
  {P}ropagation in $^{3}\mathrm{He}$-${B}$}},\ }\href
  {https://doi.org/10.1103/PhysRevLett.45.266} {\bibfield  {journal} {\bibinfo
  {journal} {Phys. Rev. Lett.}\ }\textbf {\bibinfo {volume} {45}},\ \bibinfo
  {pages} {266} (\bibinfo {year} {1980})}\BibitemShut {NoStop}%
\bibitem [{\citenamefont {Avenel}\ \emph {et~al.}(1980)\citenamefont {Avenel},
  \citenamefont {Varoquaux},\ and\ \citenamefont {Ebisawa}}]{ave80}%
  \BibitemOpen
  \bibfield  {author} {\bibinfo {author} {\bibfnamefont {O.}~\bibnamefont
  {Avenel}}, \bibinfo {author} {\bibfnamefont {E.}~\bibnamefont {Varoquaux}},\
  and\ \bibinfo {author} {\bibfnamefont {H.}~\bibnamefont {Ebisawa}},\
  }\bibfield  {title} {\bibinfo {title} {{Field {S}plitting of the {N}ew
  {S}ound {A}ttenuation {P}eak in $^{3}\mathrm{He}$-${B}$}},\ }\href
  {https://doi.org/10.1103/PhysRevLett.45.1952} {\bibfield  {journal} {\bibinfo
   {journal} {Phys. Rev. Lett.}\ }\textbf {\bibinfo {volume} {45}},\ \bibinfo
  {pages} {1952} (\bibinfo {year} {1980})}\BibitemShut {NoStop}%
\bibitem [{\citenamefont {Movshovich}\ \emph {et~al.}(1988)\citenamefont
  {Movshovich}, \citenamefont {Varoquaux}, \citenamefont {Kim},\ and\
  \citenamefont {Lee}}]{mov88}%
  \BibitemOpen
  \bibfield  {author} {\bibinfo {author} {\bibfnamefont {R.}~\bibnamefont
  {Movshovich}}, \bibinfo {author} {\bibfnamefont {E.}~\bibnamefont
  {Varoquaux}}, \bibinfo {author} {\bibfnamefont {N.}~\bibnamefont {Kim}},\
  and\ \bibinfo {author} {\bibfnamefont {D.~M.}\ \bibnamefont {Lee}},\
  }\bibfield  {title} {\bibinfo {title} {{Splitting of the Squashing Collective
  Mode of Superfluid $^{3}\mathrm{He}\ensuremath{-}B$ by a Magnetic Field}},\
  }\href {https://doi.org/10.1103/PhysRevLett.61.1732} {\bibfield  {journal}
  {\bibinfo  {journal} {Phys. Rev. Lett.}\ }\textbf {\bibinfo {volume} {61}},\
  \bibinfo {pages} {1732} (\bibinfo {year} {1988})}\BibitemShut {NoStop}%
\bibitem [{\citenamefont {Lee}\ \emph {et~al.}(1999)\citenamefont {Lee},
  \citenamefont {Haard}, \citenamefont {Halperin},\ and\ \citenamefont
  {Sauls}}]{lee99}%
  \BibitemOpen
  \bibfield  {author} {\bibinfo {author} {\bibfnamefont {Y.}~\bibnamefont
  {Lee}}, \bibinfo {author} {\bibfnamefont {T.~M.}\ \bibnamefont {Haard}},
  \bibinfo {author} {\bibfnamefont {W.~P.}\ \bibnamefont {Halperin}},\ and\
  \bibinfo {author} {\bibfnamefont {J.~A.}\ \bibnamefont {Sauls}},\ }\bibfield
  {title} {\bibinfo {title} {{Discovery of the acoustic {F}araday effect in
  superfluid $^3${H}e-{B}}},\ }\href {https://doi.org/10.1038/22712} {\bibfield
   {journal} {\bibinfo  {journal} {Nature}\ }\textbf {\bibinfo {volume}
  {400}},\ \bibinfo {pages} {431} (\bibinfo {year} {1999})}\BibitemShut
  {NoStop}%
\bibitem [{\citenamefont {Davis}\ \emph {et~al.}(2006)\citenamefont {Davis},
  \citenamefont {Choi}, \citenamefont {Pollanen},\ and\ \citenamefont
  {Halperin}}]{dav06}%
  \BibitemOpen
  \bibfield  {author} {\bibinfo {author} {\bibfnamefont {J.~P.}\ \bibnamefont
  {Davis}}, \bibinfo {author} {\bibfnamefont {H.}~\bibnamefont {Choi}},
  \bibinfo {author} {\bibfnamefont {J.}~\bibnamefont {Pollanen}},\ and\
  \bibinfo {author} {\bibfnamefont {W.~P.}\ \bibnamefont {Halperin}},\
  }\bibfield  {title} {\bibinfo {title} {{Collective Modes and $f$-Wave Pairing
  Interactions in Superfluid $^{3}\mathrm{He}$}},\ }\href
  {https://doi.org/10.1103/PhysRevLett.97.115301} {\bibfield  {journal}
  {\bibinfo  {journal} {Phys. Rev. Lett.}\ }\textbf {\bibinfo {volume} {97}},\
  \bibinfo {pages} {115301} (\bibinfo {year} {2006})}\BibitemShut {NoStop}%
\bibitem [{\citenamefont {Davis}\ \emph {et~al.}(2008)\citenamefont {Davis},
  \citenamefont {Pollanen}, \citenamefont {Choi}, \citenamefont {Sauls},\ and\
  \citenamefont {Halperin}}]{dav08}%
  \BibitemOpen
  \bibfield  {author} {\bibinfo {author} {\bibfnamefont {J.~P.}\ \bibnamefont
  {Davis}}, \bibinfo {author} {\bibfnamefont {J.}~\bibnamefont {Pollanen}},
  \bibinfo {author} {\bibfnamefont {H.}~\bibnamefont {Choi}}, \bibinfo {author}
  {\bibfnamefont {J.~A.}\ \bibnamefont {Sauls}},\ and\ \bibinfo {author}
  {\bibfnamefont {W.~P.}\ \bibnamefont {Halperin}},\ }\bibfield  {title}
  {\bibinfo {title} {{Discovery of an excited pair state in superfluid
  $^3$He}},\ }\href {https://doi.org/10.1038/nphys969} {\bibfield  {journal}
  {\bibinfo  {journal} {Nat. Phys.}\ }\textbf {\bibinfo {volume} {4}},\
  \bibinfo {pages} {571} (\bibinfo {year} {2008})}\BibitemShut {NoStop}%
\bibitem [{\citenamefont {Zavjalov}\ \emph {et~al.}(2016)\citenamefont
  {Zavjalov}, \citenamefont {Autti}, \citenamefont {Eltsov}, \citenamefont
  {Heikkinen},\ and\ \citenamefont {Volovik}}]{zav16}%
  \BibitemOpen
  \bibfield  {author} {\bibinfo {author} {\bibfnamefont {V.~V.}\ \bibnamefont
  {Zavjalov}}, \bibinfo {author} {\bibfnamefont {S.}~\bibnamefont {Autti}},
  \bibinfo {author} {\bibfnamefont {V.~B.}\ \bibnamefont {Eltsov}}, \bibinfo
  {author} {\bibfnamefont {P.~J.}\ \bibnamefont {Heikkinen}},\ and\ \bibinfo
  {author} {\bibfnamefont {G.~E.}\ \bibnamefont {Volovik}},\ }\bibfield
  {title} {\bibinfo {title} {{Light {H}iggs channel of the resonant decay of
  magnon condensate in superfluid $^3${H}e-{B}}},\ }\href
  {https://www.nature.com/articles/ncomms10294} {\bibfield  {journal} {\bibinfo
   {journal} {Nat. Commun.}\ }\textbf {\bibinfo {volume} {7}},\ \bibinfo
  {pages} {10294} (\bibinfo {year} {2016})}\BibitemShut {NoStop}%
\bibitem [{\citenamefont {Koch}\ and\ \citenamefont {W\"olfle}(1981)}]{koc81}%
  \BibitemOpen
  \bibfield  {author} {\bibinfo {author} {\bibfnamefont {V.~E.}\ \bibnamefont
  {Koch}}\ and\ \bibinfo {author} {\bibfnamefont {P.}~\bibnamefont
  {W\"olfle}},\ }\bibfield  {title} {\bibinfo {title} {{Coupling of {N}ew
  {O}rder-{P}arameter {C}ollective {M}odes to {S}ound {W}aves in {S}uperfluid
  $^{3}${H}e}},\ }\href {https://doi.org/10.1103/PhysRevLett.46.486} {\bibfield
   {journal} {\bibinfo  {journal} {Phys. Rev. Lett.}\ }\textbf {\bibinfo
  {volume} {46}},\ \bibinfo {pages} {486} (\bibinfo {year} {1981})}\BibitemShut
  {NoStop}%
\bibitem [{sau()}]{sau00}%
  \BibitemOpen
  \href@noop {} {}\bibinfo {note} {J. A. Sauls, Broken Symmetry and
  Non-Equilibrium Superfluid $^3$He, in {\it Topological Defects and
  Non-Equilibrium Symmetry Breaking Phase Transitions, Lecture Notes for the
  1999 Les Houches Winter School}, edited by H. Godfrin and Y. Bunkov
  (Elsevier, Amsterdam, 2000), pp. 239-265.}\BibitemShut {Stop}%
\bibitem [{\citenamefont {Sauls}\ and\ \citenamefont
  {Mizushima}(2017)}]{sau17}%
  \BibitemOpen
  \bibfield  {author} {\bibinfo {author} {\bibfnamefont {J.~A.}\ \bibnamefont
  {Sauls}}\ and\ \bibinfo {author} {\bibfnamefont {T.}~\bibnamefont
  {Mizushima}},\ }\bibfield  {title} {\bibinfo {title} {On the {N}ambu
  fermion-boson relations for superfluid $^{3}\mathrm{He}$},\ }\href
  {https://doi.org/10.1103/PhysRevB.95.094515} {\bibfield  {journal} {\bibinfo
  {journal} {Phys. Rev. B}\ }\textbf {\bibinfo {volume} {95}},\ \bibinfo
  {pages} {094515} (\bibinfo {year} {2017})}\BibitemShut {NoStop}%
\bibitem [{\citenamefont {Radzihovsky}\ and\ \citenamefont
  {Vishwanath}(2009)}]{rad09}%
  \BibitemOpen
  \bibfield  {author} {\bibinfo {author} {\bibfnamefont {L.}~\bibnamefont
  {Radzihovsky}}\ and\ \bibinfo {author} {\bibfnamefont {A.}~\bibnamefont
  {Vishwanath}},\ }\bibfield  {title} {\bibinfo {title} {{Quantum Liquid
  Crystals in an Imbalanced Fermi Gas: Fluctuations and Fractional Vortices in
  Larkin-Ovchinnikov States}},\ }\href
  {https://doi.org/10.1103/PhysRevLett.103.010404} {\bibfield  {journal}
  {\bibinfo  {journal} {Phys. Rev. Lett.}\ }\textbf {\bibinfo {volume} {103}},\
  \bibinfo {pages} {010404} (\bibinfo {year} {2009})}\BibitemShut {NoStop}%
\bibitem [{\citenamefont {Edge}\ and\ \citenamefont {Cooper}(2009)}]{edg09}%
  \BibitemOpen
  \bibfield  {author} {\bibinfo {author} {\bibfnamefont {J.~M.}\ \bibnamefont
  {Edge}}\ and\ \bibinfo {author} {\bibfnamefont {N.~R.}\ \bibnamefont
  {Cooper}},\ }\bibfield  {title} {\bibinfo {title} {{Signature of the
  Fulde-Ferrell-Larkin-Ovchinnikov Phase in the Collective Modes of a Trapped
  Ultracold Fermi Gas}},\ }\href
  {https://doi.org/10.1103/PhysRevLett.103.065301} {\bibfield  {journal}
  {\bibinfo  {journal} {Phys. Rev. Lett.}\ }\textbf {\bibinfo {volume} {103}},\
  \bibinfo {pages} {065301} (\bibinfo {year} {2009})}\BibitemShut {NoStop}%
\bibitem [{\citenamefont {Edge}\ and\ \citenamefont {Cooper}(2010)}]{edg10}%
  \BibitemOpen
  \bibfield  {author} {\bibinfo {author} {\bibfnamefont {J.~M.}\ \bibnamefont
  {Edge}}\ and\ \bibinfo {author} {\bibfnamefont {N.~R.}\ \bibnamefont
  {Cooper}},\ }\bibfield  {title} {\bibinfo {title} {{Collective modes as a
  probe of imbalanced Fermi gases}},\ }\href
  {https://doi.org/10.1103/PhysRevA.81.063606} {\bibfield  {journal} {\bibinfo
  {journal} {Phys. Rev. A}\ }\textbf {\bibinfo {volume} {81}},\ \bibinfo
  {pages} {063606} (\bibinfo {year} {2010})}\BibitemShut {NoStop}%
\bibitem [{\citenamefont {Radzihovsky}(2011)}]{rad11}%
  \BibitemOpen
  \bibfield  {author} {\bibinfo {author} {\bibfnamefont {L.}~\bibnamefont
  {Radzihovsky}},\ }\bibfield  {title} {\bibinfo {title} {{Fluctuations and
  phase transitions in Larkin-Ovchinnikov liquid-crystal states of a
  population-imbalanced resonant Fermi gas}},\ }\href
  {https://doi.org/10.1103/PhysRevA.84.023611} {\bibfield  {journal} {\bibinfo
  {journal} {Phys. Rev. A}\ }\textbf {\bibinfo {volume} {84}},\ \bibinfo
  {pages} {023611} (\bibinfo {year} {2011})}\BibitemShut {NoStop}%
\bibitem [{\citenamefont {Koinov}\ \emph {et~al.}(2011)\citenamefont {Koinov},
  \citenamefont {Mendoza},\ and\ \citenamefont {Fortes}}]{koi11}%
  \BibitemOpen
  \bibfield  {author} {\bibinfo {author} {\bibfnamefont {Z.}~\bibnamefont
  {Koinov}}, \bibinfo {author} {\bibfnamefont {R.}~\bibnamefont {Mendoza}},\
  and\ \bibinfo {author} {\bibfnamefont {M.}~\bibnamefont {Fortes}},\
  }\bibfield  {title} {\bibinfo {title} {{Rotonlike Fulde-Ferrell Collective
  Excitations of an Imbalanced Fermi Gas in a Two-Dimensional Optical
  Lattice}},\ }\href {https://doi.org/10.1103/PhysRevLett.106.100402}
  {\bibfield  {journal} {\bibinfo  {journal} {Phys. Rev. Lett.}\ }\textbf
  {\bibinfo {volume} {106}},\ \bibinfo {pages} {100402} (\bibinfo {year}
  {2011})}\BibitemShut {NoStop}%
\bibitem [{\citenamefont {Heikkinen}\ and\ \citenamefont
  {T\"orm\"a}(2011)}]{hei11}%
  \BibitemOpen
  \bibfield  {author} {\bibinfo {author} {\bibfnamefont {M.~O.~J.}\
  \bibnamefont {Heikkinen}}\ and\ \bibinfo {author} {\bibfnamefont
  {P.}~\bibnamefont {T\"orm\"a}},\ }\bibfield  {title} {\bibinfo {title}
  {{Collective modes and the speed of sound in the
  Fulde-Ferrell-Larkin-Ovchinnikov state}},\ }\href
  {https://doi.org/10.1103/PhysRevA.83.053630} {\bibfield  {journal} {\bibinfo
  {journal} {Phys. Rev. A}\ }\textbf {\bibinfo {volume} {83}},\ \bibinfo
  {pages} {053630} (\bibinfo {year} {2011})}\BibitemShut {NoStop}%
\bibitem [{\citenamefont {Radzihovsky}(2012)}]{rad12}%
  \BibitemOpen
  \bibfield  {author} {\bibinfo {author} {\bibfnamefont {L.}~\bibnamefont
  {Radzihovsky}},\ }\bibfield  {title} {\bibinfo {title} {{Quantum
  liquid-crystal order in resonant atomic gases}},\ }\href
  {https://doi.org/https://doi.org/10.1016/j.physc.2012.04.014} {\bibfield
  {journal} {\bibinfo  {journal} {Physica C}\ }\textbf {\bibinfo {volume}
  {481}},\ \bibinfo {pages} {189} (\bibinfo {year} {2012})}\BibitemShut
  {NoStop}%
\bibitem [{\citenamefont {Dutta}\ and\ \citenamefont {Mueller}(2017)}]{dut17}%
  \BibitemOpen
  \bibfield  {author} {\bibinfo {author} {\bibfnamefont {S.}~\bibnamefont
  {Dutta}}\ and\ \bibinfo {author} {\bibfnamefont {E.~J.}\ \bibnamefont
  {Mueller}},\ }\bibfield  {title} {\bibinfo {title} {{Collective Modes of a
  Soliton Train in a Fermi Superfluid}},\ }\href
  {https://doi.org/10.1103/PhysRevLett.118.260402} {\bibfield  {journal}
  {\bibinfo  {journal} {Phys. Rev. Lett.}\ }\textbf {\bibinfo {volume} {118}},\
  \bibinfo {pages} {260402} (\bibinfo {year} {2017})}\BibitemShut {NoStop}%
\bibitem [{\citenamefont {Boyack}\ \emph {et~al.}(2017)\citenamefont {Boyack},
  \citenamefont {Wu}, \citenamefont {Anderson},\ and\ \citenamefont
  {Levin}}]{boy17}%
  \BibitemOpen
  \bibfield  {author} {\bibinfo {author} {\bibfnamefont {R.}~\bibnamefont
  {Boyack}}, \bibinfo {author} {\bibfnamefont {C.-T.}\ \bibnamefont {Wu}},
  \bibinfo {author} {\bibfnamefont {B.~M.}\ \bibnamefont {Anderson}},\ and\
  \bibinfo {author} {\bibfnamefont {K.}~\bibnamefont {Levin}},\ }\bibfield
  {title} {\bibinfo {title} {{Collective mode contributions to the Meissner
  effect: Fulde-Ferrell and pair-density wave superfluids}},\ }\href
  {https://doi.org/10.1103/PhysRevB.95.214501} {\bibfield  {journal} {\bibinfo
  {journal} {Phys. Rev. B}\ }\textbf {\bibinfo {volume} {95}},\ \bibinfo
  {pages} {214501} (\bibinfo {year} {2017})}\BibitemShut {NoStop}%
\bibitem [{\citenamefont {Agterberg}\ \emph {et~al.}(2020)\citenamefont
  {Agterberg}, \citenamefont {Davis}, \citenamefont {Edkins}, \citenamefont
  {Fradkin}, \citenamefont {Van~Harlingen}, \citenamefont {Kivelson},
  \citenamefont {Lee}, \citenamefont {Radzihovsky}, \citenamefont {Tranquada},\
  and\ \citenamefont {Wang}}]{agt20}%
  \BibitemOpen
  \bibfield  {author} {\bibinfo {author} {\bibfnamefont {D.~F.}\ \bibnamefont
  {Agterberg}}, \bibinfo {author} {\bibfnamefont {J.~S.}\ \bibnamefont
  {Davis}}, \bibinfo {author} {\bibfnamefont {S.~D.}\ \bibnamefont {Edkins}},
  \bibinfo {author} {\bibfnamefont {E.}~\bibnamefont {Fradkin}}, \bibinfo
  {author} {\bibfnamefont {D.~J.}\ \bibnamefont {Van~Harlingen}}, \bibinfo
  {author} {\bibfnamefont {S.~A.}\ \bibnamefont {Kivelson}}, \bibinfo {author}
  {\bibfnamefont {P.~A.}\ \bibnamefont {Lee}}, \bibinfo {author} {\bibfnamefont
  {L.}~\bibnamefont {Radzihovsky}}, \bibinfo {author} {\bibfnamefont {J.~M.}\
  \bibnamefont {Tranquada}},\ and\ \bibinfo {author} {\bibfnamefont
  {Y.}~\bibnamefont {Wang}},\ }\bibfield  {title} {\bibinfo {title} {The
  physics of pair-density waves: Cuprate superconductors and beyond},\ }\href
  {https://doi.org/https://doi.org/10.1146/annurev-conmatphys-031119-050711}
  {\bibfield  {journal} {\bibinfo  {journal} {Annu. Rev. Condens. Matter
  Phys.}\ }\textbf {\bibinfo {volume} {11}},\ \bibinfo {pages} {231} (\bibinfo
  {year} {2020})}\BibitemShut {NoStop}%
\bibitem [{\citenamefont {Huang}\ \emph {et~al.}(2022)\citenamefont {Huang},
  \citenamefont {Ting}, \citenamefont {Zhu},\ and\ \citenamefont
  {Lin}}]{hua22}%
  \BibitemOpen
  \bibfield  {author} {\bibinfo {author} {\bibfnamefont {Z.}~\bibnamefont
  {Huang}}, \bibinfo {author} {\bibfnamefont {C.~S.}\ \bibnamefont {Ting}},
  \bibinfo {author} {\bibfnamefont {J.-X.}\ \bibnamefont {Zhu}},\ and\ \bibinfo
  {author} {\bibfnamefont {S.-Z.}\ \bibnamefont {Lin}},\ }\bibfield  {title}
  {\bibinfo {title} {{Gapless Higgs mode in the
  Fulde-Ferrell-Larkin-Ovchinnikov state of a superconductor}},\ }\href
  {https://doi.org/10.1103/PhysRevB.105.014502} {\bibfield  {journal} {\bibinfo
   {journal} {Phys. Rev. B}\ }\textbf {\bibinfo {volume} {105}},\ \bibinfo
  {pages} {014502} (\bibinfo {year} {2022})}\BibitemShut {NoStop}%
\bibitem [{\citenamefont {Samokhin}(2010)}]{sam10}%
  \BibitemOpen
  \bibfield  {author} {\bibinfo {author} {\bibfnamefont {K.~V.}\ \bibnamefont
  {Samokhin}},\ }\bibfield  {title} {\bibinfo {title} {{Goldstone modes in
  Larkin-Ovchinnikov-Fulde-Ferrell superconductors}},\ }\href
  {https://doi.org/10.1103/PhysRevB.81.224507} {\bibfield  {journal} {\bibinfo
  {journal} {Phys. Rev. B}\ }\textbf {\bibinfo {volume} {81}},\ \bibinfo
  {pages} {224507} (\bibinfo {year} {2010})}\BibitemShut {NoStop}%
\bibitem [{\citenamefont {Samokhin}(2011)}]{sam11}%
  \BibitemOpen
  \bibfield  {author} {\bibinfo {author} {\bibfnamefont {K.~V.}\ \bibnamefont
  {Samokhin}},\ }\bibfield  {title} {\bibinfo {title} {{Spectrum of Goldstone
  modes in Larkin-Ovchinnikov-Fulde-Ferrell superfluids}},\ }\href
  {https://doi.org/10.1103/PhysRevB.83.094514} {\bibfield  {journal} {\bibinfo
  {journal} {Phys. Rev. B}\ }\textbf {\bibinfo {volume} {83}},\ \bibinfo
  {pages} {094514} (\bibinfo {year} {2011})}\BibitemShut {NoStop}%
\bibitem [{\citenamefont {Soto-Garrido}\ \emph {et~al.}(2017)\citenamefont
  {Soto-Garrido}, \citenamefont {Wang}, \citenamefont {Fradkin},\ and\
  \citenamefont {Cooper}}]{sot17}%
  \BibitemOpen
  \bibfield  {author} {\bibinfo {author} {\bibfnamefont {R.}~\bibnamefont
  {Soto-Garrido}}, \bibinfo {author} {\bibfnamefont {Y.}~\bibnamefont {Wang}},
  \bibinfo {author} {\bibfnamefont {E.}~\bibnamefont {Fradkin}},\ and\ \bibinfo
  {author} {\bibfnamefont {S.~L.}\ \bibnamefont {Cooper}},\ }\bibfield  {title}
  {\bibinfo {title} {{Higgs modes in the pair density wave superconducting
  state}},\ }\href {https://doi.org/10.1103/PhysRevB.95.214502} {\bibfield
  {journal} {\bibinfo  {journal} {Phys. Rev. B}\ }\textbf {\bibinfo {volume}
  {95}},\ \bibinfo {pages} {214502} (\bibinfo {year} {2017})}\BibitemShut
  {NoStop}%
\bibitem [{\citenamefont {Jian}\ \emph {et~al.}(2020)\citenamefont {Jian},
  \citenamefont {Scherer},\ and\ \citenamefont {Yao}}]{jia20}%
  \BibitemOpen
  \bibfield  {author} {\bibinfo {author} {\bibfnamefont {S.-K.}\ \bibnamefont
  {Jian}}, \bibinfo {author} {\bibfnamefont {M.~M.}\ \bibnamefont {Scherer}},\
  and\ \bibinfo {author} {\bibfnamefont {H.}~\bibnamefont {Yao}},\ }\bibfield
  {title} {\bibinfo {title} {{Mass hierarchy in collective modes of
  pair-density-wave superconductors}},\ }\href
  {https://doi.org/10.1103/PhysRevResearch.2.013034} {\bibfield  {journal}
  {\bibinfo  {journal} {Phys. Rev. Res.}\ }\textbf {\bibinfo {volume} {2}},\
  \bibinfo {pages} {013034} (\bibinfo {year} {2020})}\BibitemShut {NoStop}%
\bibitem [{\citenamefont {Ba\ifmmode~\mbox{\c{s}}\else \c{s}\fi{}ar}\ and\
  \citenamefont {Dunne}(2008{\natexlab{a}})}]{bas1}%
  \BibitemOpen
  \bibfield  {author} {\bibinfo {author} {\bibfnamefont {G.}~\bibnamefont
  {Ba\ifmmode~\mbox{\c{s}}\else \c{s}\fi{}ar}}\ and\ \bibinfo {author}
  {\bibfnamefont {G.~V.}\ \bibnamefont {Dunne}},\ }\bibfield  {title} {\bibinfo
  {title} {{Self-Consistent Crystalline Condensate in Chiral Gross-Neveu and
  Bogoliubov--de Gennes Systems}},\ }\href
  {https://doi.org/10.1103/PhysRevLett.100.200404} {\bibfield  {journal}
  {\bibinfo  {journal} {Phys. Rev. Lett.}\ }\textbf {\bibinfo {volume} {100}},\
  \bibinfo {pages} {200404} (\bibinfo {year} {2008}{\natexlab{a}})}\BibitemShut
  {NoStop}%
\bibitem [{\citenamefont {Ba\ifmmode~\mbox{\c{s}}\else \c{s}\fi{}ar}\ and\
  \citenamefont {Dunne}(2008{\natexlab{b}})}]{bas2}%
  \BibitemOpen
  \bibfield  {author} {\bibinfo {author} {\bibfnamefont {G.}~\bibnamefont
  {Ba\ifmmode~\mbox{\c{s}}\else \c{s}\fi{}ar}}\ and\ \bibinfo {author}
  {\bibfnamefont {G.~V.}\ \bibnamefont {Dunne}},\ }\bibfield  {title} {\bibinfo
  {title} {{Twisted kink crystal in the chiral Gross-Neveu model}},\ }\href
  {https://doi.org/10.1103/PhysRevD.78.065022} {\bibfield  {journal} {\bibinfo
  {journal} {Phys. Rev. D}\ }\textbf {\bibinfo {volume} {78}},\ \bibinfo
  {pages} {065022} (\bibinfo {year} {2008}{\natexlab{b}})}\BibitemShut
  {NoStop}%
\bibitem [{\citenamefont {Yoshii}\ \emph {et~al.}(2011)\citenamefont {Yoshii},
  \citenamefont {Tsuchiya}, \citenamefont {Marmorini},\ and\ \citenamefont
  {Nitta}}]{yos11}%
  \BibitemOpen
  \bibfield  {author} {\bibinfo {author} {\bibfnamefont {R.}~\bibnamefont
  {Yoshii}}, \bibinfo {author} {\bibfnamefont {S.}~\bibnamefont {Tsuchiya}},
  \bibinfo {author} {\bibfnamefont {G.}~\bibnamefont {Marmorini}},\ and\
  \bibinfo {author} {\bibfnamefont {M.}~\bibnamefont {Nitta}},\ }\bibfield
  {title} {\bibinfo {title} {{Spin imbalance effect on the
  Larkin-Ovchinnikov-Fulde-Ferrel state}},\ }\href
  {https://doi.org/10.1103/PhysRevB.84.024503} {\bibfield  {journal} {\bibinfo
  {journal} {Phys. Rev. B}\ }\textbf {\bibinfo {volume} {84}},\ \bibinfo
  {pages} {024503} (\bibinfo {year} {2011})}\BibitemShut {NoStop}%
\bibitem [{\citenamefont {Takahashi}\ \emph {et~al.}(2012)\citenamefont
  {Takahashi}, \citenamefont {Tsuchiya}, \citenamefont {Yoshii},\ and\
  \citenamefont {Nitta}}]{dat12}%
  \BibitemOpen
  \bibfield  {author} {\bibinfo {author} {\bibfnamefont {D.~A.}\ \bibnamefont
  {Takahashi}}, \bibinfo {author} {\bibfnamefont {S.}~\bibnamefont {Tsuchiya}},
  \bibinfo {author} {\bibfnamefont {R.}~\bibnamefont {Yoshii}},\ and\ \bibinfo
  {author} {\bibfnamefont {M.}~\bibnamefont {Nitta}},\ }\bibfield  {title}
  {\bibinfo {title} {{Fermionic solutions of chiral Gross-Neveu and
  Bogoliubov-de Gennes systems in nonlinear Schr\"{o}dinger hierarchy}},\
  }\href {https://doi.org/https://doi.org/10.1016/j.physletb.2012.10.058}
  {\bibfield  {journal} {\bibinfo  {journal} {Phys. Lett. B}\ }\textbf
  {\bibinfo {volume} {718}},\ \bibinfo {pages} {632} (\bibinfo {year}
  {2012})}\BibitemShut {NoStop}%
\bibitem [{\citenamefont {Takahashi}\ and\ \citenamefont
  {Nitta}(2013)}]{dat13}%
  \BibitemOpen
  \bibfield  {author} {\bibinfo {author} {\bibfnamefont {D.~A.}\ \bibnamefont
  {Takahashi}}\ and\ \bibinfo {author} {\bibfnamefont {M.}~\bibnamefont
  {Nitta}},\ }\bibfield  {title} {\bibinfo {title} {{Self-Consistent Multiple
  Complex-Kink Solutions in Bogoliubov--de Gennes and Chiral Gross-Neveu
  Systems}},\ }\href {https://doi.org/10.1103/PhysRevLett.110.131601}
  {\bibfield  {journal} {\bibinfo  {journal} {Phys. Rev. Lett.}\ }\textbf
  {\bibinfo {volume} {110}},\ \bibinfo {pages} {131601} (\bibinfo {year}
  {2013})}\BibitemShut {NoStop}%
\bibitem [{\citenamefont {Frantzeskakis}(2010)}]{fra10}%
  \BibitemOpen
  \bibfield  {author} {\bibinfo {author} {\bibfnamefont {D.~J.}\ \bibnamefont
  {Frantzeskakis}},\ }\bibfield  {title} {\bibinfo {title} {{Dark solitons in
  atomic Bose-Einstein condensates: from theory to experiments}},\ }\href
  {https://doi.org/10.1088/1751-8113/43/21/213001} {\bibfield  {journal}
  {\bibinfo  {journal} {J. Phys. A}\ }\textbf {\bibinfo {volume} {43}},\
  \bibinfo {pages} {213001} (\bibinfo {year} {2010})}\BibitemShut {NoStop}%
\bibitem [{\citenamefont {Takayama}\ \emph {et~al.}(1980)\citenamefont
  {Takayama}, \citenamefont {Lin-Liu},\ and\ \citenamefont {Maki}}]{tak80}%
  \BibitemOpen
  \bibfield  {author} {\bibinfo {author} {\bibfnamefont {H.}~\bibnamefont
  {Takayama}}, \bibinfo {author} {\bibfnamefont {Y.~R.}\ \bibnamefont
  {Lin-Liu}},\ and\ \bibinfo {author} {\bibfnamefont {K.}~\bibnamefont
  {Maki}},\ }\bibfield  {title} {\bibinfo {title} {{Continuum model for
  solitons in polyacetylene}},\ }\href
  {https://doi.org/10.1103/PhysRevB.21.2388} {\bibfield  {journal} {\bibinfo
  {journal} {Phys. Rev. B}\ }\textbf {\bibinfo {volume} {21}},\ \bibinfo
  {pages} {2388} (\bibinfo {year} {1980})}\BibitemShut {NoStop}%
\bibitem [{\citenamefont {Anderson}(1963)}]{and63}%
  \BibitemOpen
  \bibfield  {author} {\bibinfo {author} {\bibfnamefont {P.~W.}\ \bibnamefont
  {Anderson}},\ }\bibfield  {title} {\bibinfo {title} {{Plasmons, Gauge
  Invariance, and Mass}},\ }\href {https://doi.org/10.1103/PhysRev.130.439}
  {\bibfield  {journal} {\bibinfo  {journal} {Phys. Rev.}\ }\textbf {\bibinfo
  {volume} {130}},\ \bibinfo {pages} {439} (\bibinfo {year}
  {1963})}\BibitemShut {NoStop}%
\bibitem [{\citenamefont {Higgs}(1964)}]{higgs}%
  \BibitemOpen
  \bibfield  {author} {\bibinfo {author} {\bibfnamefont {P.~W.}\ \bibnamefont
  {Higgs}},\ }\bibfield  {title} {\bibinfo {title} {{Broken Symmetries and the
  Masses of Gauge Bosons}},\ }\href
  {https://doi.org/10.1103/PhysRevLett.13.508} {\bibfield  {journal} {\bibinfo
  {journal} {Phys. Rev. Lett.}\ }\textbf {\bibinfo {volume} {13}},\ \bibinfo
  {pages} {508} (\bibinfo {year} {1964})}\BibitemShut {NoStop}%
\bibitem [{\citenamefont {Pekker}\ and\ \citenamefont {Varma}(2015)}]{varma}%
  \BibitemOpen
  \bibfield  {author} {\bibinfo {author} {\bibfnamefont {D.}~\bibnamefont
  {Pekker}}\ and\ \bibinfo {author} {\bibfnamefont {C.~M.}\ \bibnamefont
  {Varma}},\ }\bibfield  {title} {\bibinfo {title} {{Amplitude/{H}iggs {M}odes
  in {C}ondensed {M}atter {P}hysics}},\ }\href
  {https://doi.org/10.1146/annurev-conmatphys-031214-014350} {\bibfield
  {journal} {\bibinfo  {journal} {Annu. Rev. Condens. Matter Phys.}\ }\textbf
  {\bibinfo {volume} {6}},\ \bibinfo {pages} {269} (\bibinfo {year}
  {2015})}\BibitemShut {NoStop}%
\bibitem [{\citenamefont {Klein}\ and\ \citenamefont
  {Dierker}(1984)}]{klein84}%
  \BibitemOpen
  \bibfield  {author} {\bibinfo {author} {\bibfnamefont {M.~V.}\ \bibnamefont
  {Klein}}\ and\ \bibinfo {author} {\bibfnamefont {S.~B.}\ \bibnamefont
  {Dierker}},\ }\bibfield  {title} {\bibinfo {title} {{Theory of Raman
  scattering in superconductors}},\ }\href
  {https://doi.org/10.1103/PhysRevB.29.4976} {\bibfield  {journal} {\bibinfo
  {journal} {Phys. Rev. B}\ }\textbf {\bibinfo {volume} {29}},\ \bibinfo
  {pages} {4976} (\bibinfo {year} {1984})}\BibitemShut {NoStop}%
\bibitem [{\citenamefont {Devereaux}\ and\ \citenamefont
  {Hackl}(2007)}]{dev07}%
  \BibitemOpen
  \bibfield  {author} {\bibinfo {author} {\bibfnamefont {T.~P.}\ \bibnamefont
  {Devereaux}}\ and\ \bibinfo {author} {\bibfnamefont {R.}~\bibnamefont
  {Hackl}},\ }\bibfield  {title} {\bibinfo {title} {{Inelastic light scattering
  from correlated electrons}},\ }\href
  {https://doi.org/10.1103/RevModPhys.79.175} {\bibfield  {journal} {\bibinfo
  {journal} {Rev. Mod. Phys.}\ }\textbf {\bibinfo {volume} {79}},\ \bibinfo
  {pages} {175} (\bibinfo {year} {2007})}\BibitemShut {NoStop}%
\bibitem [{\citenamefont {Cea}\ \emph {et~al.}(2016)\citenamefont {Cea},
  \citenamefont {Castellani},\ and\ \citenamefont {Benfatto}}]{cea16}%
  \BibitemOpen
  \bibfield  {author} {\bibinfo {author} {\bibfnamefont {T.}~\bibnamefont
  {Cea}}, \bibinfo {author} {\bibfnamefont {C.}~\bibnamefont {Castellani}},\
  and\ \bibinfo {author} {\bibfnamefont {L.}~\bibnamefont {Benfatto}},\
  }\bibfield  {title} {\bibinfo {title} {{Nonlinear optical effects and
  third-harmonic generation in superconductors: Cooper pairs versus Higgs mode
  contribution}},\ }\href {https://doi.org/10.1103/PhysRevB.93.180507}
  {\bibfield  {journal} {\bibinfo  {journal} {Phys. Rev. B}\ }\textbf {\bibinfo
  {volume} {93}},\ \bibinfo {pages} {180507} (\bibinfo {year}
  {2016})}\BibitemShut {NoStop}%
\bibitem [{\citenamefont {Jujo}(2018)}]{juj18}%
  \BibitemOpen
  \bibfield  {author} {\bibinfo {author} {\bibfnamefont {T.}~\bibnamefont
  {Jujo}},\ }\bibfield  {title} {\bibinfo {title} {{Quasiclassical Theory on
  Third-Harmonic Generation in Conventional Superconductors with Paramagnetic
  Impurities}},\ }\href {https://doi.org/10.7566/JPSJ.87.024704} {\bibfield
  {journal} {\bibinfo  {journal} {J. Phys. Soc. Jpn.}\ }\textbf {\bibinfo
  {volume} {87}},\ \bibinfo {pages} {024704} (\bibinfo {year}
  {2018})}\BibitemShut {NoStop}%
\bibitem [{\citenamefont {Silaev}(2019)}]{sil19}%
  \BibitemOpen
  \bibfield  {author} {\bibinfo {author} {\bibfnamefont {M.}~\bibnamefont
  {Silaev}},\ }\bibfield  {title} {\bibinfo {title} {Nonlinear electromagnetic
  response and higgs-mode excitation in bcs superconductors with impurities},\
  }\href {https://doi.org/10.1103/PhysRevB.99.224511} {\bibfield  {journal}
  {\bibinfo  {journal} {Phys. Rev. B}\ }\textbf {\bibinfo {volume} {99}},\
  \bibinfo {pages} {224511} (\bibinfo {year} {2019})}\BibitemShut {NoStop}%
\bibitem [{\citenamefont {Tsuji}\ and\ \citenamefont {Nomura}(2020)}]{tsu20}%
  \BibitemOpen
  \bibfield  {author} {\bibinfo {author} {\bibfnamefont {N.}~\bibnamefont
  {Tsuji}}\ and\ \bibinfo {author} {\bibfnamefont {Y.}~\bibnamefont {Nomura}},\
  }\bibfield  {title} {\bibinfo {title} {{Higgs-mode resonance in third
  harmonic generation in NbN superconductors: Multiband electron-phonon
  coupling, impurity scattering, and polarization-angle dependence}},\ }\href
  {https://doi.org/10.1103/PhysRevResearch.2.043029} {\bibfield  {journal}
  {\bibinfo  {journal} {Phys. Rev. Res.}\ }\textbf {\bibinfo {volume} {2}},\
  \bibinfo {pages} {043029} (\bibinfo {year} {2020})}\BibitemShut {NoStop}%
\end{thebibliography}%

\end{document}